\documentclass[preprint2,times,tighten]{aastex6}

\usepackage{amssymb,amsmath,graphicx,natbib,hyperref}

\usepackage{color}

\begin{document}
\newcommand{\hi}          {\mbox{\rm H{\small I}}}
\newcommand{\HI}         {\hi}
\newcommand{\hii}         {\mbox{\rm H{\small II}}}
\newcommand{\htwo}        {\mbox{H$_{2}$}}
\newcommand{\jone}        {\mbox{$J=1-0$}}
\newcommand{\jtwo}        {\mbox{$J=2-1$}}
\newcommand{\jthree}        {\mbox{$J=3-2$}}
\newcommand{\jfour}        {\mbox{$J=4-3$}}
\newcommand{\um}          {$\mu$m}
\newcommand{\ha}          {H$\alpha$}
\newcommand{\kmpers}      {\mbox{\rm km~s$^{-1}$}}
\newcommand{\kms} {\kmpers}
\newcommand{\percmcu}     {\mbox{\rm cm$^{-3}$}}
\newcommand{\msun}        {\mbox{\rm M$_\odot$}}
\newcommand{\msunperpcsq} {\mbox{\rm M$_\odot$~pc$^{-2}$}}
\newcommand{\msunperpcsqyr} {\mbox{\rm M$_\odot$~pc$^{-2}$~yr$^{-1}$}}
\newcommand{\msunperyr}   {\mbox{\rm M$_\odot$~yr$^{-1}$}}
\newcommand{\msunperpccu} {\mbox{\rm M$_\odot$~pc$^{-3}$}}
\newcommand{\msunperyrkpcsq} {\mbox{\rm M$_\odot$~yr$^{-1}$~kpc$^{-2}$}}
\newcommand{\xco}         {\mbox{$X_{\rm CO}$}}
\newcommand{\xcot}         {\mbox{$X_{\rm CO,20}$}}
\newcommand{\aco}         {\mbox{$\alpha_{\rm CO}$}}
\newcommand{\xcounits}    {\mbox{\rm cm$^{-2}$(K km s$^{-1}$)$^{-1}$}}
\newcommand{\acounits}  {\mbox{\rm M$_\odot$ (K km s$^{-1}$ pc$^2$)$^{-1}$}}
\newcommand{\Lcounits}  {\mbox{\rm K km s$^{-1}$ pc$^2$}}
\newcommand{\Kkmpers}     {\mbox{\rm K km s$^{-1}$}}
\newcommand{\Kkmperspcsq} {\mbox{\rm K km s$^{-1}$ pc$^2$}}
\newcommand{\co}          {\mbox{$^{12}$CO}}
\newcommand{\cothree}          {\mbox{$^{13}$CO}}
\newcommand{\Ico}         {\mbox{I$_{\rm CO}$}}
\newcommand{\av}          {\mbox{$A_V$}}
\newcommand{\percmsq}     {\mbox{cm$^{-2}$}}
\newcommand{\cii}         {\mbox{\rm [C{\small II}]}}
\newcommand{\nii}         {\mbox{\rm [N{\small II}]}}
\newcommand{\ci}         {\mbox{\rm [C{\small I}]}}
\newcommand{\wco}         {\mbox{\rm W(CO)}}
\newcommand{\fscii}       {($^2$P$_{3/2}\rightarrow^2$P$_{1/2}$)}
\newcommand{\Smol}        {\mbox{$\Sigma_{\rm mol}$}}
\newcommand{\Mmol}        {\mbox{${\rm M}_{\rm mol}$}}
\newcommand{\Ssfr}        {\mbox{$\Sigma_{\rm SFR}$}}
\newcommand{\Sgmc}        {\mbox{$\Sigma_{\rm GMC}$}}
\newcommand{\Sstar}       {\mbox{$\Sigma_*$}}
\newcommand{\Lco}         {\mbox{$L_{\rm CO}$}}
\newcommand{\Rhmol}       {\mbox{$R^{\rm mol}_{\rm 1/2}$}}
\newcommand{\Rhst}       {\mbox{$R^*_{\rm 1/2}$}}
\newcommand{\ag}{\mbox{ \raisebox{-.4ex}{$\stackrel{\textstyle >}{\sim}$} }}
\newcommand{\al}{\mbox{ \raisebox{-.4ex}{$\stackrel{\textstyle <}{\sim}$} }}
\newcommand{\dgr}         {\mbox{$\delta_{\rm DGR}$}}
\newcommand{\dgrp}         {\mbox{$\delta_{\rm DGR}'$}}
\newcommand{\rco}{R_{\rm CO}}
\newcommand{\rht}{R_{\rm H_2}}
\newcommand{\davdg}{\Delta A_{V}}
\newcommand{\rhogas}{\mbox{$\rho_{\rm gas}$}}
\newcommand{\tsf}{\mbox{$t_{\rm SF}$}}
\newcommand{\tj}{\mbox{$t_{\rm dyn}$}}
\newcommand{\tdep}{\mbox{$\tau_{\rm dep,mol}$}}
\newcommand{\siggas}{\mbox{$\Sigma_{\rm gas}$}}
\newcommand{\sigsfr}{\mbox{$\Sigma_{\rm SFR}$}}
\newcommand{\sighi}{\mbox{$\Sigma_{\rm HI}$}}
\newcommand{\sightwo}{\mbox{$\Sigma_{\rm H_2}$}}
\newcommand{\eightmu}{\mbox{8\,$\mu$m}}
\newcommand{\xcou}{\mbox{cm$^{-2}$~(K~km~s$^{-1}$)$^{-1}$}}
\newcommand{\Rtt}{${\cal R}_{12/13}$}
\newcommand{\tdepmol}{\mbox{$\tau_{\rm dep,mol}$}}

\title{The EDGE-CALIFA Survey: Interferometric Observations of 126 Galaxies with CARMA} 

\author{Alberto D. Bolatto\altaffilmark{1}, Tony Wong\altaffilmark{2}, Dyas Utomo\altaffilmark{3}, Leo Blitz\altaffilmark{3}, Stuart N. Vogel\altaffilmark{1}, Sebasti\'an F. S\'anchez\altaffilmark{4}, Jorge Barrera-Ballesteros\altaffilmark{5}, Yixian Cao\altaffilmark{2}, Dario Colombo\altaffilmark{6}, Helmut Dannerbauer\altaffilmark{7}, Rub\'en Garc\'{\i}a-Benito\altaffilmark{8}, Rodrigo Herrera-Camus\altaffilmark{9}, 
Bernd Husemann\altaffilmark{10}, 
Veselina Kalinova\altaffilmark{6}, Adam K. Leroy\altaffilmark{11}, Gigi Leung\altaffilmark{10}, Rebecca C. Levy\altaffilmark{1}, Dami\'an Mast\altaffilmark{12}, Eve Ostriker\altaffilmark{13}, Erik Rosolowsky\altaffilmark{14}, Karin M. Sandstrom\altaffilmark{15}, Peter Teuben\altaffilmark{1}, Glenn van de Ven\altaffilmark{10}, \& Fabian Walter\altaffilmark{10}}

\shorttitle{The CARMA EDGE-CALIFA Survey}
\shortauthors{A. D. Bolatto, T. Wong,  et al.}

\altaffiltext{1}{Department of Astronomy, University of Maryland, College Park, MD 20742, USA (email: bolatto@umd.edu)}
\altaffiltext{2}{Department of Astronomy, University of Illinois, Urbana, IL 61801, USA}
\altaffiltext{3}{Department of Astronomy, University of California, Berkeley, CA 94720, USA}
\altaffiltext{4}{Instituto de Astronom\'\i a, Universidad Nacional Aut\'onoma de M\'exico, A.P. 70-264, 04510 M\'exico, D.F.,  Mexico}
\altaffiltext{5}{Department of Physics and Astronomy, Johns Hopinks University, Baltimore, MD 21218, USA}
\altaffiltext{6}{Max Planck Institut f\"ur Radioastronomie, D-53010 Bonn, Germany}
\altaffiltext{7}{Instituto de Astrof\'{\i}sica de Canarias, E-38205 La Laguna, and Departamento de Astrof\'{\i}sica, Universidad de La Laguna, E-38206 La Laguna, Tenerife, Spain}
\altaffiltext{8}{Instituto de Astrof\'{\i}sica de Andaluc\'{\i}a, CSIC, E-18008 Granada, Spain}
\altaffiltext{9}{Max Planck Institute f\"ur Extraterrestrische Physik, D-85741 Garching bei M\"unchen, Germany}
\altaffiltext{10}{Max-Planck-Institut f\"ur Astronomie, K\"onigstuhl 17, D-69117
Heidelberg, Germany}
\altaffiltext{11}{Department of Astronomy, The Ohio State University, Columbus, OH 43210, USA}
\altaffiltext{12}{Observatorio Astron\'omico de C\'ordoba, 5000 C\'ordoba, C\'ordoba, Argentina}
\altaffiltext{13}{Department of Astrophysical Sciences, Princeton University, Princeton, New Jersey 08544, USA}
\altaffiltext{14}{Department of Physics, University of Alberta, Edmonton, AB, Canada}
\altaffiltext{15}{Department of Physics, University of California, San Diego, CA 92093, USA}

\begin{abstract}
We present interferometric CO observations made with the Combined Array for Millimeter-wave Astronomy (CARMA) of galaxies from the Extragalactic Database for Galaxy Evolution survey (EDGE). These galaxies are selected from the Calar Alto Legacy Integral Field Area (CALIFA) sample, mapped with optical integral field spectroscopy.
EDGE provides good quality CO data (3$\sigma$ sensitivitity $\Sigma_{\rm mol}\sim11$\,\msunperpcsq\ before inclination correction, resolution $\sim1.4$\,kpc) for 126 galaxies, constituting the largest interferometric CO survey of galaxies in the nearby universe. We describe the survey, the data characteristics, the data products, and present initial science results. We find that the exponential scale-lengths of the molecular, stellar, and star-forming disks are approximately equal, and galaxies that are more compact in molecular gas than in stars tend to show signs of interaction. We characterize the molecular to stellar ratio as a function of Hubble type and stellar mass, present preliminary results on the resolved relations between the molecular gas, stars, and star formation rate, and discuss the dependence of the resolved molecular depletion time on stellar surface density, nebular extinction, and gas metallicity. EDGE provides a key dataset to address outstanding topics regarding gas and its role in star formation and galaxy evolution, which will be publicly available on completion of the quality assessment. 
%
%
\end{abstract}


\section{Introduction}

Over the past few decades, increasingly powerful optical surveys of galaxies have been used to study the process of structure formation in the Universe. These surveys investigate how the universe started from the very smooth state imprinted on the Cosmic Microwave Background radiation and evolved into the ``cosmic web,'' primitive galaxies, and ultimately galaxies as they exist today. Spectroscopic surveys in particular have revealed clear trends in star formation, metal enrichment, stellar populations, and nuclear activity.  Yet large-scale spectroscopic surveys largely neglect the internal structure of galaxies, which is key to their evolution.  An era of integral field unit (IFU) spectroscopy is now upon us, providing simultaneous spectral and spatial coverage and resolution. These data allow us to map gas and stellar metallicities, ionized gas and stellar dynamics, extinctions, extinction-corrected star formation rates, stellar mass densities, and ages.  Coupling these results with imaging spectroscopy of molecular gas from millimeter interferometers offers a new window for studying the baryon cycle in galaxies at redshift $z=0-1$, the epoch that anchors our understanding of galaxy evolution and during which the overall cosmic star formation rate declined significantly \citep{Madau:14}.

We undertook EDGE (the Extragalactic Database for Galaxy Evolution) to explore the complementary potential of combined optical IFU surveys and millimeter-wave interferometry. EDGE was
one of a few large, ambitious legacy programs 
completed by the CARMA\footnote{Combined Array for Research in Millimeter-wave Astronomy}  interferometer \citep{Bock:06} between November 2014 and the closing of operations on April 2015. The CARMA EDGE survey comprises 126 infrared-selected galaxies from the Calar Alto Legacy Integral Field Area (CALIFA) IFU sample \citep{Sanchez:12} imaged in $^{12}$CO and $^{13}$CO at good sensitivity and with an angular resolution and field-of-view matched to CALIFA. It thus joins the power of one of the pre-eminent optical integral field area surveys with the largest and most uniform CO interferometric survey in existence.


\begin{figure*}[t] 
\begin{center}
\includegraphics[height=0.96\textheight]{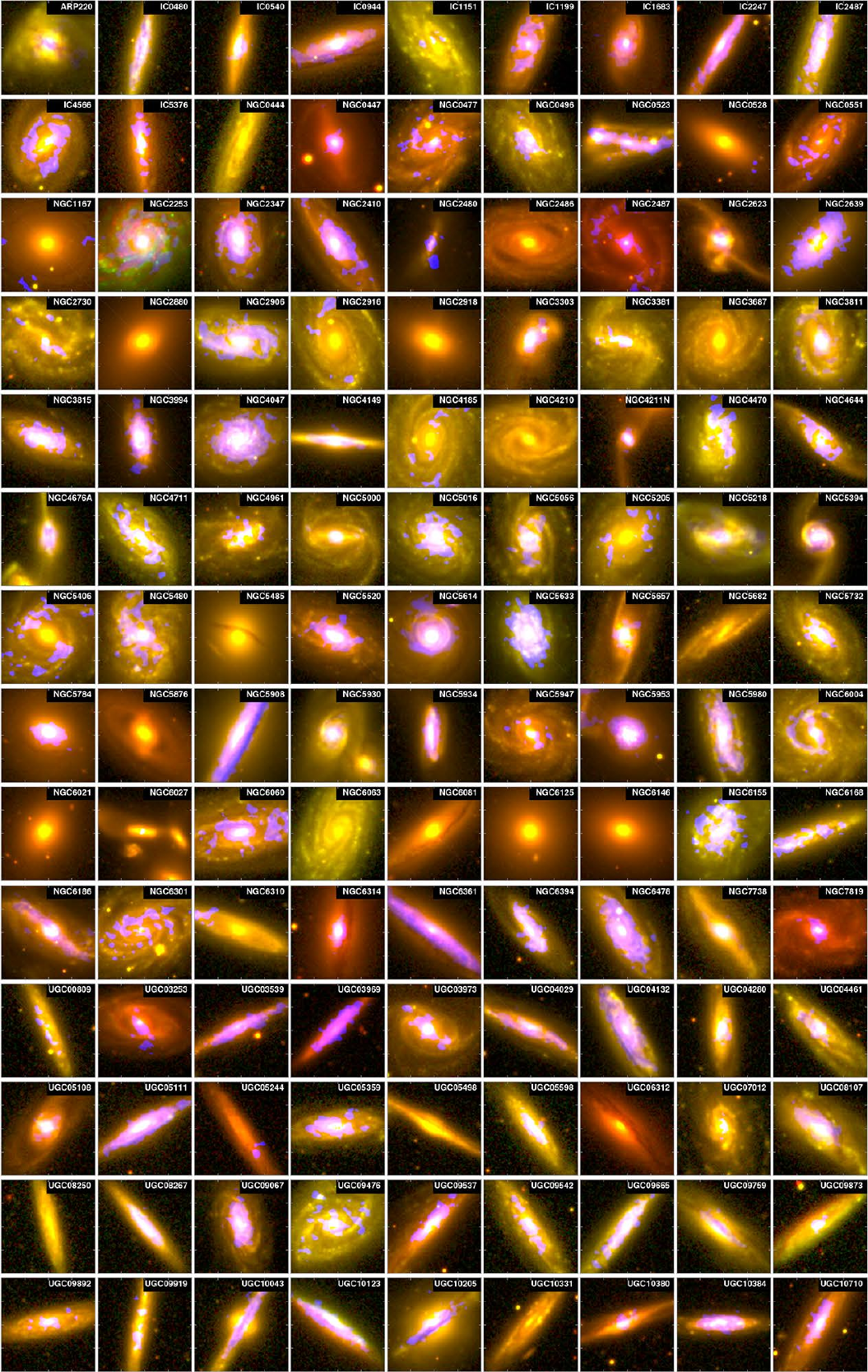} 
\end{center}
\caption{A menagerie of stellar and molecular distributions present in local
        galaxies. SDSS $g$ (green channel) and $i$ (red channel) composite images
        for the 126 high-resolution CO galaxies in CARMA EDGE. The CO
        intensity map is shown in the blue channel. Local galaxies are very
        diverse in their CO properties, and understanding the mechanisms behind
       that diversity is the goal of EDGE.  
\label{fig:sdssim}} 
\end{figure*}

\subsection{Motivation for a Survey}

The evolution of galaxies is fundamentally linked to the rate and efficiency of
star formation: some of the clearest trends along the Hubble sequence are trends
in gas fraction and star formation history. Since star formation occurs in the
dense, molecular (H$_2$) phase of the interstellar medium (ISM), it is
essential to identify the physical processes that govern the molecular
gas content of galaxies and the rate at which that gas is converted into stars.
Despite considerable progress in this area \citep[e.g.,][]{Leroy:08}, our conceptual understanding has been hampered by the difficulty of connecting global studies, which sample large
numbers of galaxies but are unresolved, with spatially-resolved studies, which
are frequently limited to a handful of prototypes and do not span a significant
range of masses, environments, and types.  The combination of CALIFA and EDGE is
designed to overcome these limitations, providing a unique opportunity to address
key outstanding questions in galaxy structure and evolution. Some of the topics that can be uniquely investigated using this combination of millimeter-wave interferometry and IFU with complete optical spectral coverage are:

\paragraph{What local factors regulate the conversion of H$_2$ into stars?}

In discussing the conversion of gas into stars, a useful concept is the gas
depletion time $\tau_{\rm dep} \equiv \Sigma_{\rm gas}/\dot{\Sigma}_*$, where $\Sigma_{\rm gas}$
and $\dot{\Sigma}_*$ refer to the surface densities of gas mass and star
formation rate respectively. Since star formation involves gravitational
collapse, it is natural to associate $\tau_{\rm dep}$ with some dynamical time
set by gravity. On the scale of a molecular cloud, this dynamical time is often the
free-fall time $\tau_{\rm ff}=(G\rho)^{-1/2}$, while for galaxy scales a more
appropriate time scale might be the orbital time, $\tau_{\rm orb}=\Omega^{-1}$,
or the vertical dynamical time, $\tau_{\rm ver}=(G\,\rho_{\rm tot})^{-1/2}$,
with $\rho_{\rm tot}=\rho_*+\rho_{\rm gas}$ where $\rho_*$ and $\rho_{\rm gas}$ are the mean mid-plane stellar and gas volume densities respectively, as determined by the respective surface densities and velocity dispersions
\citep[e.g.,][]{KKO:11}. A long-standing conundrum is that these
dynamical times are all much shorter than the typical values of $\tau_{\rm dep}$
observed in normal galaxies, even when only the molecular gas is included in $\Sigma_{\rm gas}$ (thus resulting in \tdepmol).
This reflects the ``inefficiency'' of star formation and suggests that feedback
mechanisms throttle the rate of star formation well below its ``natural''
rate. We still lack an understanding of the spatial and temporal scales on
which these mechanisms operate, and what are their environmental dependencies.  By
combining extinction-corrected star formation rates from IFU spectroscopy with
measurements of gas surface densities, velocity dispersions, and disk kinematics
for a representative galaxy sample, EDGE constitutes a key dataset for testing
galaxy-scale star formation models.

\paragraph{Is molecular gas structured differently in extreme environments?}

At any given time, only a small fraction of the ISM is cold and dense enough to
collapse into stars \citep[e.g.,][]{Fukui:10}.  Traditionally H$_2$, and particularly CO-emitting H$_2$,
has been identified with gravitationally bound giant molecular clouds (GMCs). It has become increasingly clear, however, that not all CO
emission may originate from dense gas, and that a significant portion of the
molecular ISM may lie outside the gravitationally-bound structures that are the
sites for star formation \citep{Liszt:10,Pety:13}. Moreover, studies of extreme
environments such as the Central Molecular Zone (CMZ) of the Milky Way
\citep{Kruijssen:14} and actively star-forming galaxies at high redshift
\citep{Swinbank:11} suggest that in those environments very high gas densities must be reached for
gravitational collapse to ensue ($n_{\rm H} \sim 10^7$--$10^8$ cm$^{-3}$
compared to $\sim$10$^{5}$ cm$^{-3}$ in nearby GMCs). This is also observed in nearby galaxies, where dense gas fractions are higher in the central regions, but the star formation per unit dense gas appears to drop significantly in the same regions \citep{Usero:15,Bigiel:16}.
Diffuse, unbound
molecular gas not immediately available to star formation may dominate both the mass and volume of the ISM in these extreme
environments.  By including gas that is not directly involved in star formation,
CO measurements may be overestimating $\tau_{\rm dep}$ and thus inferring too
low a star formation efficiency.  Using resolved measurements of the
$^{12}$CO/$^{13}$CO ratio from EDGE as an indicator of the CO line opacity, and spatially resolved \hi\ measurements to
compute \hi/CO ratios as an indicator of the atomic/molecular phase balance, we can
characterize the cold ISM structure on kilo-parsec scales
across a wide variety of galactic systems. Only a small fraction of the EDGE galaxies have interferometric \hi\ observations available
at this time, but many of the sources south of $\delta=+30^\circ$ will be available to the Australia Square Kilometer Array Pathfinder (ASKAP) \hi\ All-Sky Survey WALLABY \citep{Koribalski:12}, and all of them can be observed with the Jansky Very Large Array (JVLA).

\begin{figure*}[t]
\begin{center}
\includegraphics[width=\textwidth]{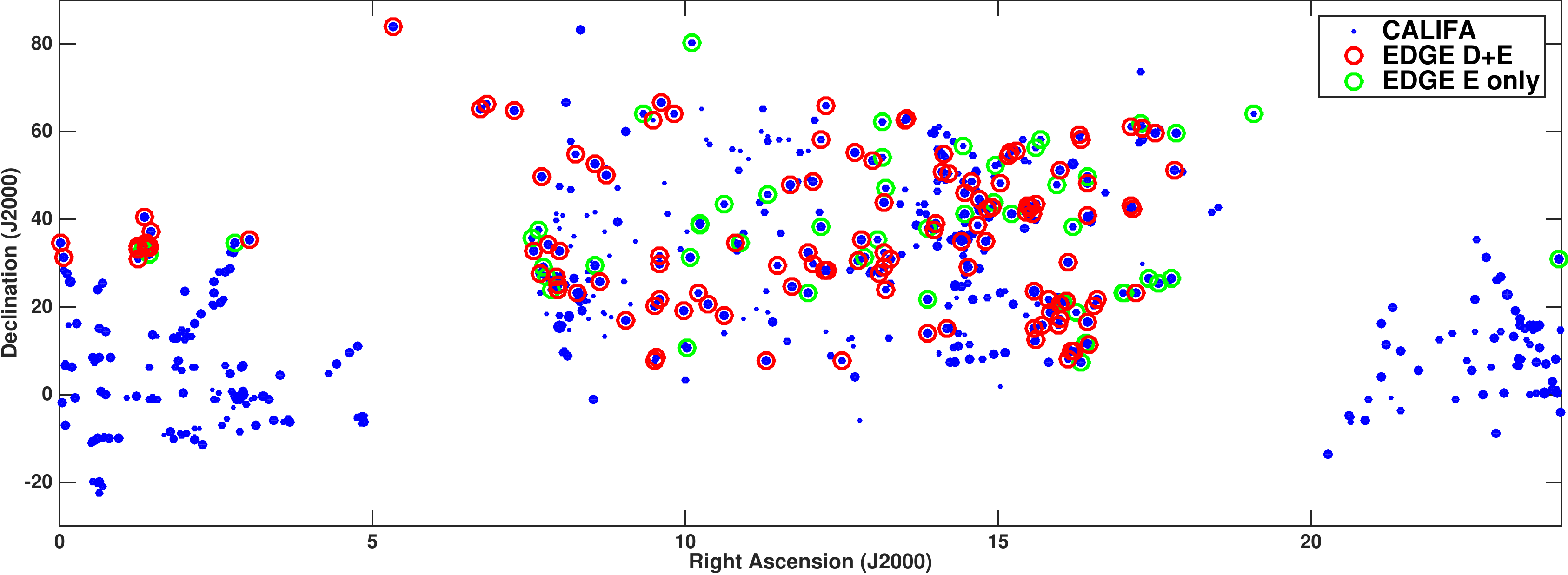}
\end{center}
\caption{The distribution of EDGE and CALIFA galaxies in the sky. The filled circles indicate the distribution of observed CALIFA galaxies (data release 3). The size of the symbol is proportional to the logarithm of the SFR, as determined from extinction-corrected H$\alpha$ (using the Balmer decrement). The red circles indicate the galaxies with EDGE D+E observations. The green circles show the galaxies with only short CARMA E array observations, which were not followed up in CARMA D array because they did not show obvious CO $J = 1 - 0$ emission.}
\label{fig:radec}
\end{figure*}

\begin{figure*}[t] 
\begin{center}
\includegraphics[width=\textwidth]{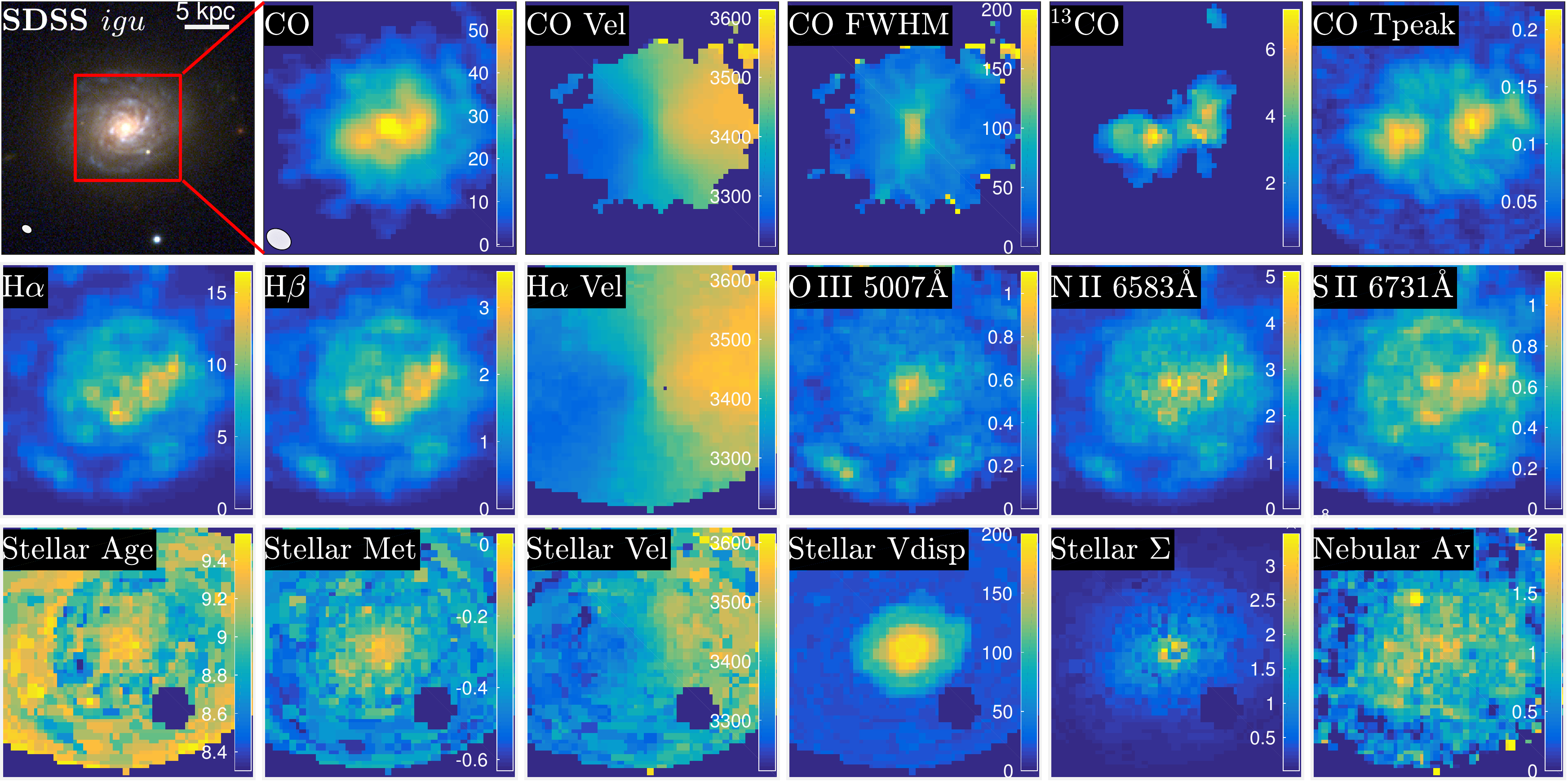} 
\end{center}

\caption{Example of the broad range of data available for each galaxy
        in the EDGE sample. The top left panel shows the $2\arcmin\times2\arcmin$ multicolor {\em igu}
        composite for NGC~4047, with a 50\arcsec\ box in red zoomed in for the rest of the panels. The CARMA
        synthesized beam (4.4\arcsec) is shown in the bottom left corner of the
        second panel. The EDGE data products include CO intensity (\Kkmpers), CO
        velocity (\kmpers), CO line-width (\kmpers), $^{13}$CO intensity (\Kkmpers), and CO peak temperature (K).
        A handful of the CALIFA measurements (from {\sc Pipe3D}) are displayed in the
        second row: line fluxes for H$\alpha$, H$\beta$, [OIII] 5007\AA, [NII] 6583\AA, and [SII] 6731\AA\ (in $10^{-16}$\,erg\,s$^{-1}$\,cm$^{-2}$), together with the H$\alpha$ velocity field (\kmpers).  The third row shows a few of the CALIFA {\sc Pipe3D} data products: luminosity-weighted age of the stellar population (logarithmic years), luminosity-weighted stellar metallicity (logarithmic solar units), stellar velocity field (\kmpers), stellar velocity dispersion (\kmpers), extinction corrected stellar mass surface density (logarithmic \msunperpcsq), and nebular extinction computed from the Balmer decrement (mag). 
\label{fig:multiNGC4047}}
\end{figure*}

\paragraph{How do galaxies grow and age?}

The conventional view that galaxies begin their lives as gas-rich disks that
slowly convert their baryonic mass to stars, only later to be transformed into
``red and dead'' ellipticals by major mergers, has been increasingly displaced
by the view that major mergers may not be an important avenue of galaxy growth,
whereas continued accretion of gas is vital for maintaining star formation
\citep[e.g.,][]{Lilly:13}. At the same time, gas can be driven outwards by a
powerful starburst or active galactic nucleus (AGN) before star formation can
run to completion, leading to the relative rarity of very massive galaxies \citep[e.g.,][]{Baldry:08}.  In this emerging picture, the conversion of gas into stars
is ultimately regulated by net flows in or out of the galaxy.  
It has been proposed that star-formation in galaxies either halts smoothly through a slow aging process as cold gas inflow decreases slowly and systems become chemically old, or it quenches on short time-scales in objects in high-density environments, creating a bimodality in the stellar population properties \citep[e.g.,][]{casado15}. This scenario is perhaps too simple, as it ignores the fact that star-formation is a local process in which aging and quenching depend on galactocentric distance, generating an internal age bimodality \citep[e.g.,][]{Zibetti:17} and local relations between the observed parameters \citep[e.g.][]{rosales12,sanchez13,cano16}. With the EDGE and CALIFA data we can explore locally the star-formation process in comparison with the star-formation and metal enrichment histories, looking to determine if the aging/quenching dichotomy is due to a smooth decrease or a fast removal of the gas content, or perhaps it is induced by local processes, such as gas heating \cite[e.g.,][]{Forbes:16}.

Large single-dish single-ponting CO and \HI\ surveys have made seminal
contributions to understanding how gas content varies with galactic mass \citep[FCRAO Survey, COLD GASS;][]{Young:95,Saintonge:11a}. But
spatially resolved data are critical for understanding how gas actually enters
and exits the galaxy, how it is transported within galaxies, and how it locally leads to star formation. These are questions central to efforts such as THINGS/HERACLES, which have played a key role in our current understanding of these processes in local galaxies \citep{Walter:08,Leroy:13b}.  The combination of CALIFA and EDGE brings a full suite of diagnostics to
bear on these subjects, including both gas and stellar metallicities (which are
sensitive to the balance between star formation and gas accretion and outflow)
as well as gas and stellar kinematics, mass surface densities, and spectral
diagnostics of star formation and nuclear activity.

\begin{figure*}[t]
\begin{center}
\includegraphics[width=\textwidth]{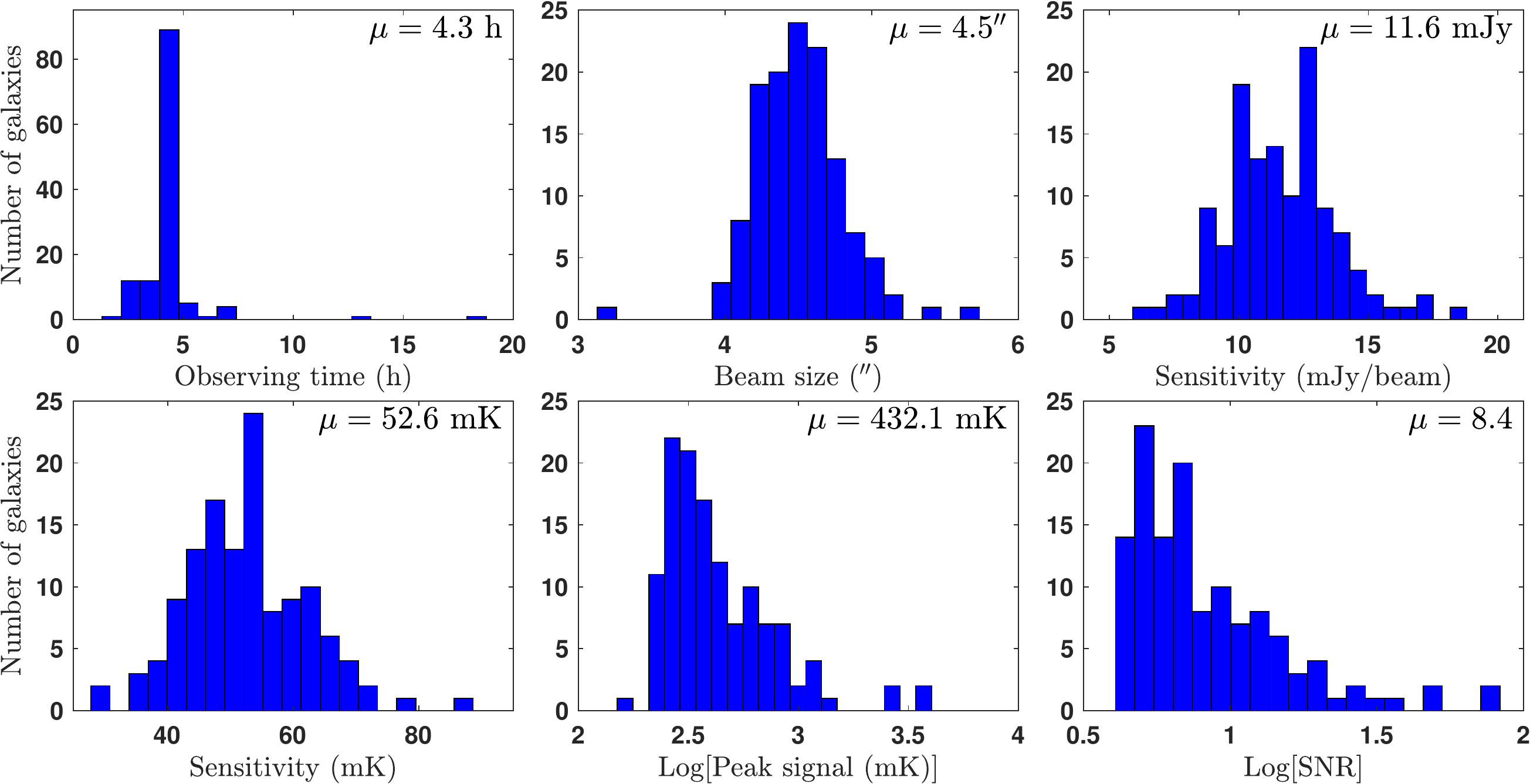}
\end{center}
\caption{Characteristics of the EDGE observations. Distributions of the observing time per source, the synthesized beam size for the resulting \co\ cube, flux density sensitivity in 10\,\kmpers-wide channels, the Rayleigh-Jeans brightness sensitivity, the signal in the brightest beam in the cube, and the signal-to-noise in the brightest beam in the cube. The legend in the top right corner shows the mean value of the parameter.}
\label{fig:surveychar}
\end{figure*}

\section{Sample, Observations, Ancillary Data, and Methods}\label{sec:edge}

\subsection{Sample Selection}

The CARMA EDGE CO $J=1-0$ survey is based on the CALIFA sample (see \S\ref{sec:califa}), but with an emphasis on infrared (IR) bright galaxies in light of the well-known correlation between IR and CO luminosity.  
The initial sample, culled from the 457 galaxies which had complete CALIFA observations in both optical gratings as of October 2014, consisted of 177 galaxies selected for high {\it WISE} 22\,$\mu$m flux and mostly concentrated close to 12 hours of right ascension.
These 177 galaxies were initially observed in the most compact (E) configuration of CARMA (typically sampling $3-30$\,k$\lambda$, corresponding to projected baselines $8-81$\,m) using snapshots with 40 minutes of integration per object. 
A sub-sample of 125 galaxies was subsequently observed in the more extended D configuration of CARMA (typically sampling $7-54$\,k$\lambda$, corresponding to baselines of $19-146$\,m).
The galaxies selected for D-array imaging included the 77 E-array targets exhibiting peak SNR in the CO line of $>$5, with the remainder chosen based on visual inspection of the E-array maps to assess the likelihood that a signal was detected (this determination favored galaxies where the brightness peaks in the maps were centrally located and/or spatially aligned). 
By themselves, the very short E-array observations suffer from limited sensitivity and coverage of the $uv$ plane, and are primarily useful to establish the likelihood of bright emission. Therefore in this paper we will not discuss further the 52 galaxies that were only observed in E-array. To the sample of 125 objects we added another CALIFA galaxy, NGC~7738, which was the object of a pilot study conducted in September 2014 for which we also obtained D+E observations, for a total of 126 galaxies. A color composite for the sample observed in D+E configurations is presented in Figure \ref{fig:sdssim}, to illustrate the variety of morphologies. We note here that NGC~5953 is part of an interacting galaxy pair, and our observations also include the companion NGC~5954 (which is also part of CALIFA), although we are not counting it here as a separate galaxy. This also occurs for NGC~5930 and its companion, NGC~5929, or in galaxy pairs or multiples such as NGC~4211, NGC~4676A, NGC~6027, and UGC~5498.

Figure \ref{fig:radec} shows the distribution of EDGE surveyed galaxies on the sky, compared with the distribution of galaxies in the parent CALIFA sample. For reasons of scheduling efficiency the EDGE galaxies are concentrated toward RA$\sim12^{\rm h}$, in the Virgo region. This allowed CARMA to carry out Galactic surveys in parallel, while most of the observing time that is almost purely extragalactic was dedicated to EDGE.

Henceforth when discussing our results we will refer to the sub-sample of 126 galaxies having both D- and E-array observations. The characteristics of the galaxies in the D+E sample are summarized in Table \ref{tab:table1}, and their optical and infrared photometry is tabulated in Table \ref{tab:table2}.  These tables will be discussed in more detail in \S3.

\begin{figure*}[t!]
\begin{center}
\includegraphics[width=\textwidth]{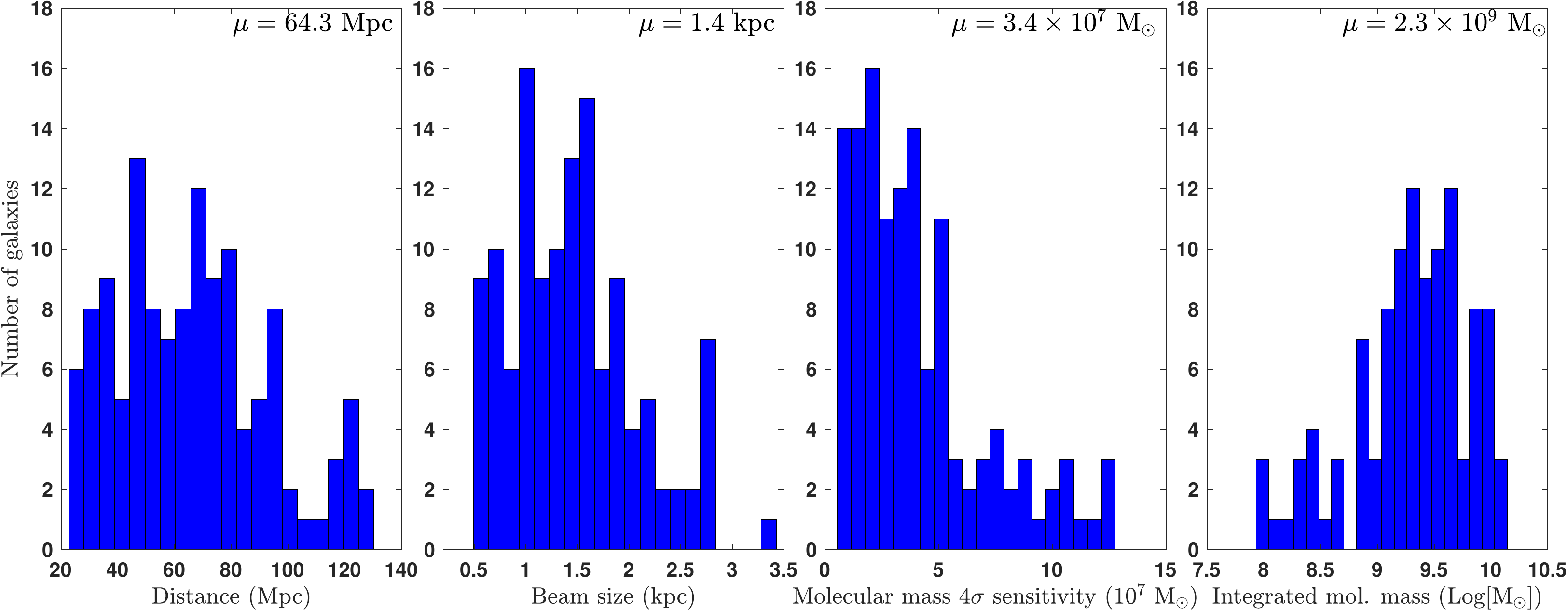}
\end{center}
\caption{Physical parameters associated with the CARMA EDGE observations. Median values for the parameters are indicated in the top right corner. {\bf Left to right:} 1) galaxy distance in Mpc, according to the luminosity distance computed from the CALIFA redshift (see \S\ref{sec:results}). 2) Physical beam size in kpc.  3) Resolved molecular mass 4$\sigma$ sensitivity. We take the typical line-width of a molecular cloud complex to be 30\,\kmpers\ on the spatial scales we observe, and assume a Galactic \aco=4.36\,\acounits\ for this calculation. 4) Distribution of integrated molecular masses for the EDGE detected galaxies. 
\label{fig:surveyphyspar}}
\end{figure*}

\subsection{Observations and Data Reduction}
\label{sec:observations}
Observations of the original 177 galaxies sample sample were conducted in CARMA's E-array in late 2014, integrating 40 minutes per galaxy. 
With the exception of 5 galaxies which were observed in the first week of observations (NGC 3106, 5029, 5485, 5520, and 5947), for which only a single central pointing was observed in E-array, all targets were observed in a 7-point hexagonal mosaic with pointings separated by 27\arcsec\ (half the primary beam width of the 10~m telescopes), yielding a half-power field-of-view (FOV) with radius $\approx$50\arcsec. 
As noted above, 125 galaxies were subsequently (from 2014 December to 2015 March) observed in the more extended D configuration, with an additional $\sim3.5$ hours of integration per target.  
For the D-array observations, a 7-point hexagonal mosaic was employed for all galaxies.

We observed in CARMA's ``snapshot'' mode, in which a list of targets is priority-ordered and the system automatically selects the highest priority target that is over a given elevation and needs integration time. Observations then proceed on a target and a nearby phase-referencing quasar until it falls below the elevation threshold, or until the requested integration time has been completed. 
To facilitate rapid switching between targets, the sources were divided into three groups based on optical redshift (1500--4000 \kms, 4000--6500 \kms, and 6500--9000 \kms), and a common, fixed tuning and correlator setup was adopted for each group.  During a typical 4-hour observation only sources in one redshift group would be observed, along with passband and flux calibrators.  The CARMA correlator was configured with five 250-MHz windows covering the $^{12}$CO line with 3.4 \kms\ resolution and a 3000 \kms\ velocity range, and three 500-MHz windows covering the $^{13}$CO line with 14.3 \kms\ resolution and a 3800 \kms\ velocity range.

The visibility data were calibrated in MIRIAD using an automated pipeline based on scripts developed for the STING galaxy survey \citep{Rahman:11,Rahman:12,Wong:13}.
Frequency-dependent (passband) gains were determined using observations of a bright ($>$8 Jy) quasar, usually 3C273 or OJ 287 (J0854+201).  The planets Mars and Uranus were the primary flux calibrators; when neither was available the compact \hii\ region MWC349 (with an adopted flux of 1.2 Jy) or the quasar 3C273 (with an adopted flux of 8--12 Jy, based on analyzing tracks on adjacent days during which both a planet and 3C273 were observed) was used for flux calibration.  Since the antenna gains are approximately known, CARMA data are provided with a default calibration, which flux calibration adjusts by a factor $\sim$1.  The distribution of derived flux calibration factors is Gaussian, with a width measured by the standard deviation of $\sigma=0.10$. This width provides a measure of the systematic uncertainty in the flux calibration scale, which would be $\pm10\%$. In cases where a nuclear continuum source was detected (ARP 220, NGC 1167, NGC 2639, NGC 6146), a first-order spectral baseline was subtracted from the visibility spectra before imaging.

The calibrated visibilities were imaged in MIRIAD and deconvolved using an implementation of SDI CLEAN designed for mosaics (task {\sc mossdi2}).  CLEAN components were searched for down to the 2.5$\sigma$ level over the entire region where the sensitivity was within a factor of 2.5 of the FOV center.  Cubes were generated with 1\arcsec\ pixels and 10 and 20 \kms\ channel spacing across a default velocity range of 860 \kms\ (extended as needed to cover broader lines in two galaxies, NGC 6027 and ARP 220). The resulting CO integrated intensity maps are shown in Fig. \ref{fig:sdssim} together with the SDSS images of the EDGE targets.

\begin{figure*}[t]
\begin{center}
\includegraphics[width=\textwidth]{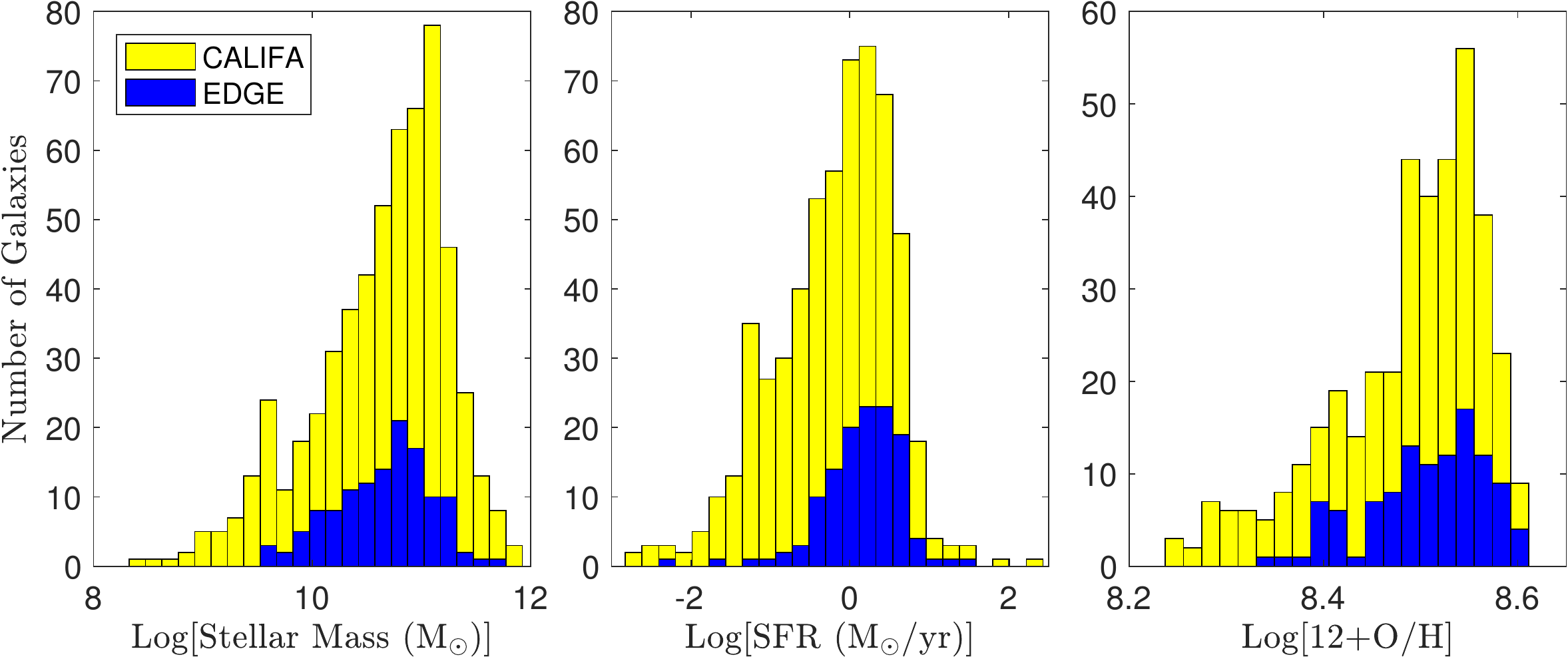}
\end{center}
\caption{Comparison between the distributions of galaxy parameters in CARMA EDGE and CALIFA DR3. The histograms show the distributions of galaxy stellar mass, star formation rate, and metallicity at the equivalent radius for both samples. EDGE is approximately representative of the larger sample, with the caveats that its range of masses is narrower, and its SFR and metallicity are biased toward the higher end of the CALIFA distribution. 
\label{fig:surveymass}}
\end{figure*}

\begin{figure*}[t] 
\begin{center}
\includegraphics[height=2.75in]{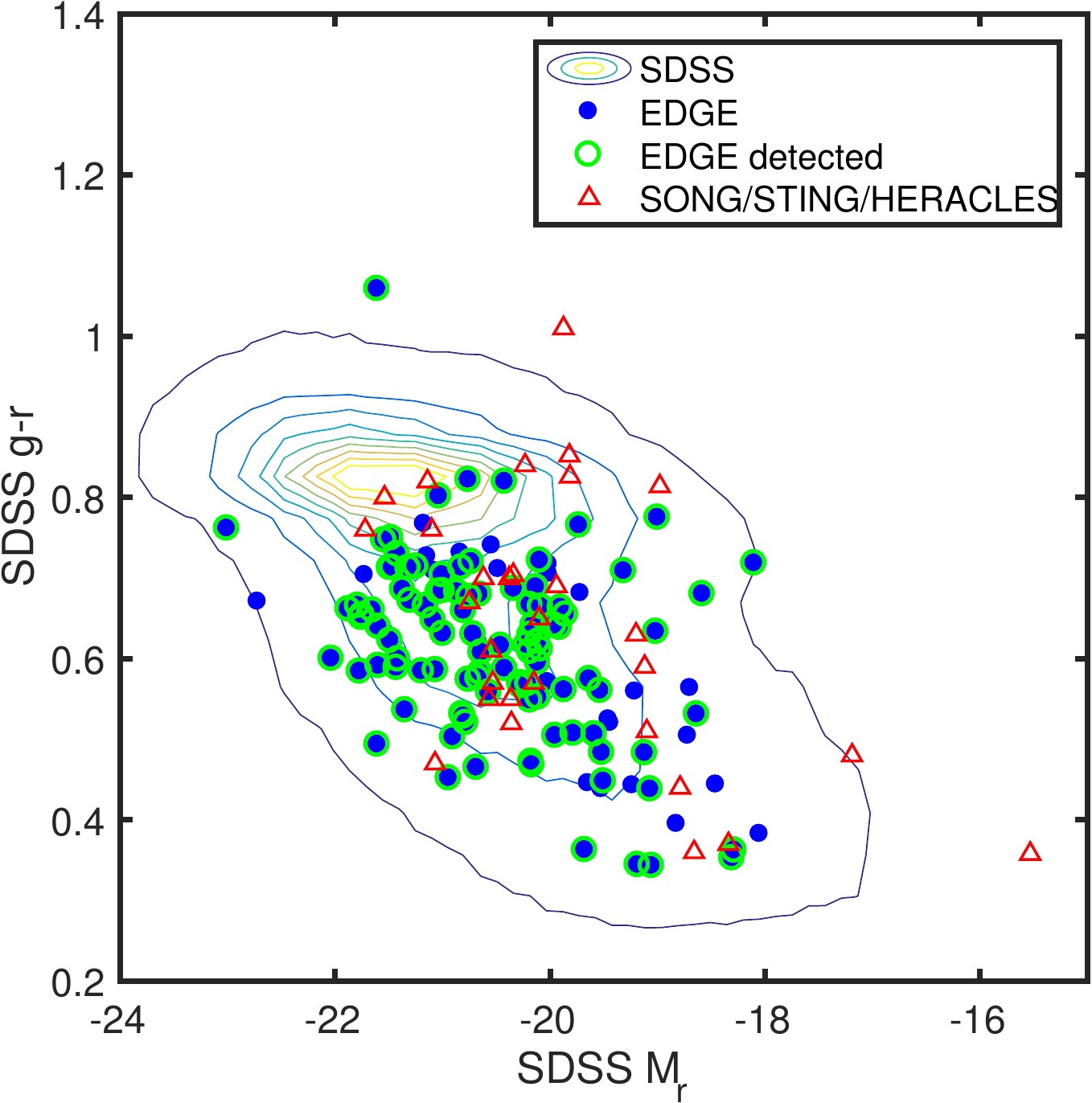}
\includegraphics[height=2.75in]{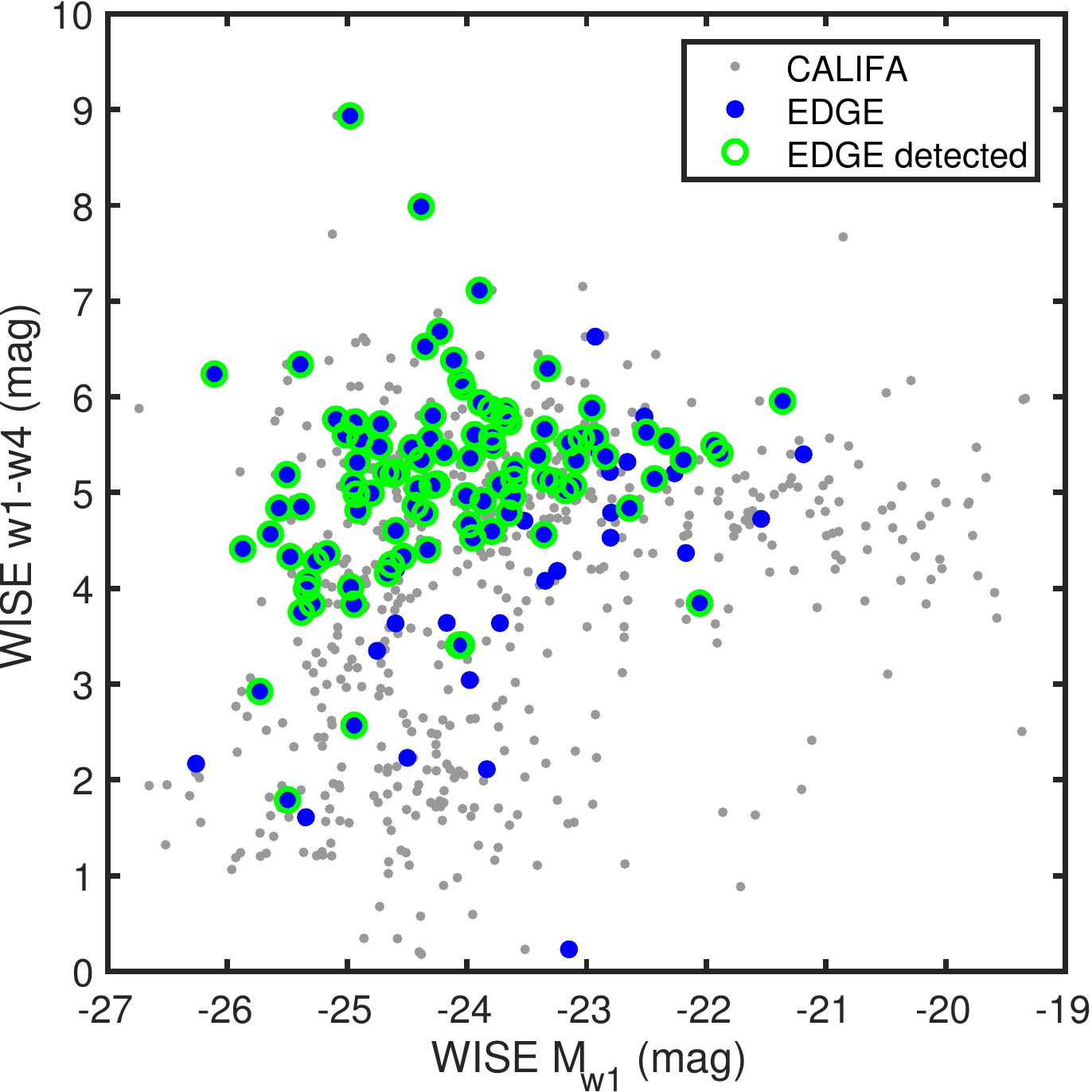} 
\end{center}
\caption{Placing the CARMA EDGE survey in the context of the entire population of galaxies. {\bf Left:} Location of EDGE galaxies (blue filled dots) in SDSS color-magnitude space. EDGE is compared with other large CO samples (open triangles, from BIMA SONG, CARMA STING, and HERACLES, for galaxies with available SDSS photometry) as well as the bulk population of SDSS galaxies \citep[contours;][]{Simard:11}.  Circled dots indicate EDGE galaxies with $>5\sigma$ detections of CO.  The CARMA EDGE galaxies do a very good job at statistically sampling the general galaxy population of blue galaxies, but do not include the red sequence. {\bf Right:} Location of EDGE galaxies (blue dots) in {\em WISE} color-magnitude space (circled dots indicate CO detections). The x-axis corresponds to absolute 3.4 $\mu$m magnitude (a proxy for stellar mass).  The y-axis corresponds to the 3.4 $\mu$m to 22 $\mu$m flux ratio (a proxy for specific star formation activity). The small dots show the CALIFA sample, which is designed to be statistically representative of the entire population. Low mass galaxies and galaxies with little 22 $\mu$m emission are underrepresented in EDGE.  
\label{fig:sample}} 
\end{figure*}

\begin{figure*}[t] 
\begin{center}
\includegraphics[width=0.95\textwidth]{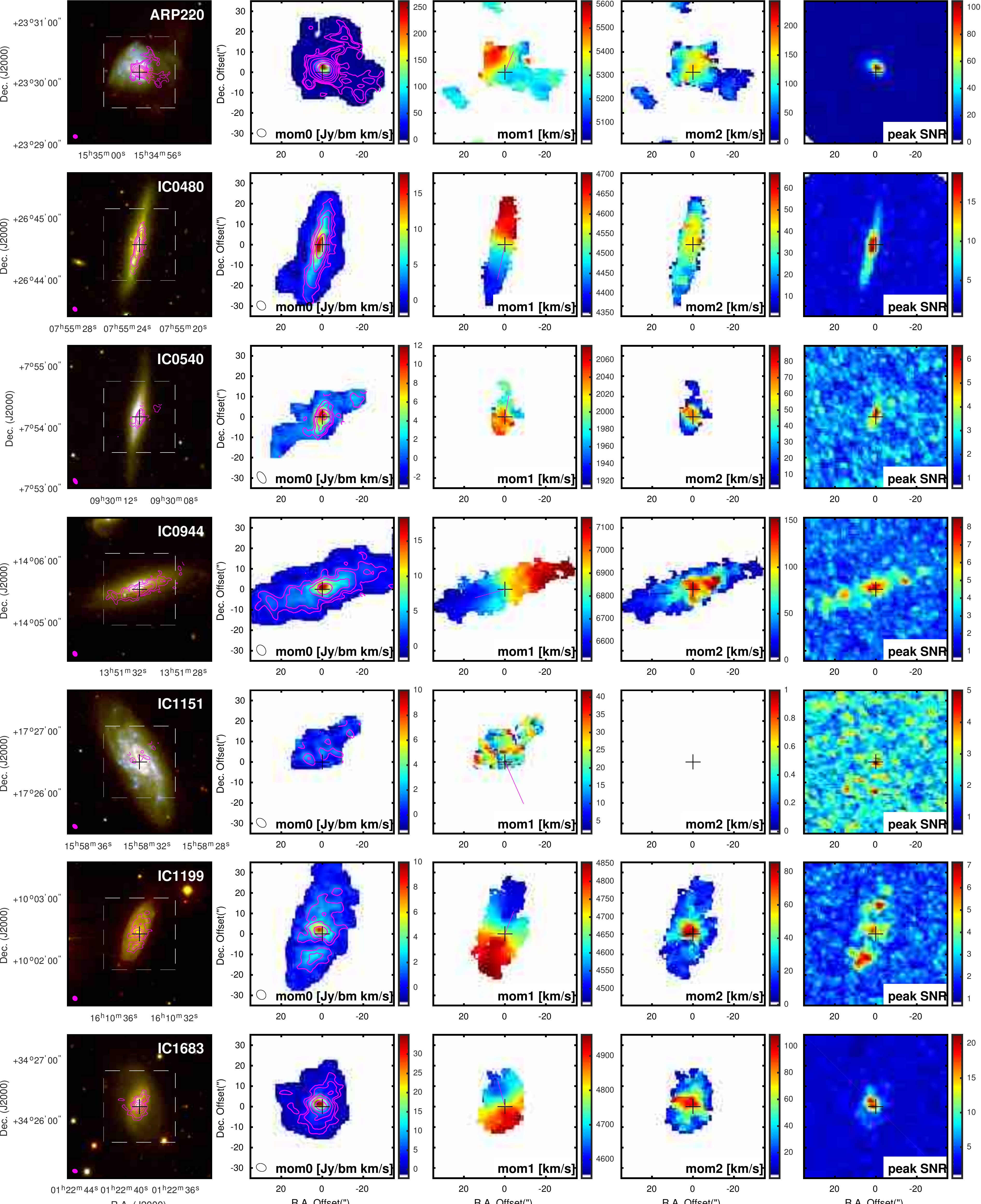}\\ 
\end{center}
\caption{EDGE data products for each galaxy. The first panel shows the SDSS {\em igu} multicolor image with contours from our integrated intensity masked map overlaid. Contours are fixed and logarithmically spaced: 1, 2.2, 4.6, 10, 22, 46, 100\,Jy\,\kmpers\,beam$^{-1}$. The following panels zoom into the $70\arcsec\times70\arcsec$ size region represented by the white square. They show: 1) the same contours overlaid on the ``smooth-mask'' moment zero, 2) the ``dilated-mask'' moment 1 (line-of-sight velocity), magenta line shows preferred PA, 3) the ``dilated-mask'' moment 2 (velocity dispersion), and 4) the peak signal-to-noise map. Panels for the remainder of the survey can be found in Appendix A.
\label{fig:multipanel}} 
\end{figure*}

\subsection{Optical IFU Data: CALIFA}\label{sec:califa}

The CALIFA survey \citep{Sanchez:12}
observed $\sim$700 diameter-selected galaxies with the Postdam Multi-Aperture Spectrophotometer (PMAS) spectrograph and the PMAS fiber PAcK (PPAK) IFU at the 3.5m
telescope of the Calar Alto Observatory in Spain \citep[see
also][]{Garcia-Benito:15,Sanchez:16}. The CALIFA sample is selected from Sloan Digital Sky Survey (SDSS) within a particular redshift range (1500--9000 \kms), but not restricted to the SDSS spectroscopic survey.
It is intended to reflect the present-day galaxy population in a statistically meaningful way for
galaxies in the stellar mass range $\log[{\rm M_*/M_\odot}]=9.4-11.4$
\citep{Walcher:14}. CALIFA observed in both a wide (3745--7500\AA; $R\sim850$) and a narrow (3400--4840\AA; $R\sim1650$) spectral setting to ensure a broad wavelength coverage containing critical ionized gas emission and stellar absorption lines as well
as the Balmer break, while attaining a high enough resolution to enable good kinematic determinations for both the stars and the ionized gas \citep[centroid errors $\lesssim 20$ \kms,][]{Barrera-Ballesteros:14}.

The angular diameter selection criterion of CALIFA
($45\arcsec<D_{25}<80\arcsec$), the final spatial resolution of the IFU observations \citep[$\sim$2.5$\arcsec$,][]{Sanchez:16}, and the lower redshift range (0.005$<z<$0.03), results in a sample that is ideally suited for CO interferometry studies. This is particularly the case with CARMA, whose optimal field-of-view is well-matched to that of PPAK ($64\arcsec\times70\arcsec$). The adopted observational scheme and sample selection imply that galaxies are observed over a wide range of galactocentric distances (out to $\sim$2.5 times the effective radius, $R_e$, enclosing half the optical light). This results in more extensive sampling of galaxy disks than surveys that emphasize galaxy centers \citep[e.g., ATLAS$^{\rm 3D}$,][]{Capellari:11}, while attaining higher spatial resolution than surveys that emphasize more distant galaxies such as SAMI \citep[][]{Croom:12,Bryant:15}, and MaNGA \citep[][]{Bundy:15}. 

There are a number of other comparative differences. ATLAS$^{\rm 3D}$, for example, observed only early-type/red galaxies mostly in the central regions ($\sim0.5-1 R_e$), covering a relatively narrow wavelength range but with a better spatial resolution than CALIFA due to the lower redshift range and smaller size of their spaxels and FWHM of the PSF ($\sim1\arcsec$). On the other hand, SAMI and MaNGA observe a wide range of galaxy types with fiber bundles of diameter 16$\arcsec$ and 32$\arcsec$ (for the largest bundle), respectively. To provide coverage out to 1 $R_e$ (SAMI) and $1.5-2.5 R_e$ (MaNGA), however, they observe galaxies at higher redshifts (between $z\sim$0.04 and $z\sim$0.17). Since both of them have an angular resolution similar to CALIFA (FWHM$\sim$2.5$\arcsec$), this results in lower physical resolution. Indeed, \citet{hector16} show that half of the galaxies in the MaNGA survey have $R_e$ smaller than two times the FWHM of the PSF and are thus almost unresolved. In summary, the galaxies in the CALIFA survey are very well-suited for interferometric studies that must trade brightness sensitivity for spatial resolution, in particular for millimeter-wave interferometers where the field of view of existing instruments is about one arc-minute (i.e., the typical size of the CALIFA galaxies). For a further comparative discussion of these surveys see, for example, \citet{Bundy:15} and \citet{sanchez2015IAU}.

The CALIFA data are reduced by a standard pipeline
\citep{bernd13,Garcia-Benito:15,Sanchez:16}, following the prescriptions included in \citet{2006AN....327..850S}. The resulting datacubes are analyzed by a second pipeline, {\sc Pipe3D} \citep{sanchez2016Pipe3DII},  based on a modified version of the {\sc FIT3D} package
\citep{Sanchez:06,sanchez16Pipe3DI} that is designed also for use in SAMI and MaNGA \citep[e.g.][]{hector16,jkbb16}. The output
of this analysis is described in detail in \citet{sanchez2016Pipe3DII}, being publicly accessible for most of the galaxies in the EDGE survey. It includes the fluxes of
all bright lines as well as the line-of-sight velocity and velocity dispersion of the ionized gas and stars, star formation history, luminosity and mass weighted stellar population age, luminosity and mass weighted stellar metallicity, dust attenuation for the stars from SED modeling, mass-to-light
ratio, and stellar mass surface density. 
The richness of the data available for every EDGE galaxy is illustrated in Figure~\ref{fig:multiNGC4047}.

\subsection{Survey Characteristics}

Figure \ref{fig:surveychar} show the overall characteristics of the survey. The typical time invested per galaxy between the two CARMA configurations used was 4.3~hours. A handful of galaxies (NGC2623, NGC 4210, UGC 8107) had observations obtained during an initial concept-test trial run which were combined with the latter observations, and so accumulated $13-19$ hours. The typical synthesized beam after combining the CARMA D and E array observations is $4\farcs5$, with most observations between $4\arcsec$ and $5\arcsec$. The RMS noise in the resulting maps is typically $\sim11.6$\,mJy\,beam$^{-1}$ measured in 10\,\kmpers\ channels, although it can be as small as $4$ and as large as $18.5$\,mJy\,beam$^{-1}$ depending on the weather conditions during the observations and the actual integration time. This RMS in flux density yields a distribution of Rayleigh-Jeans brightness temperature sensitivities centered on $53$~mK, with most values between 40 and 65~mK, depending on the synthesized beam. Most galaxies in EDGE are detected, with a logarithmically-weighted mean for the signal of $\sim430$\,mK and a logarithmically-weighted mean significance of the brightest detection peak for a galaxy of $\sim8.4\sigma$. 

The \htwo\ column density associated with our mean brightness temperature $1\sigma$ sensitivity of 52.6 mK for a Galactic conversion factor of $\xco=2\times10^{20}$ \xcounits\ \citep[][and references therein]{Bolatto:13a} is $N(\htwo)\sim1\times10^{20}$\,\percmsq\ in 10\,\kmpers, which corresponds to $\av\sim0.1$ mag for a Galactic dust-to-gas ratio \citep{Bohlin:78}, and to a mass surface density of gas (including He) of $\Sigma_{\rm mol}=2.3$\,\msunperpcsq. For a channel-width of 30\,\kmpers, more representative of the line-width in a beam for our typical spatial resolution, the $1\sigma$ mass surface density sensitivity is $\Sigma_{\rm mol}=3.9$\,\msunperpcsq. Given our distribution of distances, this translates into a typical 4$\sigma$ mass sensitivity of $3.5\times10^7$\,\msun\ for a 30\,\kmpers\ line, although there is a wide distribution for this parameter, with cubes as sensitive as $6\times10^6$\,\msun\ for the nearest galaxies (Figure \ref{fig:surveyphyspar}). Therefore we are sensitive to objects on the mass scale of Giant Molecular Associations \citep[GMAs,][]{Vogel:88}. Distances to galaxies in our sample range between 23 and 130 Mpc, and the median physical resolution of the interferometer data is 1.4\,kpc. The median integrated molecular gas for the detected EDGE galaxies is $2.3\times10^9$\,\msun, determined from the EDGE observations themselves (see \S\ref{sec:fluxrecovery}).

\subsection{Sample Demographics}

EDGE galaxies are a subsample of CALIFA, selected on the basis of their 22\,$\mu$m brightness and the ability to schedule their observations during the CARMA campaign.
Nonetheless, Figure \ref{fig:surveymass} shows that EDGE galaxies are reasonably representative of the peaks of the CALIFA distributions for M$_*$, SFR, and metallicity. They do not, however, represent well the low M$_*$, SFR, and metallicity ranges, and they also under-represent the early galaxy types. 

The optical characteristics of the CARMA EDGE sample are summarized in the left panel of Figure~\ref{fig:sample}, which shows their location on an SDSS color-magnitude diagram
($g-r$ vs.\ $M_r$).  The galaxies span a wide range in both color and
luminosity, and overall provide a much better sampling of the SDSS population
of blue galaxies than previous spatially-resolved CO surveys of nearby galaxies such as the Berkeley-Illinois-Maryland Array survey of nearby galaxies BIMA SONG \citep[44
galaxies,][]{Regan:01,Helfer:03}, HERACLES \citep[48 galaxies,][]{Leroy:09b,Leroy:13b,Schruba:12},
and CARMA STING \citep[23 galaxies,][]{Rahman:11,Rahman:12}. 
Those surveys, which have no overlap with CALIFA, targeted galaxies that are more local and thus do not probe well the larger stellar masses.
For the purposes of identifying ``detected'' galaxies, we define a detection as a cube where there is at least one beam with a 10\,\kms\ channel with a signal that is 5 times the RMS. This provides a fairly reasonable estimate of ``true'' detections, as can be verified by inspection of Figure \ref{fig:multipanel} and figures in Appendix A. By this metric the overall CO detection rate of the EDGE survey is 80\%, and those galaxies are indicated with green circles in Figure \ref{fig:sample}.

The right panel of Figure~\ref{fig:sample} shows the distribution of the observed EDGE targets in a plot of
22\,$\mu$m (WISE band $W4$) to 3.4\,$\mu$m ($W1$) flux ratio vs.\ 3.4\,$\mu$m luminosity, as measured in images from
the {\sl WISE} all-sky survey (Bitsakis et al., in prep.), shown in magnitude space.
The distribution of the entire CALIFA observed
sample is also shown for comparison; our subsample spans a range of luminosity
at 3.4\,$\mu$m (a band dominated by the old stellar population) that is
comparable to that of CALIFA, although low mass galaxies and galaxies that are
massive but have blue far-infrared colors indicating they are quiescent (typically early-types) are underrepresented. The distribution of $W_1-W_4$ color in EDGE is clustered around $4-6$ magnitudes where most galaxies are also detected in CO, while galaxies with $W_1-W_4<4$ are much more sparsely covered and have a much higher proportion of CO non-detections. By contrast with the 80\% overall detection rate, for the 20 galaxies with an infrared color $W_1-W_4\leq4$ the detection rate drops to $\sim50\%$.


\subsection{Basic Equations}
\label{sec:basicequations}

For the purposes of this work we compute extinction-corrected star formation rates (SFRs) using the nebular extinction (in magnitudes) based on the Balmer decrement

\begin{equation}
A_{H\alpha} = 5.86 \log\left({\frac{F_{H\alpha}}{2.86\,F_{H\beta}}}\right),
\label{eq:Aha}
\end{equation}

\noindent where $F_{H\alpha}$ and $F_{H\beta}$ are the fluxes of the respective Balmer lines, and the coefficients assumes a \citet{Cardelli:89} extinction curve and an unextincted flux ratio of 2.86 for case B recombination. The SFR (in \msun\,yr$^{-1}$) is then computed from \citep{Rosa-Gonzalez:02}

\begin{equation}
{\rm SFR} = 7.9\times10^{-42}\,F_{H\alpha}\,10^{A_{H\alpha}/2.5},
\label{eq:SFR}
\end{equation}

\noindent which assumes a Salpeter Initial Mass Function (IMF). The global SFRs are computed spaxel by spaxel and coadded, and only spaxels that fall in the star formation area of the BPT diagram \citep{Baldwin:81} following the demarcation by \citet{Kewley:02}, and have an H$\alpha$ equivalent width $>6$\,\AA, are included to remove areas that are primarily ionized by evolved stars or active galactic nuclei \citep{Sanchez-Menguiano:16}. 

Metallicities are computed employing the O3N2 indicator, unless otherwise noted, and using the calibration by \citet{Marino:13}

\begin{equation}
12 + \log [{\rm O/H}] = 8.533 - 0.214\times{\rm O3N2}.
\label{eq:O3N2}
\end{equation}

\noindent Where indicated, we use the N2 indicator together with the calibration (also from \citeauthor{Marino:13})

\begin{equation}
12 + \log [{\rm O/H}] = 8.743 + 0.462\times{\rm N2}.
\label{eq:N2}
\end{equation}

\noindent Global metallicities are computed in an annulus at the effective radius $R_e$ of the galaxy, which has been found to produce good representative values \citep{Sanchez:16}.

Molecular gas masses (in \msun) and surface densities are computed using

\begin{equation}
{\rm M_{mol}} = 1.05\times10^{4} \frac{S_{\rm CO}\Delta v D_L^2}{(1+z)},
\label{eq:Mmol}
\end{equation}

\noindent where $S_{\rm CO}\Delta v$ is the integrated CO \jone\ line flux (in Jy\,km\,s$^{-1}$), $D_L$ is the luminosity distance (in Mpc), and $z$ is the redshift. This assumes a CO-to-\htwo\ conversion factor \xco=$2\times10^{20}$\,\xcounits\ and includes a factor of 1.36 for the correction in mass due to the cosmic abundance of Helium \citep[c.f.,][]{Bolatto:13a}.

\section{Results}
\label{sec:results}

Tables \ref{tab:table1}, \ref{tab:table2}, and \ref{tab:table3} summarize the properties of the galaxies, the EDGE observations together with ancillary photometry, and the integrated and spatially resolved measurements respectively.

The first four columns in Table \ref{tab:table1} list the galaxy preferred name, the central coordinates according to HyperLEDA \citep{Makarov:14}, and the galaxy morphological type. The fifth column shows whether the galaxy has a bar (B), a ring (R), or is part of a multiple (M) according to HyperLEDA. The sixth column lists the V$_{\rm LSR}$ systemic velocity of the galaxy, as determined (in order of preference) from CO rotation curve fitting, CALIFA observations, or failing those the LEDA catalog. The velocity is expressed in the relativistic convention ($v=c(\nu_0^2-\nu^2)/(\nu_0^2+\nu^2)$). Column 7 lists the optical size of the galaxy as the 25$^{\rm th}$ magnitude isophotal diameter, according to HyperLEDA. The following columns, corresponding to Inclination and Position Angle are derived from CO rotation curve fitting where that is possible (Levy et al., in prep.; Leung et al., submitted), or determined from the shape of the outer isophotes \citep{Falcon-Barroso:17}, or failing that taken from HyperLEDA. Finally, the distance is determined from the redshift $z$ obtained by CALIFA for emission lines, and is computed as the luminosity distance in a flat cosmology with $H_0=70$\,\kmpers, $\Omega_m=0.27$, and $\Omega_\Lambda=0.73$.

Table \ref{tab:table2} lists the velocity integrated CO flux measured in the {\em smooth} masking approach described in \S\ref{sec:mapmaking}, the equivalent round synthesized beam size (computed as the geometric mean of the major and minor beam sizes), the noise of the spectra in the central region of the cube (calculated as the RMS in 10\,\kms\ channels), and the  molecular mass of the galaxy M$_{\rm mol}$ calculated using Eq. 3 in \citet{Bolatto:13a} assuming a Galactic CO-to-\htwo\ conversion factor of $\xco=2\times10^{20}$\,\xcounits\ and including the correction for the contribution from Helium to the mass (i.e., using $\alpha_{\rm CO}=4.36$\,\acounits). The last four columns lists the integrated fluxes in the SDSS $g$ and $r$ filters, computed from CALIFA synthetic photometry, and the fluxes in the {\em Wide-field Infrared Survey Explorer} (WISE) bands $W1$ and $W4$ computed from integrated photometry in the images (Bitsakis et al., in prep.). 

Table \ref{tab:table3} lists the galaxy stellar mass (M$_*$), star formation rate at the equivalent radius (SFR), and gas metallicity (as $12+\log[{\rm O/H}]$) computed from CALIFA observations by {\sc Pipe3D}. Both $M_*$ and SFR assume a Salpeter IMF. The last four columns list the radius enclosing 50\% of the CO flux ($R_{50}$), and the scale-lengths resulting from fitting exponential disk profiles to the distribution of molecular gas, stellar mass, and SFR as described in \S\ref{sec:stargasscalelength} and \S\ref{sec:SFRgasscalelength}.

\subsection{Data Products}
\label{sec:data}

Our final data products for each galaxy consist of two cubes per spectral line.
The ``signal'' cube has been primary-gain corrected and masked beyond the region where the sensitivity falls to half of its peak value.
A ``noise'' cube providing an estimate of the 1$\sigma$ noise (in mK or mJy bm$^{-1}$) at each pixel of the signal cube was also generated.
The final angular resolution achieved for galaxies observed in both D and E observations is about 4.5\arcsec\ (the median spatial resolution is 1.5~kpc), with a 4$\sigma$ brightness sensitivity of $\sim$200 mK per 10 \kms\ channel. For a typical line-width of 30\,\kms, the corresponding molecular surface density is $\Sigma_{\rm mol}\sim$15\,\msunperpcsq\ for a Galactic \xco, and including the Helium correction by mass.  
For the E-only cubes the typical resolution is about 8\arcsec\ (2.7 kpc) with similar brightness sensitivity in mK.
Of the 126 galaxies observed in both D and E arrays, $\approx$100 display a peak brightness temperature at the 5$\sigma$ or higher level and thus are considered secure detections; of these galaxies, $\approx$66 are also detected in the $^{13}$CO line.  

\begin{figure}[t] 
\begin{center}
\includegraphics[width=0.45\textwidth]{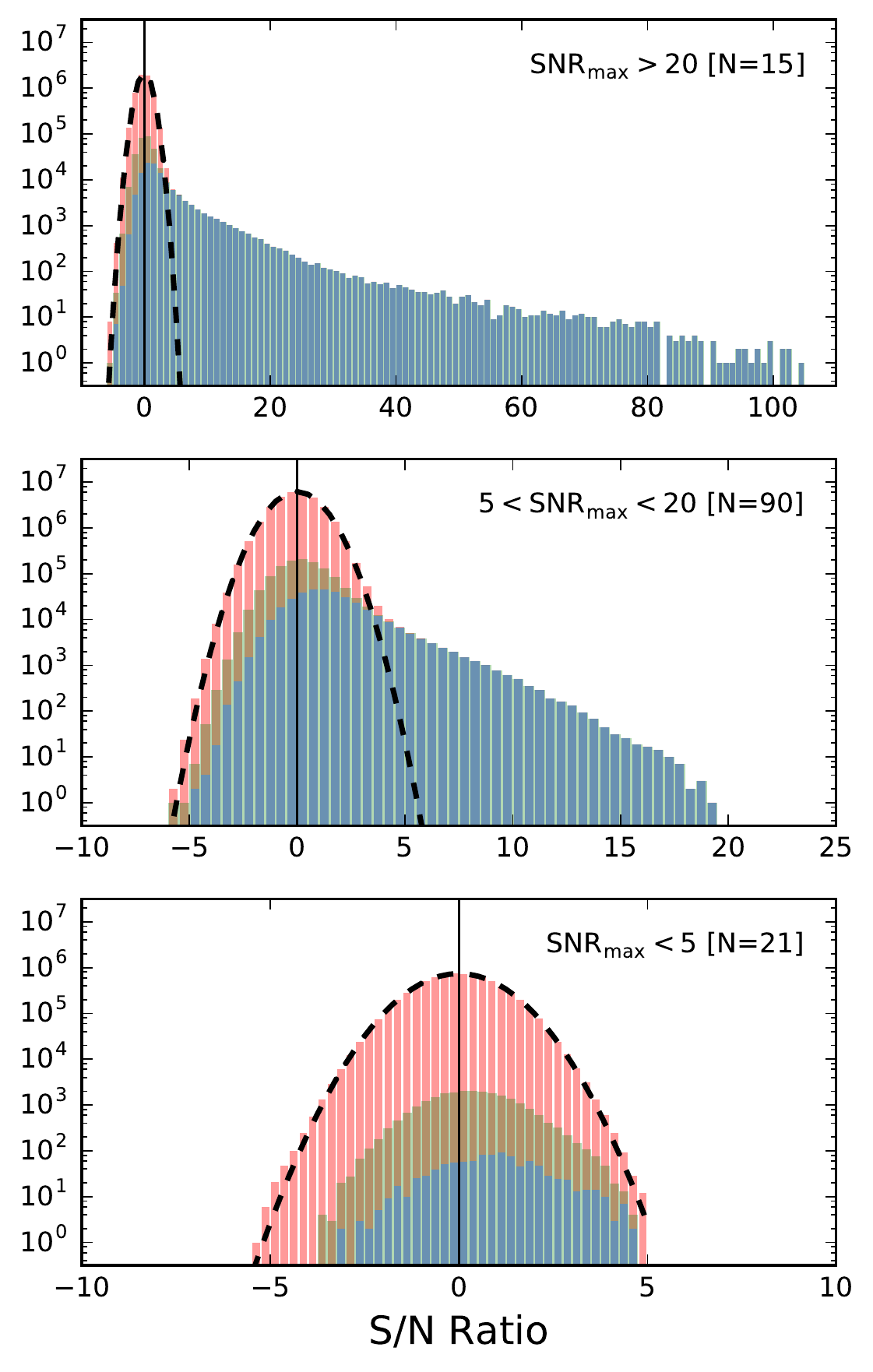}
\end{center}
\caption{
Pixel statistics of the EDGE data cubes, split across three sub-samples based on peak SNR (note that the EDGE images have a 1\arcsec\ pixel size which oversamples the beam, so not all pixels are independent).  Blue represents the signal in the dilated mask, and red bars are stacked on top and represent the noise outside the mask.  Brown bars are overlaid (not stacked) on the blue bars and represent the signal as determined by the smoothed mask, which generally encloses the dilated mask. The noise statistics follow those of Gaussian noise, and at first glance the masking strategies do a good job capturing signal in the cubes. 
\label{fig:imhist}}
\end{figure}

\subsubsection{Map Making}
\label{sec:mapmaking}

We generated moment maps of integrated intensity, intensity-weighted velocity, and velocity dispersion (moments 0, 1, and 2 of the spectral line profile) from the signal cubes of the 126 galaxies with with E+D observations.  The moment maps were generated in IDL\footnote{code available at \url{https://github.com/tonywong94/idl_mommaps}} after applying a blanking mask in order to reject noise which would otherwise overwhelm the relatively weak line emission.  The mask was created by starting at 3.5$\sigma$ (or greater) peaks in the cube, requiring that each peak span at least two velocity channels, and
expanding down to the surrounding 2$\sigma$ contour.  An additional ``guard'' band of 1 pixel in all directions around the mask was added in order to capture additional low-level emission \citep[this is a method similar to that used by][]{Rosolowsky:06}.  We refer to the resulting moment maps as the ``dilated-mask'' moment maps.  For comparison we also generated a second set of moment maps by first smoothing the cube spatially to a final resolution of 9\arcsec.  A mask was then generated by starting at 3.5$\sigma$ (or greater) peaks in the smoothed cube and expanding down to the surrounding 2$\sigma$ contour, then padding with an additional ``guard'' band of 2 pixels in all directions.  This mask was then applied to the original, unsmoothed data cube to produce the ``smoothed-mask'' moment maps. 
Finally we generated a set of unmasked moment maps by collapsing the cubes along the velocity axis with no masking.  These maps, which we call the ``unmasked'' moment maps, suffer from poor signal-to-noise because of the indiscriminate averaging of signal and noise. Because of the lack of masking, however, they are unbiased and we use them for the flux comparison described below.

The pixel statistics of the two masking methods are compared in Figure~\ref{fig:imhist}, which breaks the EDGE sample into three sets based on the peak signal-to-noise ratio (SNR).  The smoothed masks (brown bars in histograms) generally occupy a greater volume in the cubes than the dilated masks (blue bars), allowing them to capture more low-level emission.  However, the smoothed mask also suffers from greater noise contamination, as indicated by the higher counts of negative pixels.  For use by the community, we therefore recommend the dilated moment maps for visualizing the bright emission or analyzing the first and second velocity moments (due to their more strict noise masking), whereas the smooth moment-0 maps provide a more complete accounting of the total CO flux.  

\begin{figure*}[t] 
\begin{center}
\includegraphics[width=\textwidth]{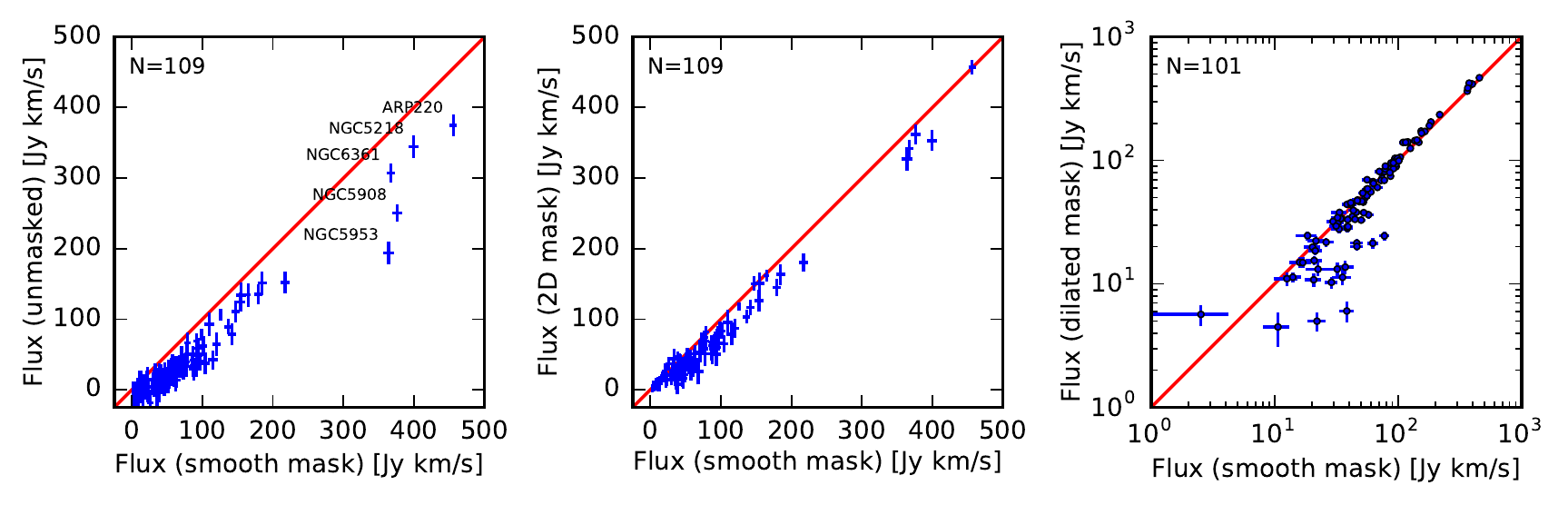}
\end{center}
\caption{
Comparison of integrated CO fluxes from three different approaches to masking the CO data cube.  We show our fiducial integrated CO flux (based on the smoothed mask) on the horizontal axis and compare it with the flux derived without masking ({\it left panel}), by projecting the full spatial extent of the smoothed mask through all velocity channels (2D mask, {\it middle panel}), and by limiting the integration to the dilated mask ({\it right panel}, on a logarithmic scale).  Error bars reflect formal 1$\sigma$ uncertainties in the total flux based on the noise in the cube and assuming the corresponding mask is correct.
\label{fig:fluxcomp}}
\end{figure*}

The systematic uncertainty in flux measurements due to masking is further explored in Figure~\ref{fig:fluxcomp}, which compares CO fluxes measured with four different masking approaches.  In addition to the dilated-mask, smoothed-mask and unmasked maps discussed above, we also measured flux by integrating the full velocity range of the cube but restricted to the sky region spanned by the smoothed mask (2D masked flux). Note that the first two panels are plotted on a linear scale while the last is shown on a logarithmic scale, to better examine the weaker sources.  As shown in the right panel, the smoothed mask tends to recover a larger flux than the dilated mask, especially for galaxies with relatively weak CO emission covering a large angular extent (e.g., NGC 4210).  However, growing the mask much further (as with the unmasked or 2D masked cases) tends to reduce the flux, as negative residuals from the deconvolution process are included.  This is especially noticeable for the strongest sources, for which deconvolution errors lead to large negative regions that reduce the flux when integrated over large areas, as shown in the left panel.  The error bars shown reflect only the formal uncertainties in the masked flux based on noise within the mask, and clearly underestimate the true uncertainties which are due in part to the choice of mask (i.e., the ambiguity in defining the extent of the emission in position and velocity space).

For four galaxies (NGC\,4676A, NGC\,6314, UGC\,3973, and UGC\,10205), the observed velocity range may not have been fully adequate to cover the line, due to the systemic velocity falling close to the edge of one of the fixed tunings. Emission is seen to continue up to the edge of the spectral coverage. For NGC\,4676A the problem is at the low velocity end, while for the other three the problem occurs at the high velocity end of the coverage. The map and position-velocity diagrams suggest that the fraction of the emission that we could be missing is small, we estimate it to be $\leq10\%$. 

Figure \ref{fig:multipanel} and the companion figures in Appendix A illustrate the results of the map-making efforts. Each row of panels correspond to one CARMA EDGE galaxy in the E+D sample. The first panel shows the SDSS {\em igu} data over an area of $140\arcsec\times140\arcsec$ centered on the catalog position of the object (marked with a cross). The overlaid magenta contours correspond to the smooth-mask integrated intensity map (values are
$S_{\rm CO}\Delta v=1, 2.2, 4.6, 10, 22, 46, 100$\,Jy\,\kmpers\,beam$^{-1}$, kept fixed for all the galaxies). The dashed white square represents the inner quarter area ($70\arcsec\times70\arcsec$), which is shown in the rest of the panels in that row. The second panel in a row shows the smooth-mask integrated intensity, with the same contours overlaid. We recommend using the masked integrated intensity images for a number of reasons. In particular, because of the fixed velocity integration interval used to make the unmasked maps, weak CO emitters and face-on galaxies (which will have narrower lines) are hard to pick up in them. The third panel shows the dilated-mask first moment map representing the line-of-sight velocity. The rotation of the galaxies is apparent in most cases, showing that the masking is doing a good job at picking up the signal and not introducing artifacts. The line starting at the center shows the PA adopted by EDGE, which is derived from (in order of preferred priority): 1) rotation curve fitting to the CO velocities, 2) ellipsoidal fitting to the outer isophotes, or 3) adopting the value quoted by HyperLEDA. The extent of the line corresponds to $R_{25}/3$ as catalogued by HyperLEDA. The fourth panel in a row shows the second moment of the CO, corresponding to the velocity dispersion, for our dilated masking. Despite the masking, by their very nature the higher order moments become increasingly more uncertain. The last panel in a row shows the peak signal-to-noise in 20\kmpers\ wide channels, derived without any masking. We obtain it by finding the highest peak in the spectrum along each spatial pixel (sometimes referred to as ``peak temperature''), and dividing by the corresponding sensitivity map. This provides a good by eye assessment of the signal to noise of the data, and the areas where emission is likely present.

\begin{figure*}[t]
\begin{center}
\includegraphics[height=4in]{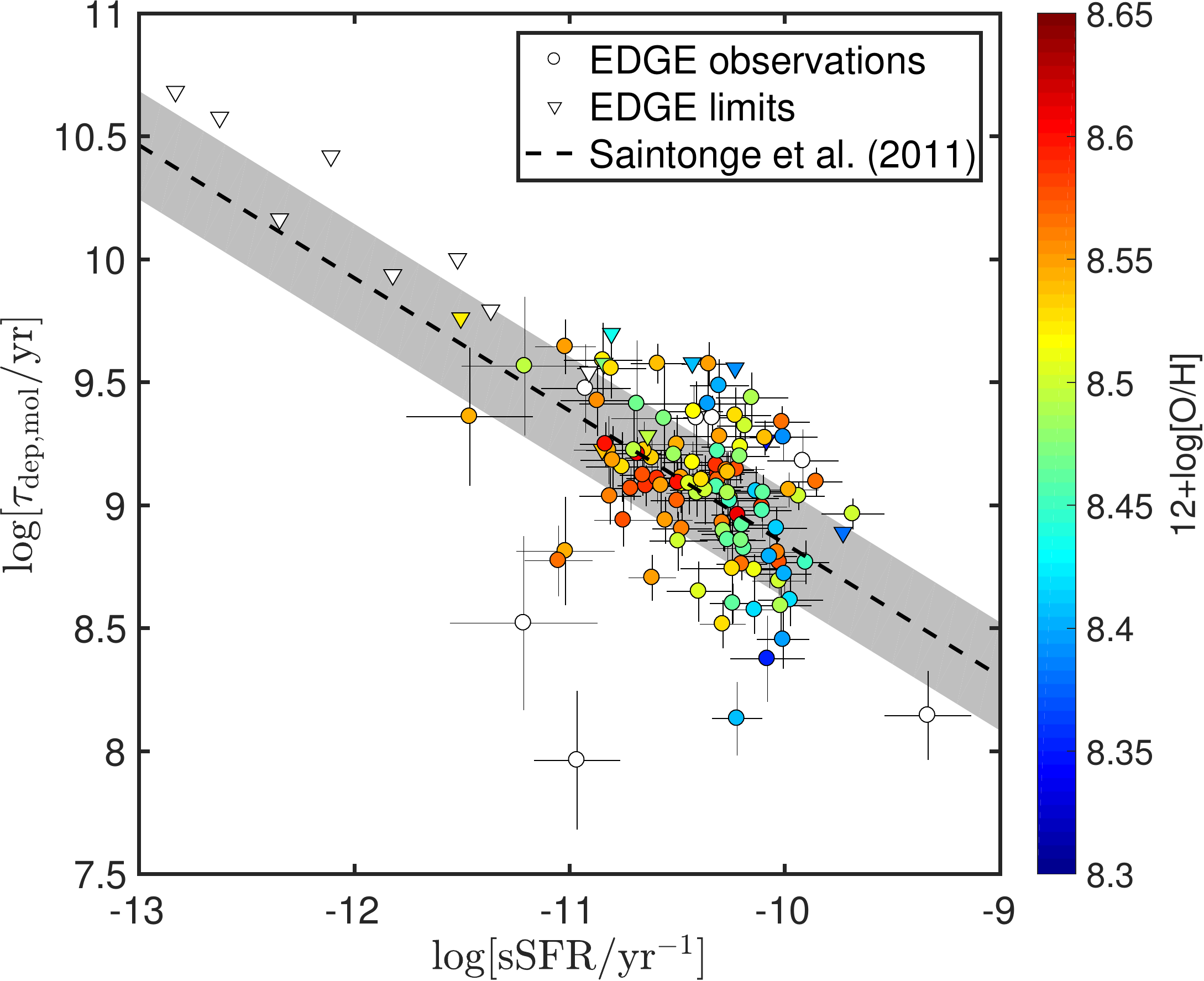}
\includegraphics[height=4in]{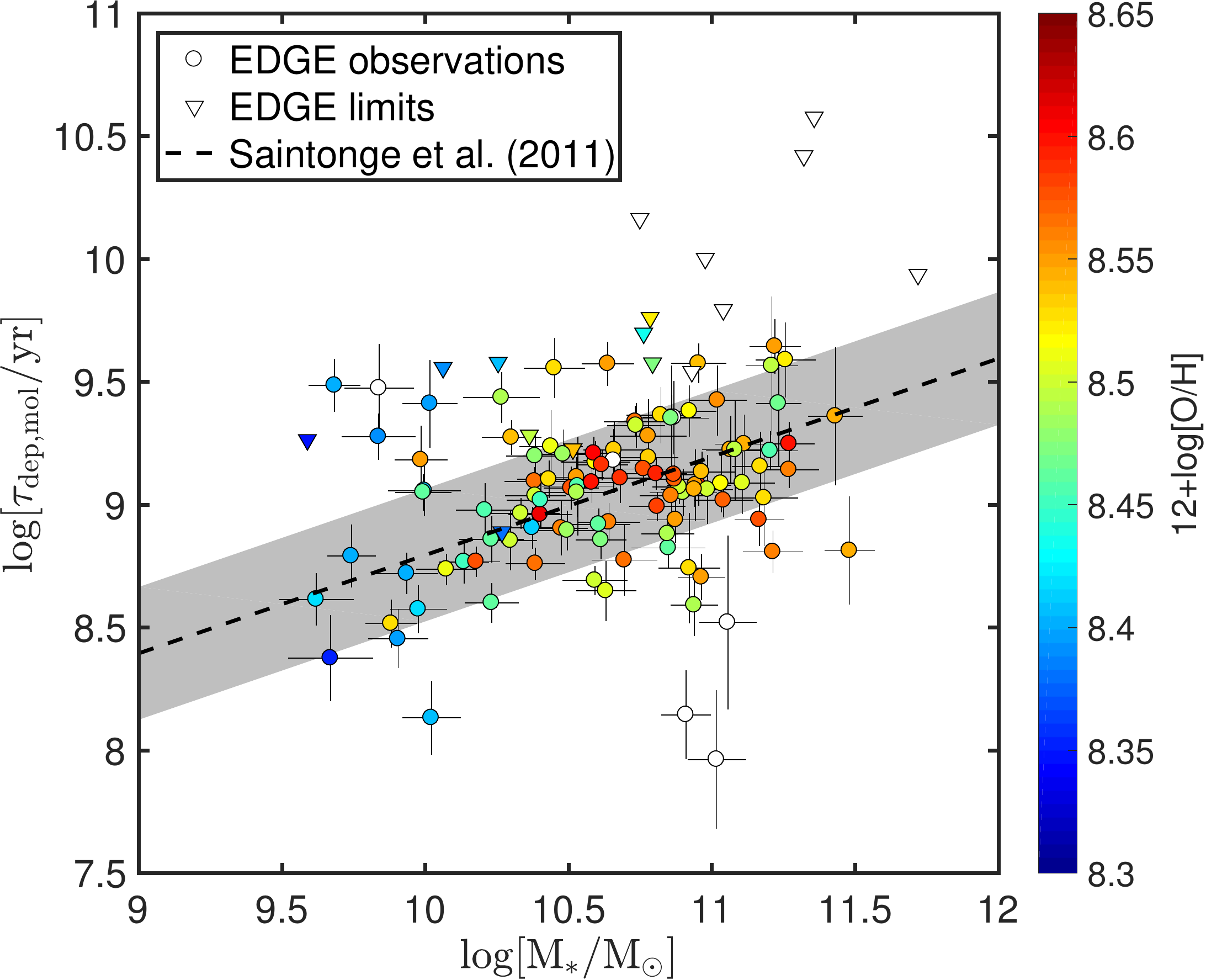}
\end{center}
\caption{Comparison of the integrated results of EDGE with the COLD GASS single-dish survey. {\bf Top:} The trend for molecular depletion time ($\tau_{\rm dep,mol}$) with specific star formation rate (sSFR) from Balmer decrement-corrected H$\alpha$. {\bf Bottom:} The trend for $\tau_{\rm dep,mol}$ with stellar mass. In both cases the dashed line and gray region show the trend and scatter present in COLD GASS \citep{Saintonge:11b}. We prefer the relation in the bottom plot, where the axes have the advantage of being independent. The mean relation and scatter in the EDGE sample are
similar to COLD GASS, suggesting that resolving out extended flux by the interferometer (which would bias our results toward shorter $\tau_{\rm dep, mol}$ and increase the scatter) is not a concern for most of the observations. The symbols are color-coded by gas phase metallicity, derived from the O3N2 indicator at 1\,R$_e$. Open symbols do not have metallicity estimation in CALIFA because they do not have line ratios corresponding to a star-forming population.}
\label{fig:tausSFR}
\end{figure*}

\subsection{Preliminary Assessment of Flux Recovery}
\label{sec:fluxrecovery}

The EDGE cubes do not include total power data from a single-dish telescope.
Note, however, that the range of spatial scales included in the combination of D+E array data is very good, 3 to 54 k$\lambda$ typically. This implies that we recover signal on 68\arcsec\ and smaller scales, which is the optical size of our objects. Also, almost all galaxies are imaged using 7-pointing hexagonal mosaics, which help produce a more complete $(u,v)$-plane sampling and improve flux recovery \citep{Helfer:02}. Nonetheless, we are in the process of comparing our CO fluxes with a subsample of galaxies observed with the IRAM 30-m telescope. We are also conducting imaging simulations to estimate the likely effect of spatial filtering and limited brightness sensitivity, both of which can limit the flux recovery of interferometer maps. Finally, we are separately considering options for acquiring single-dish total power for the entire sample. 

The comparison of our integrated fluxes with empirical relations determined from single-dish observations shows that it is unlikely that there are widespread flux recovery problems in CARMA EDGE. The relations between \tdep, sSFR,  and stellar mass derived from our integrated CO fluxes (Figure~\ref{fig:tausSFR}) are in excellent agreement with those determined for the COLD GASS single-dish sample \citep{Saintonge:11b}, suggesting that the EDGE observations by themselves already have very good flux recovery. We convert the \citeauthor{Saintonge:11b} relations to the system we use in this paper by correcting for the differences in the assumed IMF (see \S\ref{sec:mol2stellar}). The SFR requires also a conversion that is not only related to the IMF, but also to the fact that the SED modeling used by \citeauthor{Saintonge:11b} to estimate SFRs is sensitive to activity over $\sim100$\,Myr while the H$\alpha$ we use is sensitive over a much shorter time-scale of $\sim10$\,Myr \citep[e.g.,][]{Kennicutt:12}. This introduces a somewhat uncertain factor that, assuming exponentially declining star formation histories with reasonable time-scales, we estimate to be $\sim2$. The transformation results in $\log\tdepmol=-0.54(\log{\rm sSFR}+10.70)+9.16$, and $\log\tdepmol=+0.4(\log{\rm M_*}-10.90)+9.16$ respectively.  
The $1\sigma$ scatter of the EDGE observations around the $M_*-\tdepmol$ relation, for example, is 0.32 dex. This is almost identical to the measured COLD GASS scatter of 0.27 dex. The mean residual between our observations and the relation is $-0.01$ dex. It is thus very unlikely we are missing significant flux in the CARMA EDGE observations for the sample as a whole, although this is still possible for a few individual galaxies. Therefore we conclude we can use the EDGE integrated molecular masses to investigate global relations.

\begin{figure}[t]
\begin{center}
\includegraphics[width=\columnwidth]{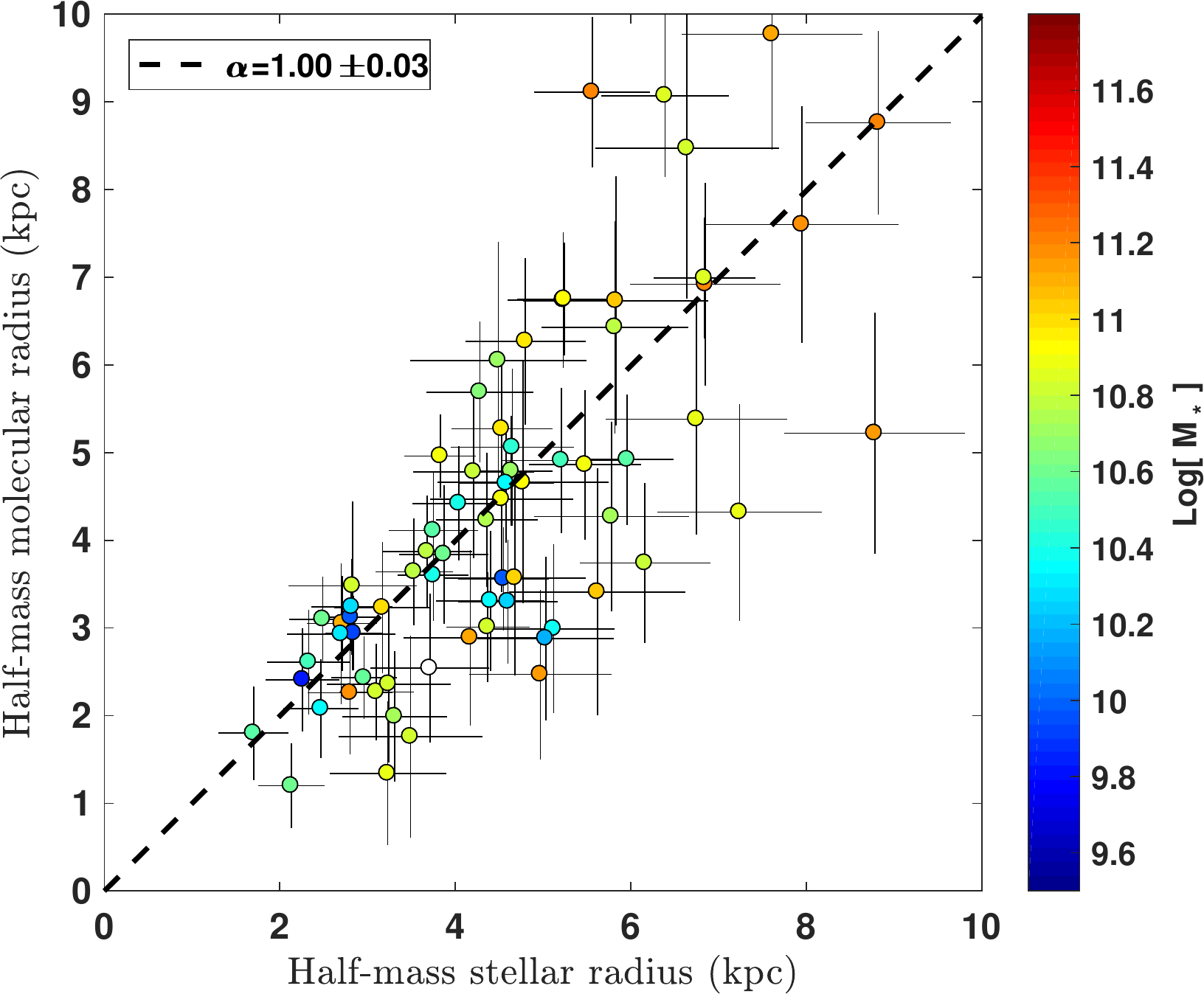}
\end{center}
\caption{Comparison of the half-mass sizes of the stellar and molecular disks. The errors take into account the spatial resolution at the distance of the galaxy. Galaxies are again color-coded by integrated stellar mass. This is a less demanding size measurement than the exponential scale-length fit, so it can be done for a larger number of galaxies (69 galaxies). The dashed line shows the line resulting from a bivariate fit through the origin. The resulting slope is $1.00\pm0.03$, with reduced $\chi^2=1.3$. }
\label{fig:diskR50}
\end{figure}

\begin{figure*}[t]
\begin{center}
\includegraphics[height=3.5in]{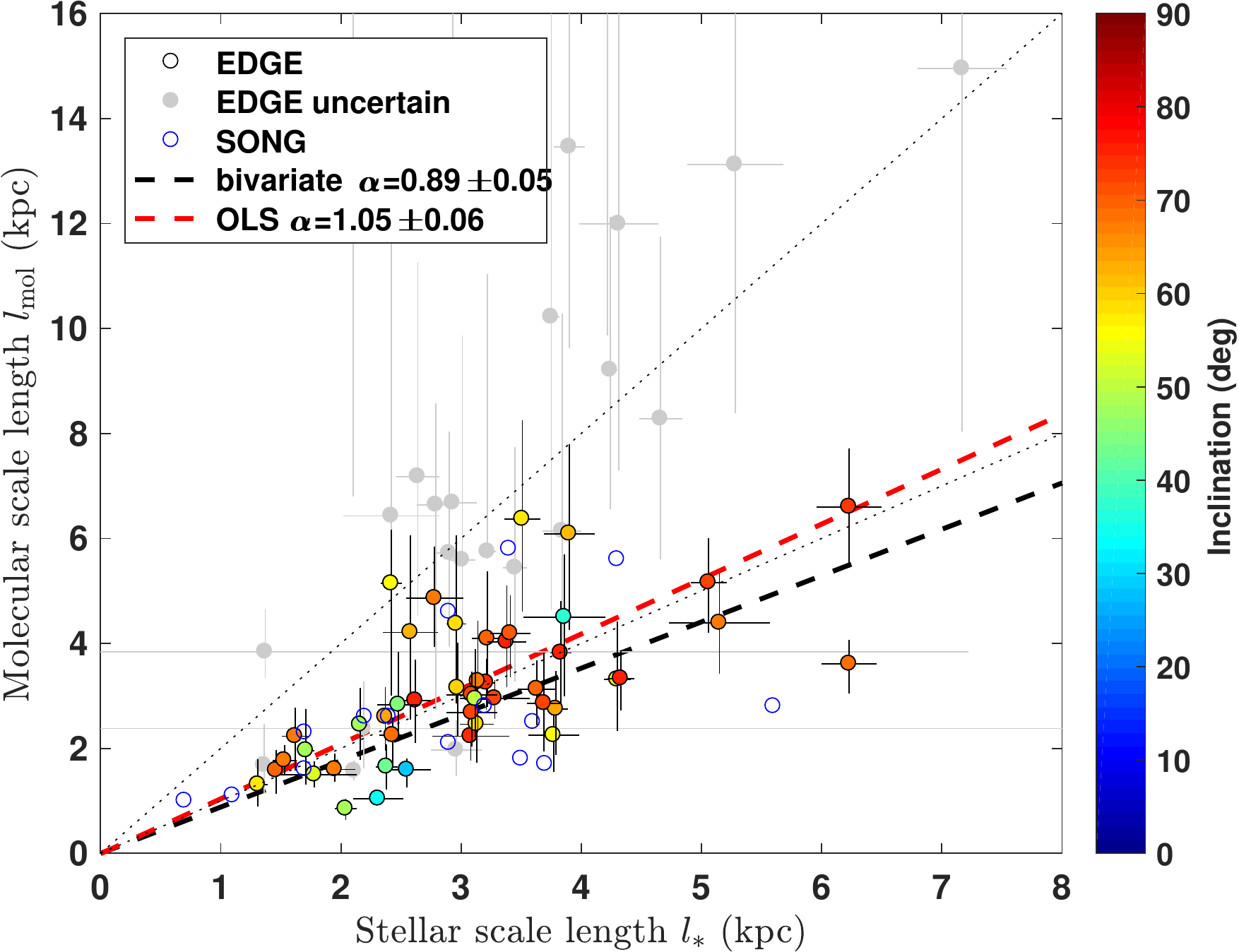}
\end{center}
\caption{Comparison of the stellar and molecular mass scale length derived from fitting an exponential to the corresponding disk profiles. The stellar profiles correspond to stellar mass as derived from SED fitting . The colored symbols show the high-quality results for the EDGE galaxies, color-coded by inclination (46 points). The open symbols show the results tabulated for BIMA SONG \citep{Regan:01}, and the gray symbols show uncertain results from EDGE (fits that have errors larger than 30\% or 2 kpc in either scale length). The dotted lines illustrate the 1:1 and 2:1 scalings. The EDGE galaxies show a distribution very similar to the SONG sample, showing that the \Smol\ and \Sstar\ exponential scale lengths are very similar. The color dashed lines show the results of fits forced through the origin: a bivariate fit weighted by the uncertainties, and an unweighted ordinary least-squares bisector. }
\label{fig:disksize}
\end{figure*}

\begin{figure*}[t]
\begin{center}
\includegraphics[width=0.9\textwidth]{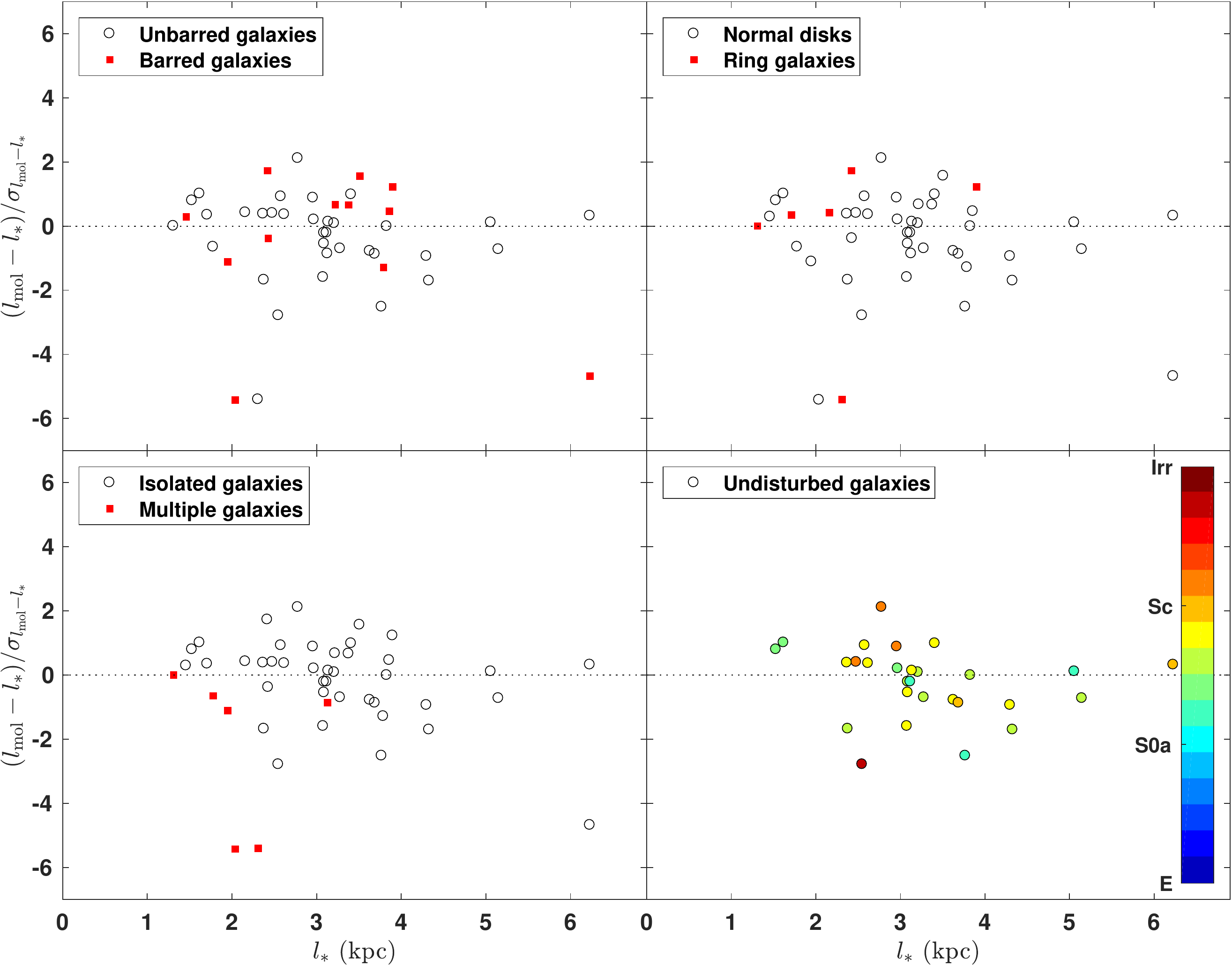}
\end{center}
\caption{Galaxies that show significant departures from the equality between stellar and molecular exponential scale lengths appear to be mostly multiple or barred systems. The panels show the difference between the molecular and stellar scale lengths divided by the corresponding error bar estimated as described in the main text to account for our systematic uncertainty. The first three panels show the locations of systems that are labeled as ``barred'', ``ring'', or ``multiple'' in the HyperLEDA database. The largest departures from $l_{\rm mol}=l_*$ occur for barred systems, and the galaxies labeled as ``multiple'' tend to show a molecular gas distribution more compact than the stellar. Once all these systems are removed (bottom right panel), $l_{\rm mol}$ tracks $l_*$ tightly, independently of whether the galaxy is an early or late-type disk.}
\label{fig:disturbed}
\end{figure*}

\begin{figure}[t]
\begin{center}
\includegraphics[width=\columnwidth]{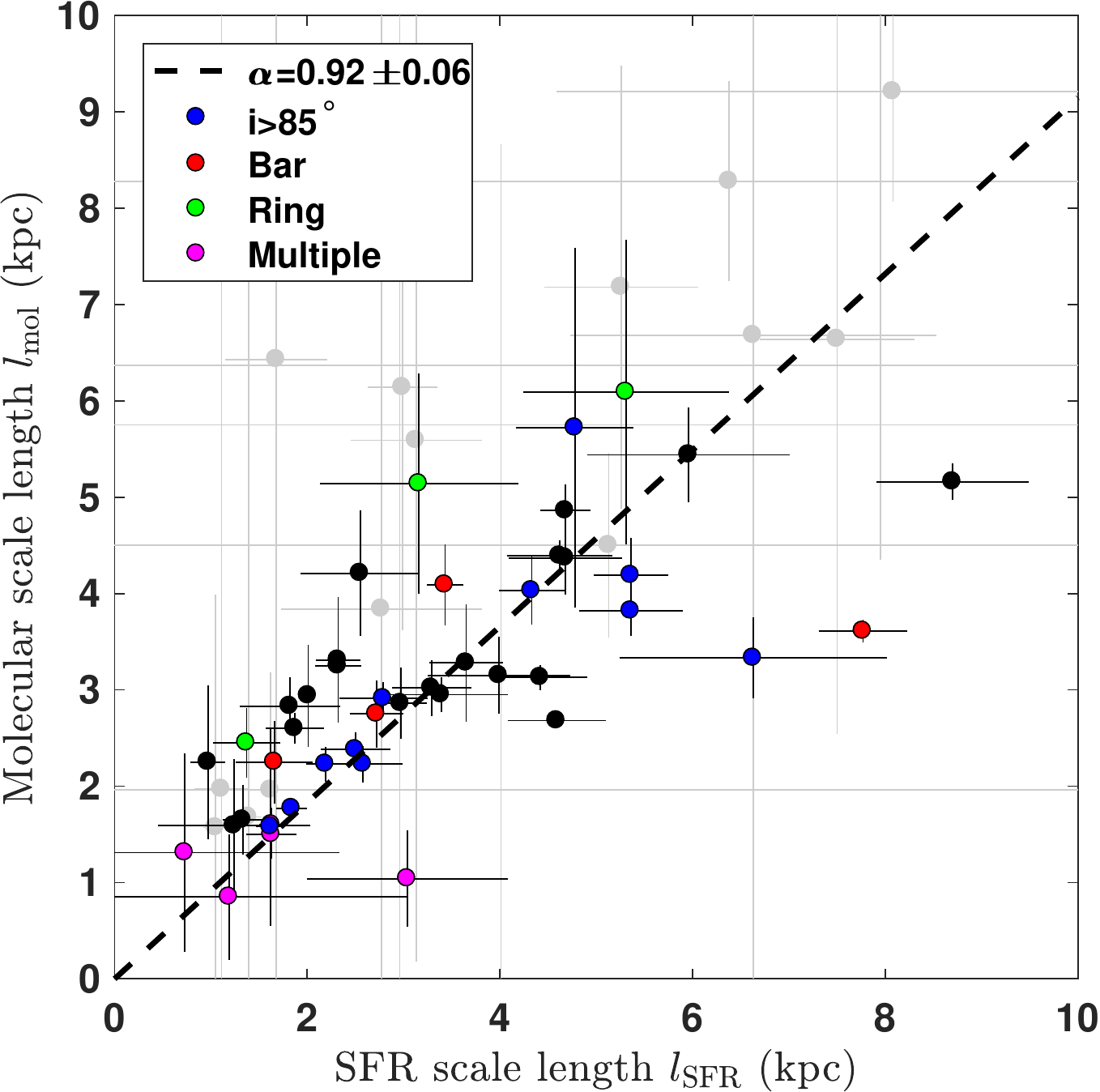}
\end{center}
\caption{Relation between the scale length of the exponential disks of molecular gas and SFR. Poor quality fits are indicated in gray, while the good quality fits are color-coded indicating highly inclined galaxies ($i>85^\circ$, blue), barred galaxies (red), ring galaxies (green), or multiple galaxies (magenta) according to the information in HyperLEDA. The best fit bivariate scaling is shown by the dashed line. 
\label{fig:coSFR}}
\end{figure}

\begin{figure*}[t]
\begin{center}
\includegraphics[height=2.85in]{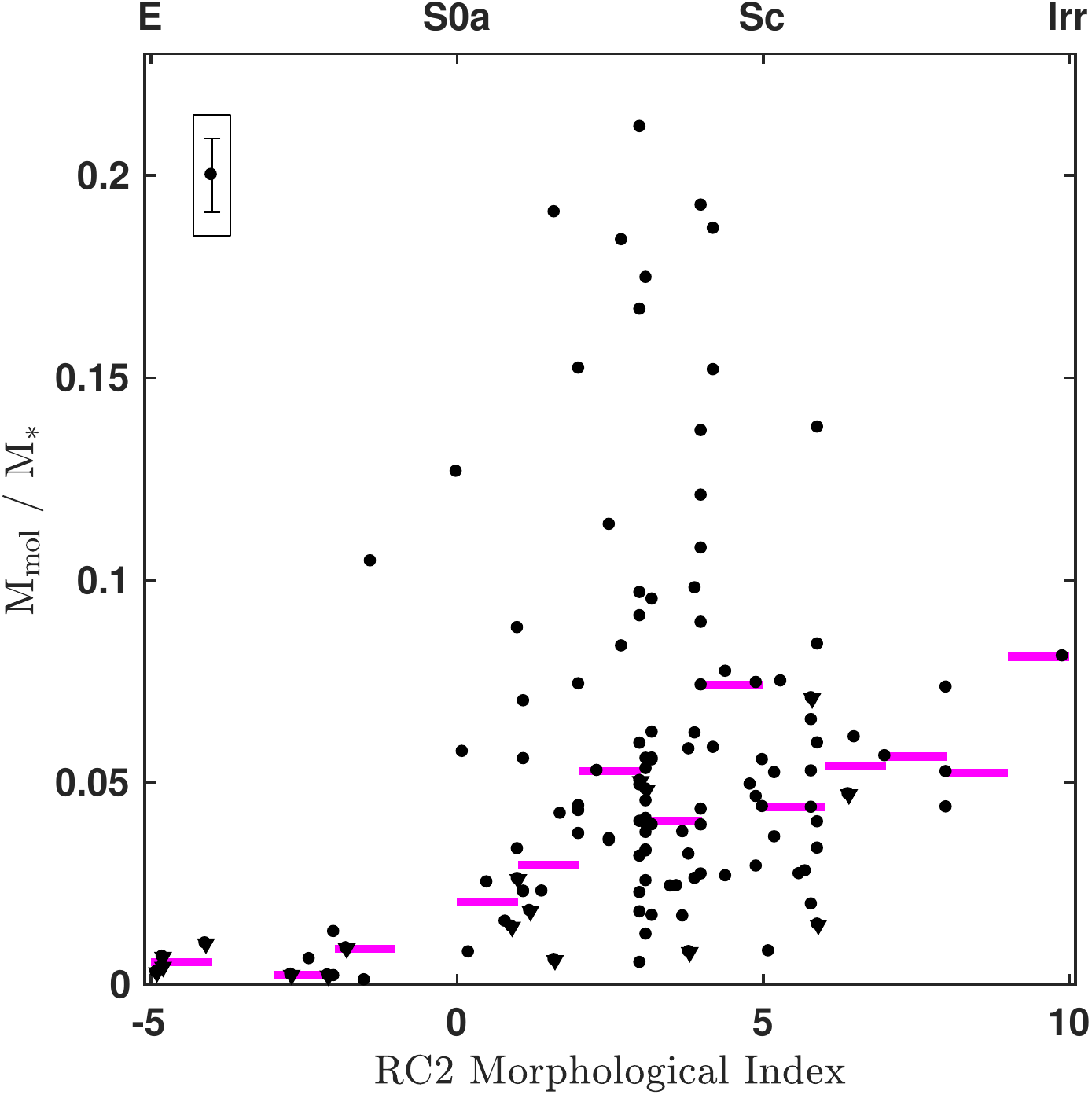}
\includegraphics[height=2.75in]{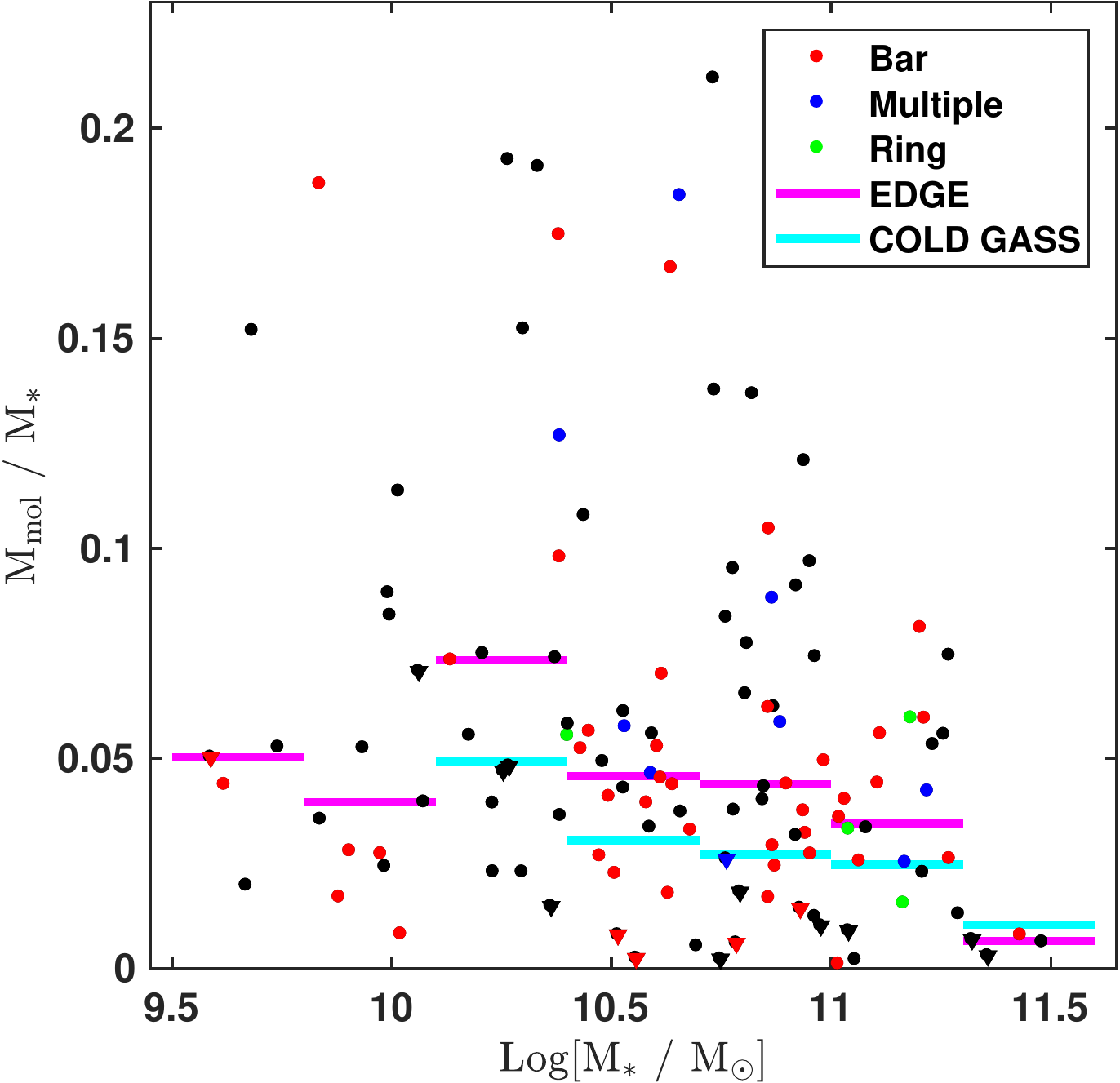}
\end{center}
\caption{The ratio of molecular to stellar mass in EDGE galaxies as a function of galaxy type and mass. {\bf Left:} molecular to stellar mass ratio versus morphological type as determined by the RC2 de Vaucouleurs morphological index provided by HyperLEDA. The upper x$-$axis indicates the mapping to the Hubble type. The stellar masses are obtained from optical SED modeling, the molecular masses from integrating the EDGE interferometric CO maps. The dots correspond to the individual measurements (Arp 220 has been removed from the sample because we assume a Galactic \xco, not applicable to a ULIRG), and triangles indicate upper limits. A representative systematic error bar is illustrated in the upper left corner, calculated assuming 26\% uncertainty in the molecular mass determination (typical error in the sample) and 20\% uncertainty in the stellar mass determination. The histogram shows the median after binning by spectral class. The median (mean) molecular-to-stellar ratio is quite constant at $4.9\%$ ($6.1\%$) for spirals of all types within the EDGE sample. Early type galaxies have a much lower molecular gas fraction $M_{\rm mol}/M_*\sim 0.6\%$. {\bf Right:} Molecular to stellar mass ratio versus stellar mass. The mean trend shows a clear decrease for $M_{stellar}\gtrsim10^{11}$\,\msun, with a marked tendency for lower molecular ratios at higher masses. The symbols have been color-coded by the presence of a bar, a companion, or a ring according to HyperLEDA. The histograms show the median $M_{\rm mol}/M_*$ in EDGE and COLD GASS detections.}
\label{fig:mmolmstars}
\end{figure*}

\subsection{Multi-wavelength dataset}

Figure \ref{fig:multiNGC4047} provides a taste of the CO data products and their relation with the optical data from SDSS and CALIFA. The galaxy shown is NGC~4047, an Sb galaxy located approximately 52\,Mpc away in the direction of Virgo. The top left panel shows the multicolor SDSS image and the placement and size of the rest of the panels, illustrated with a red box. The remainder of the panels in the top row show the EDGE data products: \co\ integrated intensity, velocity, velocity dispersion, and peak temperature, and \cothree\ integrated intensity. The CO synthesized beam size for this object is $\sim1.1$\,kpc. The panels in the second and third row show a selection of the data products available from the CALIFA observations processed with {\sc Pipe3D}: H$\alpha$ and H$\beta$ intensity, H$\alpha$ velocity, intensities of [OIII], [NII], and [SII] lines, stellar ages derived and metallicities from spectral fitting, stellar velocity and velocity dispersion, stellar surface density, and nebular extinction derived from the Balmer decrement. 

\section{Discussion}

\subsection{The Sizes of Molecular and Stellar Disks}

It is of particular interest to establish the relation between the large scale distribution of the molecular and stellar material within galaxies. We know that the distributions of CO and star formation activity in galaxies follow each other closely \citep[e.g.,][]{Leroy:13b}. The timescale for the lifetime of molecular clouds is, on the other hand, likely few to several Myrs \citep[e.g.,][]{Blitz:80,Kawamura:09,Fukui:10,Gratier:12,Meidt:15}, close to that of the stars emitting our SFR indicator (H$\alpha$) but much shorter than the lifetime of the bulk of the stellar population. So the similarity between the CO and SFR distribution is a snapshot in time, and it does not necessarily imply a similarity between the molecular and stellar distributions. 

The older stellar population that encompasses most of the mass of the stellar disk derives from successive localized episodes of star formation, so its distribution probes the distribution of the molecular material on longer timescales, at least in galaxies where most of the stellar mass is not accreted externally through mergers. In a disk that is in a secular equilibrium configuration the distribution of the molecular material, the new stars, and the older stellar population would all agree. Alternatively, if the process of raising the density of the atomic gas to convert it to molecular is highly influenced by the stellar potential \citep[e.g.,][]{blitz:04,Eve:10} we may also expect a similar distribution for the stars and the CO emission.  

The literature of resolved  molecular gas surveys shows that the  CO distribution is very similar to the stellar light. The Five Colleges Radio Astronomy Observatory (FCRAO) extragalactic CO survey \citep{Young:95} calculated isophotal diameters, defined at diameters where the integrated intensity falls to a ``typical'' outer disk level of 1 \Kkmpers, generally obtained from model fitting. They find that, for the galaxies where it could be measured, the ratio of isophotal CO diameter to the optical $D_{25}$ diameter is typically $0.5$ showing that the CO extends over half of the optical disk, with some correlation with galaxy type so that earlier-type spiral disks are more compact in molecular gas relative to stars than late-type disks. \citet{Regan:01} used the interferometric BIMA SONG CO survey to show that the CO and K-band light have very similar distributions. Comparing the scale-lengths of the exponential fits to the CO and the K-band galaxy profile data for 15 galaxies they find a typical ratio of CO to stellar scale-length of $0.88\pm0.14$. They suggest that, based on their measurement errors, there are real (and possibly systematic) galaxy-to-galaxy variations in the ratio although their sample size is not large enough to evaluate them. \citet{Leroy:08} and \citet{Schruba:11} used single-dish CO measurements in HERACLES together with 3.6\,$\mu$m images to establish that the molecular and stellar disks follow each other (note that many of galaxies are shared in HERACLES and SONG). They find that the exponential scale length for CO is $l_{\rm CO}\approx0.2\,R_{25}$, resulting in both the CO and stellar scale lengths being very similar ($l_{\rm CO}\sim l_*$). Complementary to these previous efforts, EDGE offers a larger, well characterized sample to study the systematics of CO distributions. 

\subsubsection{Half-mass sizes}

By contrast with other studies which use the distribution of light (sometimes K-band or 3.6\,$\mu$m to get away from extinction problems), we characterize the stellar disk by its distribution of mass, \Sstar, as derived by the stellar population analysis and extinction correction performed by {\sc Pipe3D}. 
The molecular distribution, on the other hand, is derived from the distribution of CO emission by assuming a constant CO-to-\htwo\ conversion factor of $\aco=4.36$\,\acounits\ \citep[i.e., the typical value in the Galactic disk,][]{Bolatto:13a}. Although this is a reasonable approximation to the molecular mass, we know that some galaxy centers exhibit a different value of the conversion factor \citep[e.g.,][]{Sandstrom:13} and this could be also true for very active galaxies \citep[e.g.,][]{Downes:98,Leroy:15}. Therefore our molecular gas distribution comes from a distribution of light assuming a constant mass-to-light ratio. Note that variations in \aco\ from galaxy to galaxy do not affect our size measurements, only variations internal to a galaxy do matter in the determination of the exponential scale length. 

Fitting exponential disk scale lengths requires sources that are well-resolved. This is not always possible, particularly when the galaxies are very distant. Even in marginally resolved sources, however, it is possible to robustly determine ``half-light'' or ``half-mass'' sizes as the radius encircling half of the emission or mass ($R_{\rm 1/2}$). This has particular interest for comparing molecular disks in local galaxies with those in main-sequence galaxies at cosmological distances \citep[e.g.][]{Tacconi:13,Bolatto:15}. This is analogous to the method employed in the comparisons carried out by early single-dish molecular surveys \citep[for example using an ``effective'' radius encircling 50\% or 70\% of the emission,][]{Young:95}. Our $R_{\rm 1/2}$ measurements are performed integrating the azimuthally-averaged radial profiles of our objects: we discuss the details of the derivation of those profiles in the following section. Note that, although there is some stellar mass at radii beyond those observed by CALIFA, it does not affect significantly the stellar half-mass radius determined from the profiles \citep{Gonzalez-Delgado:14}. Similarly, there could be molecular gas beyond the region sampled by CARMA and the sensitivity of this survey, but it is unlikely to have a significant effect on the determination of the molecular half-mass radius.

In Fig. \ref{fig:diskR50} we show the comparison of the molecular half-mass radius \Rhmol\ to the stellar half-mass radius \Rhst. The $R_{\rm 1/2}$ have been computed by integrating the azimuthally-averaged profiles described above. The error bars include in quadrature the half-width size of the beam. The symbols are color-coded by integrated stellar mass, and include all galaxies for which we were able to obtain both a molecular and a stellar size (69 galaxies). Since we are interested in the mean ratio between these two scales, we fit a line through the origin. The dashed line shows the bivariate fit for $\Rhmol=\alpha\,\Rhst$, yielding $\alpha=1.00\pm0.03$ with reduced $\chi^2=1.3$. Most of the galaxies that show the biggest departures from the mean relation tend to be more compact in molecular gas than in the stars, and we discuss this behavior in the next section. The most significant departure in the direction of $\Rhmol>\Rhst$ corresponds to NGC\,5406, a prominently barred galaxy with no detected CO in its central regions and a ring of CO emission corresponding to the bar ends.

\subsubsection{Exponential scale lengths}
\label{sec:stargasscalelength}

Several of the EDGE observations have high signal-to-noise and can be used to fit exponential disk profiles. Figure \ref{fig:disksize} shows our comparison of the sizes of molecular and stellar disks in the EDGE sample. The determination of the scale lengths comes from fitting azimuthally-averaged profiles for each galaxy. These profiles are computed by de-projecting galaxies according to their tabulated distance, position angle, and inclination (c.f., Table \ref{tab:table1}), and computing average surface densities in circular annuli. The stellar surface density is well-determined out to large galactocentric distances, but the CO emission is usually much more patchy and subject to a strong detection bias. To account for this bias we replace the regions that are non-detected in CO (those with signal $\leq2\sigma$) in a given annulus with the corresponding $1\sigma$ sensitivity of the CO observation, and compute the average molecular surface densities for annuli where at least 20\% of the area is detected at a $2\sigma$ level or larger. This choice of replacement parameters was found not to introduce an artificial break in the molecular radial profiles observed for the galaxies, and the slope results are quite insensitive to the required detection fraction. Note, however, that the systematics are included in our error bars, as described below. 

For each radial profile we fit the natural logarithm of \Smol\ versus radius $r$ using the ordinary least-squares bisector, avoiding the central region ($r\leq1.5$\,kpc) if there are measurements in the profile over $r\leq3.5$\,kpc, but including it if there are too few measurements. Avoiding the centers allows us to make our measurement as robust as possible to the aforementioned internal variations in \aco, which may be more common in the central regions \citep[e.g.,][]{Sandstrom:13}, or breaks in the exponential scale length associated with bulges \citep{Regan:01}. The uncertainty in the scale length is usually driven by the systematics in the profile, in particular the choice of the value used for the non-detections. To estimate their impact, we repeat the process replacing non-detections with zeros and with $2\sigma$. The error bars we use are the greater of: 1) the range of values obtained from this comparison, and 2) the formal error in the fit (obtained from bootstrapping).  
The derivation of the exponential length scale for \Sstar\ proceeds in a similar manner, but because the sensitivity bias for the stellar component is not as important (i.e., we have a much higher typical signal-to-noise) we do not do a replacement for the regions not detected, and only consider annuli with detections over 80\% of their area. Lowering this threshold substantially (to 40\%) only minimally affects the derived scale length.

\begin{figure*}[t] 
\begin{center}
\includegraphics[width=0.33\textwidth]{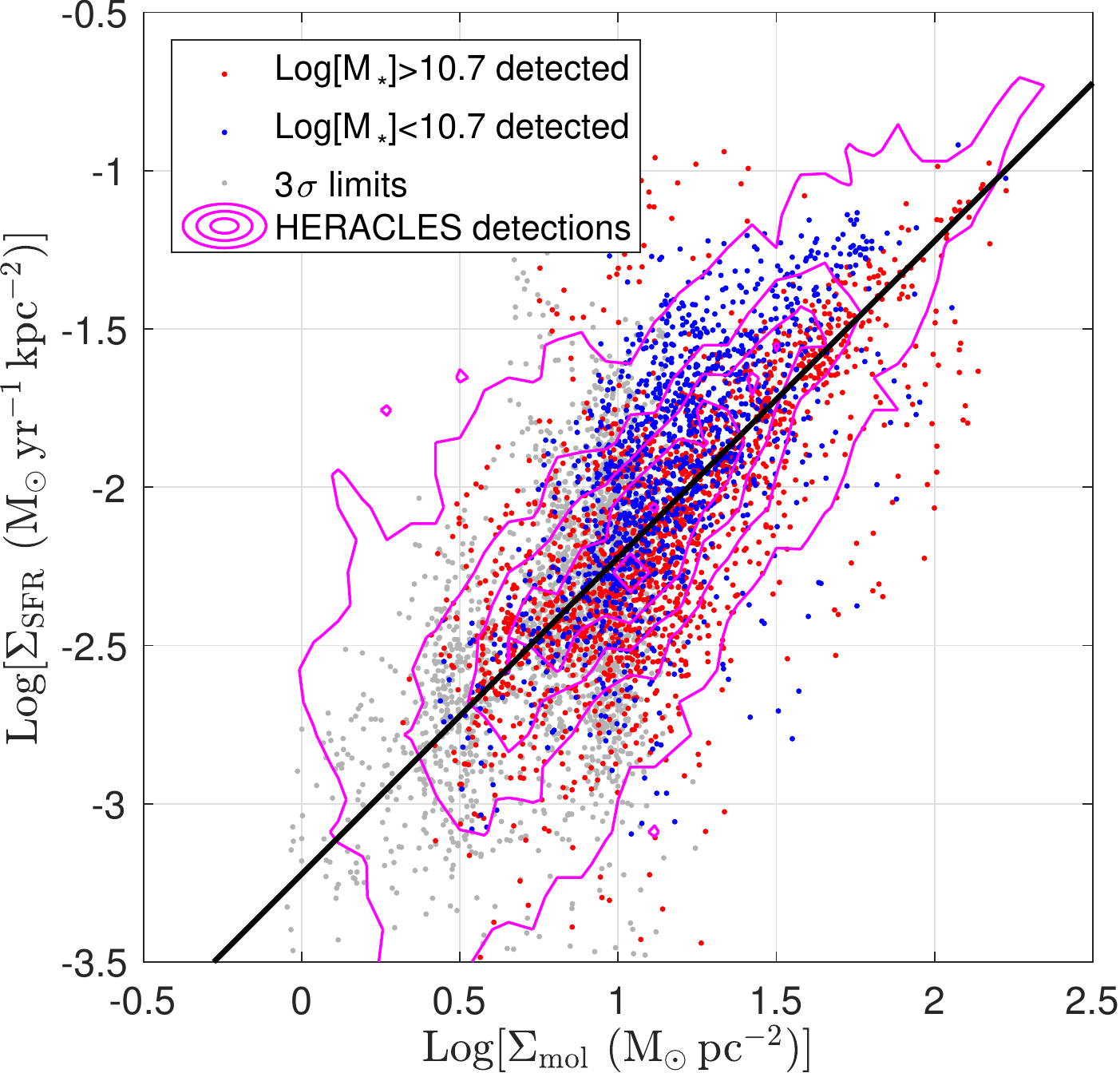} 
\includegraphics[width=0.33\textwidth]{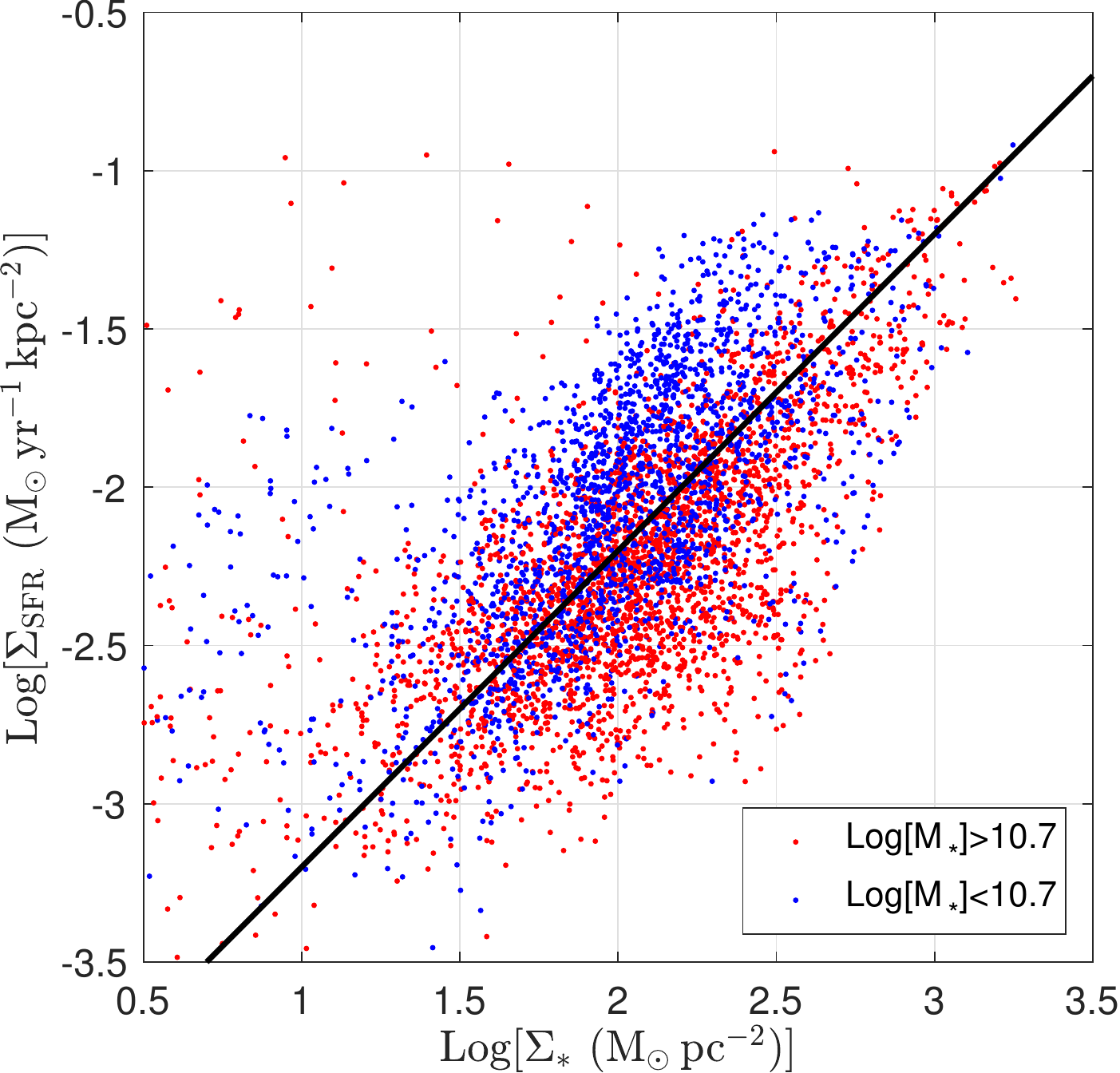} 
\includegraphics[width=0.33\textwidth]{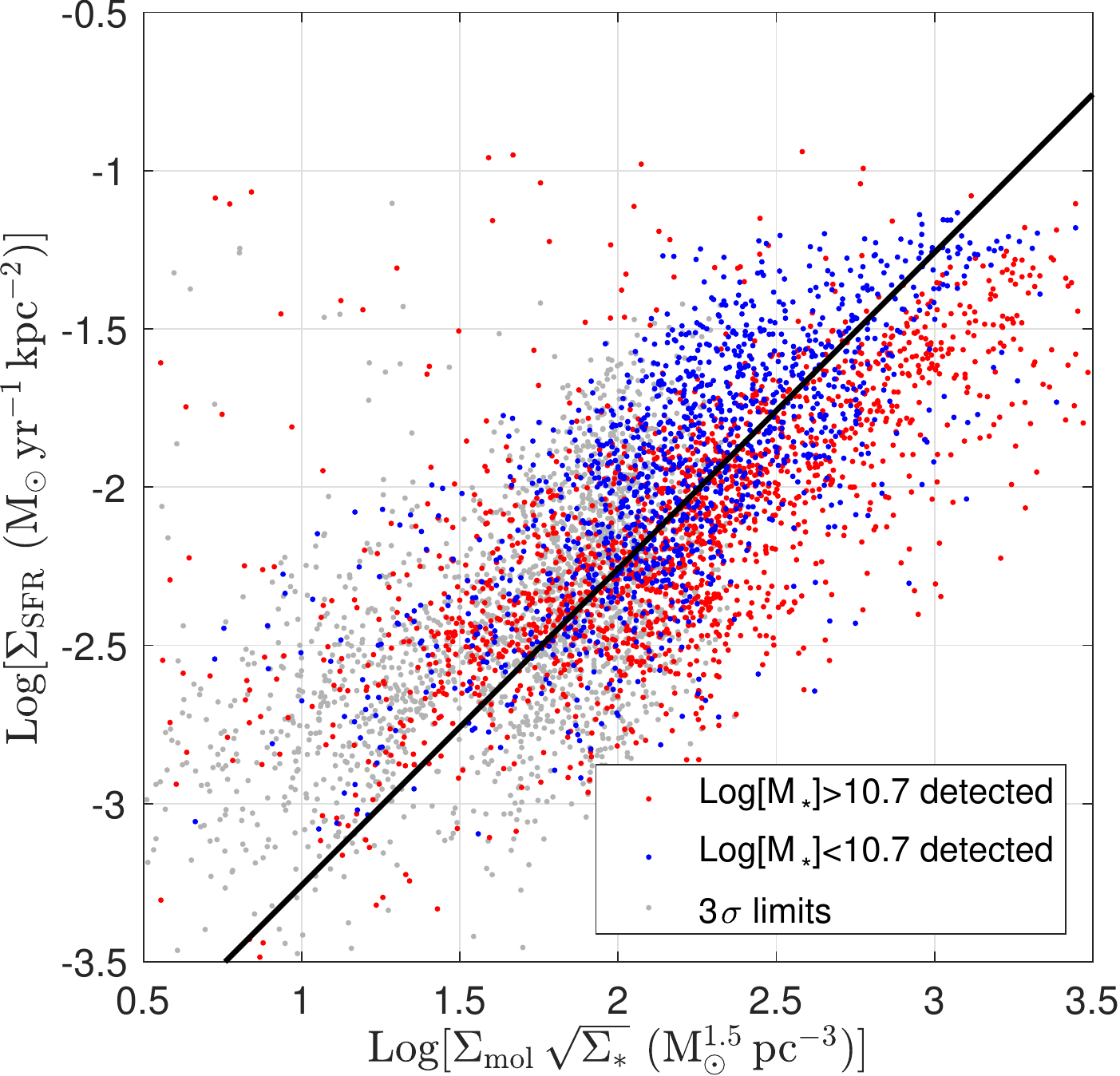} 
\end{center}
\caption{Resolved star-formation relations across the EDGE sample. The three panels correspond to the relation between molecular gas and SFR, stellar surface density and SFR, and a proxy for dynamical equilibrium pressure and SFR. The blue and red points show the results for the detected ($>3\sigma$) lines-of-sight for two galaxy stellar mass groups, while the gray points show $3\sigma$ upper limits for the CO observations within the mask in the dilated integrated intensity maps. In the central panel, which does not involve CO, all lines-of-sight are considered detected. The black line illustrates a linear relation through the center of the points: the ordinate intercepts at zero abscissa are $-3.22$, $-4.20$, and $-4.26$ for the respective plots (left to right). In the first panel, the magenta contours show the distribution for the HERACLES sample, corrected to \jone\ by assuming a constant ratio  $r_{21}=0.7$ \citep{Leroy:13b}.\label{fig:SFRvs}} 
\end{figure*}

\begin{figure*}[t] 
\begin{center}
\includegraphics[width=0.45\textwidth]{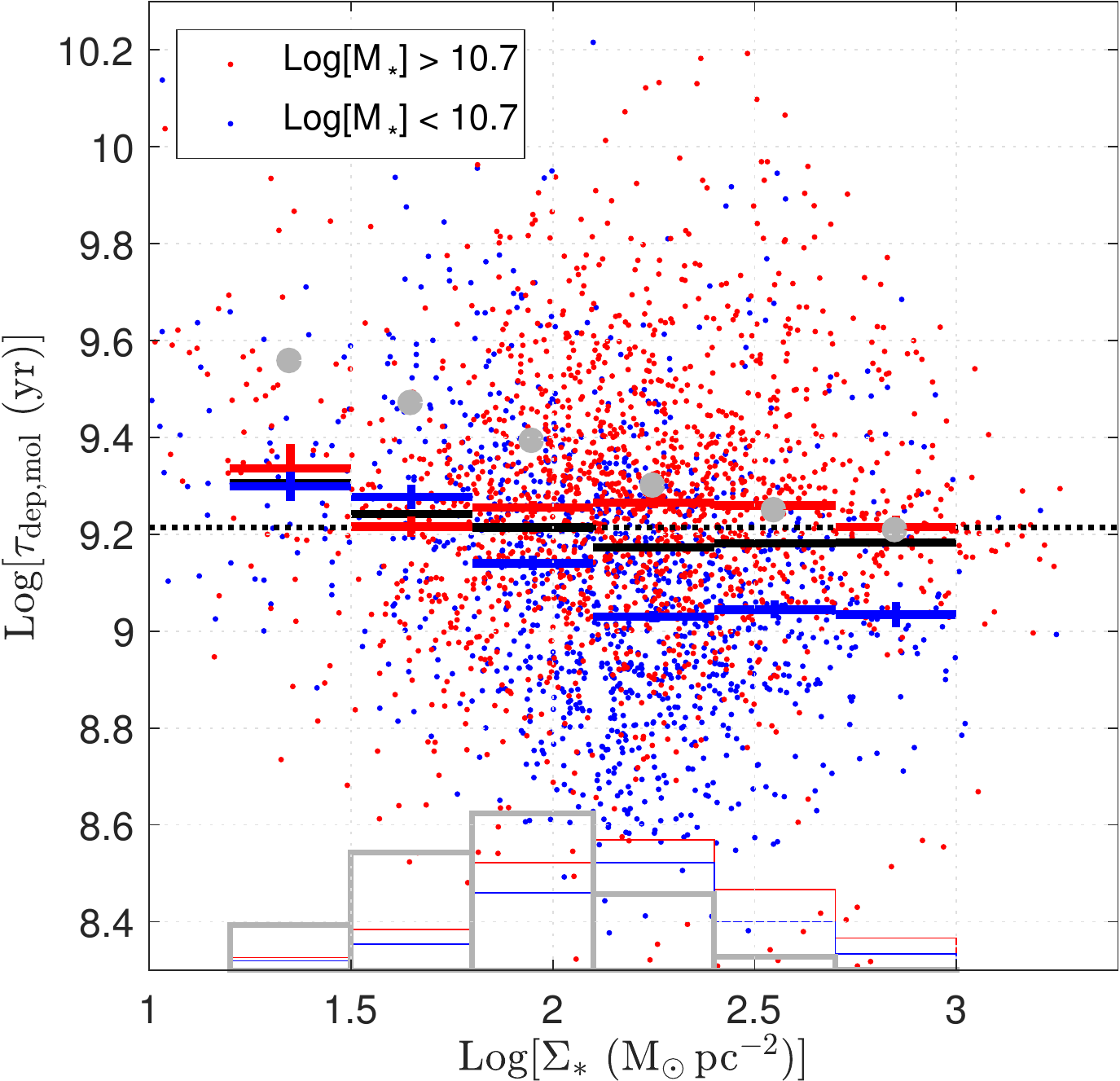} 
\includegraphics[width=0.45\textwidth]{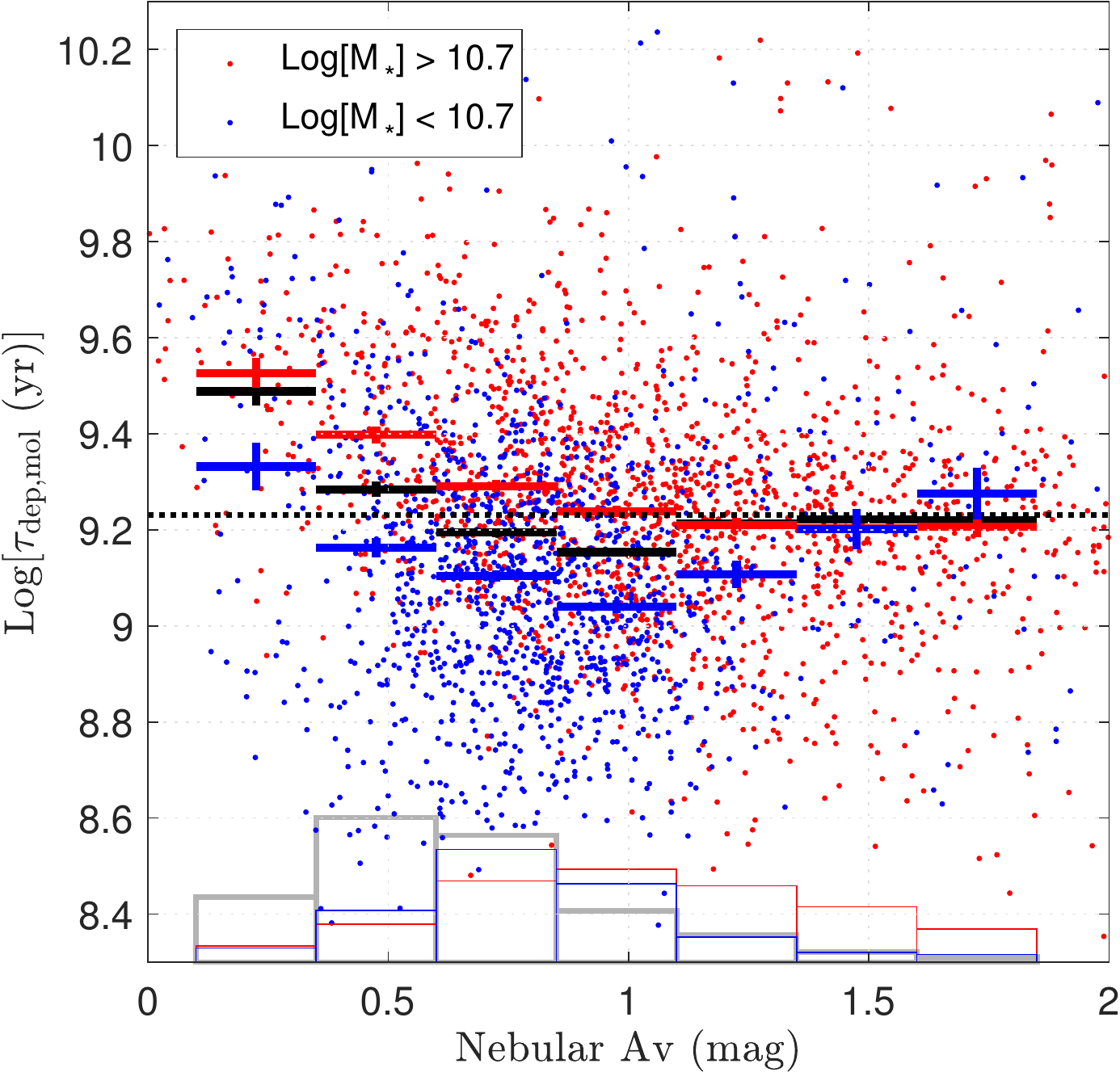}\\ 
\includegraphics[width=0.465\textwidth]{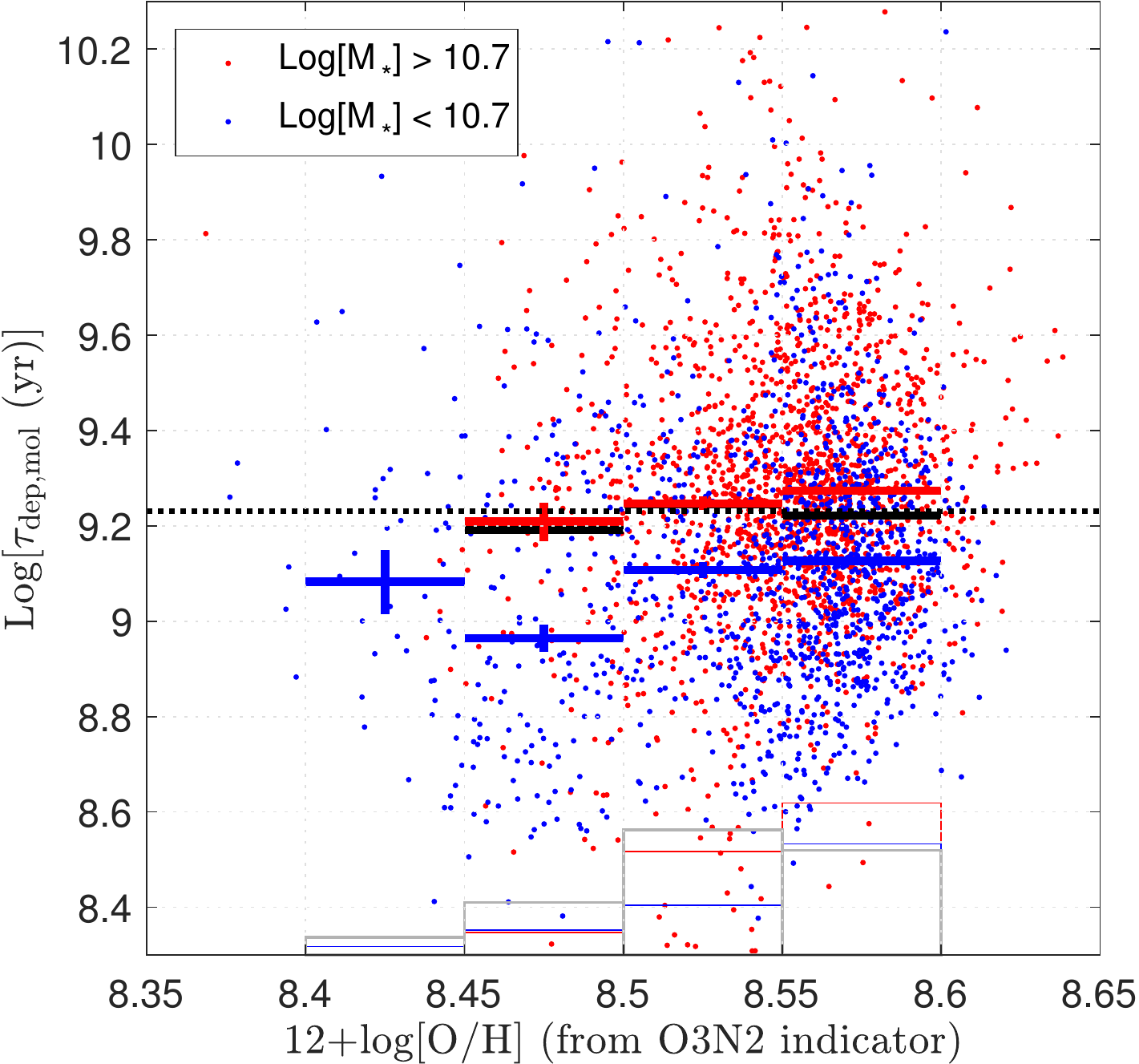} 
\includegraphics[width=0.465\textwidth]{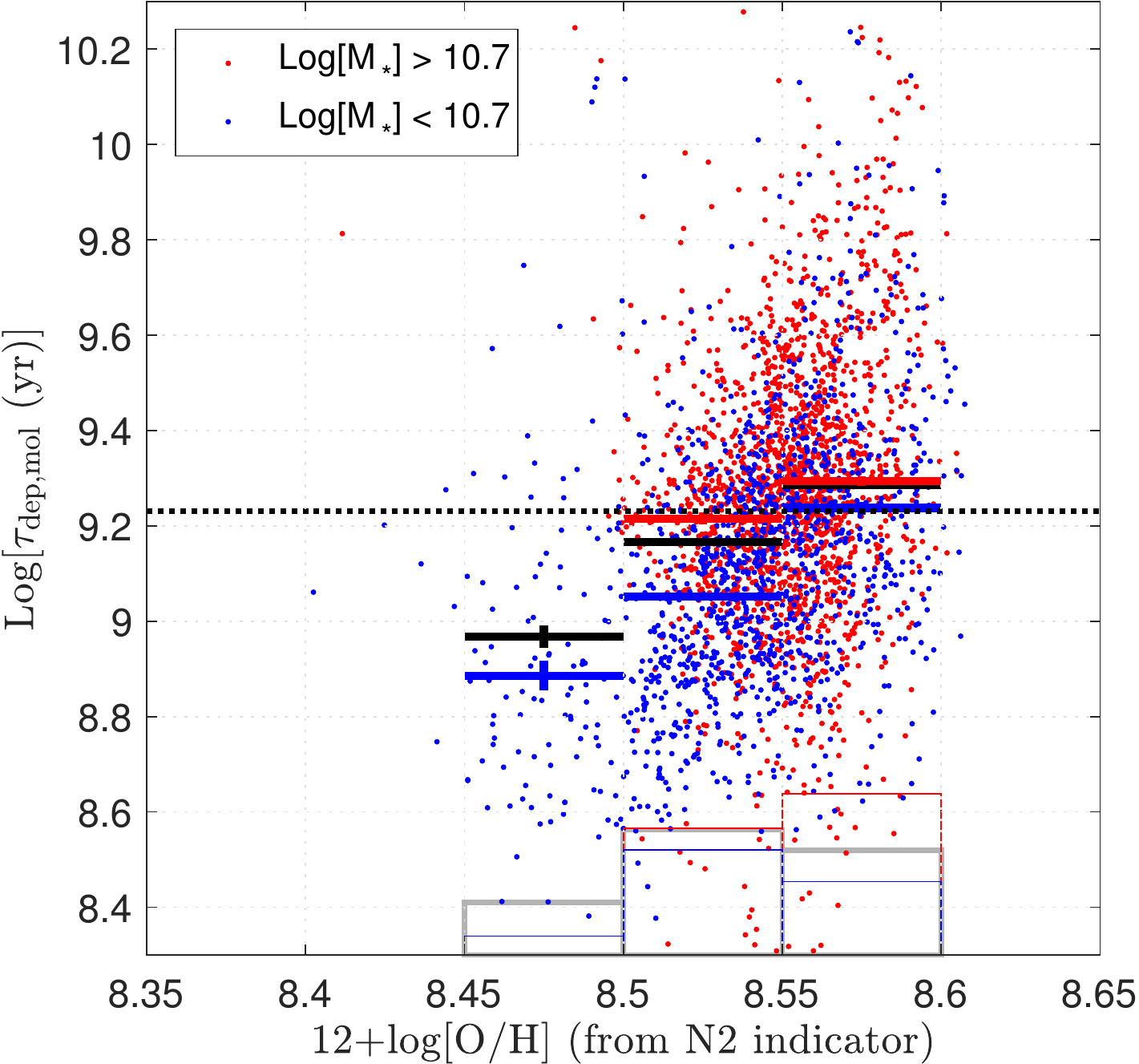} 
\end{center}
\caption{Resolved molecular depletion time in the EDGE sample, \tdepmol, as a function of different local parameters: stellar surface density, nebular extinction, and two gas metallicity indicators. The blue and red points show the results for the CO detected ($>3\sigma$) lines-of-sight for two stellar galaxy mass groups. The tick horizontal lines show the median for bins in the abscissa, color coded for mass group. The black lines show them for all the points. The bars in the bottom of each panel show the number of detected points in each galaxy mass group in their corresponding color, as well as the number of CO non-detections in gray. The horizontal dotted line shows the median \tdepmol\ for all detected lines-of-sight in the survey.\label{fig:tauvs}} 
\end{figure*}

Fig. \ref{fig:disksize} shows our results for the comparison of the molecular and stellar exponential scale lengths ($l_{\rm mol}$ and  $l_*$ respectively). The high-quality EDGE fits are shown in colored filled symbols (46 points), while the profile fits with large uncertainties either in the stellar or molecular component are in gray (large uncertainty means an error larger than 30\% or larger than 2 kpc in either scale length). The results tabulated for the SONG galaxies are shown as blue open symbols \citep[15 points from][]{Regan:01}. It is apparent that the distribution of the SONG and the EDGE results are very similar. Moreover, the results cluster around the 1:1 scaling relation (illustrated by one of the dotted lines), showing that on average the molecular and stellar scale lengths track each other, a result pointed out by other studies although with smaller statistics and using simpler tracers of the stellar distribution \citep{Young:95,Regan:01,Leroy:08}. \citet{Regan:01} also argue that a single exponential is frequently not the best description of the CO emission, while a broken exponential following the distributions of light in the bulge and disk is much better. Note, however, that our spatial resolution this is much less of a concern, since we do not resolve most bulges and in any case remove the galaxy centers from the fits. We quantify the correlation between the stellar and molecular scale lengths by fitting a relation $l_{\rm mol}=\alpha\,l_*$. A bivariate fit using uncertainties on both axes yields $\alpha=0.89\pm0.05$ (reduced $\chi^2=1.25$), while an ordinary least-squares bisector fit not weighted by the uncertainties yields $\alpha=1.05\pm0.06$; both of these are shown in Fig. \ref{fig:disksize}. The EDGE galaxies are color-coded for inclination, showing that the geometry of the system is not systematically biasing the results. 

Despite the excellent $\chi^2$, there are several systems with very small uncertainties that clearly deviate from the general trend, showing a molecular disk with a length scale considerably shorter than the stellar disk. An example is UGC\,08107, a system with a large $l_*=6.2$\,kpc, where the molecular distribution shows a much more compact $l_{\rm mol}=3.6$\,kpc many sigma away from the average relation. Inspection of Fig. \ref{fig:sdssim} shows that many of these systems are disturbed and show signs of a recent interaction. Another example is NGC\,7819, a system with a very strong bar and a central concentration of CO emission resulting in $l_{\rm mol}=1.4$\,kpc and $l_*=3.8$\,kpc. In this particular case inspection of the density profiles reveals that a single exponential is not a good description of the distribution, although the stellar and molecular profiles track each other very well in the central region. Moreover, while the CO emission is seen along the inner regions of the arms, it is much fainter there and lost in the azimuthal averaging, so the mismatch between $l_{\rm mol}$ and $l_*$ is also due in part to the limitations of our analysis. It appears that out of equilibrium situations where strong gas flows are caused by bars or interactions may frequently result in differences between $l_{\rm mol}$ and $l_*$.Figure \ref{fig:disturbed} illustrates the statistics of our sample by plotting in the ordinate the significance of the measured difference between the scale lengths $l_{\rm mol}$ and $l_*$, $\Delta l/\sigma_{\Delta l}$. Barred galaxies, for example, show an increased scatter of $2.3\sigma$ around the mean. Galaxies that are in multiple systems, on the other hand, show tantalizing systematic departures in the direction of $l_{\rm mol}<l_*$. Because there are only 6 such systems, however, the statistical significance of the difference is only marginal: a Kolmogorov-Smirnov test assigns a probability $p=0.074$ that both datasets are extracted from the same parent population. Once galaxies that have a bar, a close companion, or a ring \citep[as classified by HyperLEDA;][]{Makarov:14} are removed from the sample, the remaining 28 galaxies are  normal-distributed symmetrically around the $l_{\rm mol}=l_*$ with a standard deviation of $1.1\sigma$ (that is, their scatter is entirely compatible with the measurement errors).


\subsubsection{Molecular gas and star formation activity}
\label{sec:SFRgasscalelength}

As already mentioned, previous studies have shown that molecular gas and recent star formation activity trace each other fairly well \citep[e.g.,][]{Leroy:13b}. The large sample-size of EDGE combined with the excellent SFR determinations from extinction-corrected H$\alpha$ in CALIFA allow us to revisit the size comparison of the star-forming and molecular disks over a statistically significant sample of galaxies. 

Figure \ref{fig:coSFR} shows the comparison of the exponential fits to the SFR and molecular distributions. As with the analysis described above we exclude the $r\leq1.5$~kpc region unless the galaxy has too few radial measurements. We also use the same definition of what is a ``good quality'' fit. The comparison of scale lengths shows that $l_{\rm mol}$ and $l_{\rm SFR}$ track each other very well, so that $l_{\rm mol}=\alpha l_{\rm SFR}$ with a scaling factor $\alpha=0.92\pm0.06$. There do not appear to be strong biases associated with inclination or barredness, although good quality fits to galaxies that have companions seem to systematically yield more compact systems. But, as with the stellar disk comparison, there are strong departures for some systems. For example, IC\,0944 appears to have a $l_{\rm SFR}=8.7\pm0.8$\,kpc, measurably larger than its $l_{\rm mol}=5.2\pm0.2$\,kpc or $l_*=5.1\pm0.2$\,kpc. Indeed, inspection of its radial profiles shows that the SFR profile declines more slowly than the stellar or molecular mass profiles out to $r\sim12$~kpc, where it shows a turnover with faster decline. Its stellar mass ($\log[M_*/\msun]=11.26$) and molecular depletion time ($\log[\tdepmol/{\rm yr}]=9.6$) place it 0.4 dex over the $\tdepmol-M_*$ relation discussed in \S\ref{sec:fluxrecovery}. So as a whole it appears to be forming stars slower than its molecular mass would allow. It may be possible that something is stabilizing the molecular gas in its inner regions, lowering the efficiency of star formation and resulting in the observed $l_{\rm SFR} > l_{\rm mol}\approx l_*$. But this galaxy in particular is somewhat abnormal in that it hosts a large bulge in the center of an extremely dusty disk with prominent dust lanes and a significant inclination ($i\sim70^\circ$). Inspection of the Balmer decrement extinction correction map shows that the correction saturates (and fails) due to the non-detection of H$\beta$. So a more likely explanation is that H$\alpha$ is extincted by more than 2 mag over large areas of the disk, and the large $l_{\rm SFR}$ and depressed global SFR is simply an artifact of the incomplete extinction correction. Very few of the CALIFA galaxies are as dusty as IC\,0994.

\subsection{The Molecular to Stellar Ratio Across Galaxy Types}
\label{sec:mol2stellar}

Although the strength of EDGE, as an interferometric CO survey, is the study of the molecular gas properties and structure in connection with  the optical spectroscopic indicators,  the combination of a large number of galaxies and the well-characterize sampling of the nearby universe universe afforded by CALIFA makes its integrated quantities also interesting. In particular, although at present we have limited single-dish comparisons, independent evidence shows that the sampling of the $uv$ plane afforded by the combination of the two most compact CARMA configurations is sufficient to yield good integrated CO masses (\S\ref{sec:fluxrecovery}).  

As we introduced in \S\ref{sec:fluxrecovery}, Figure \ref{fig:tausSFR} shows that the EDGE galaxies follow the same integral relations  found in the COLD GASS survey \citep{Saintonge:11b}. On the top panel is the relation between molecular depletion time, $\tdepmol\equiv M_{\rm mol}/{\rm SFR}$, and specific star formation rate, ${\rm sSFR}\equiv{\rm SFR}/M_*$. The line and gray region show the locus and dispersion of the relation identified by COLD GASS. The bottom panel shows the relation between $\tdepmol$ and stellar mass including the trend measured by COLD GASS. This combination has the advantage of having independent variables in each axis, a reason why the relation is more scattered. The main trends and even the approximate scatter agrees well between EDGE and COLD GASS, strongly suggesting that poor total flux recovery in an integral sense is not a problem for the EDGE interferometric observations, and that other integral relations with $M_{\rm mol}$ should be well behaved. This does not mean that EDGE may not have problems recovering flux in outer disks, for example, due to signal-to-noise limitations. It simply means that it does recover the CO flux well in the region that dominates the integrated luminosity. 

With this knowledge, we can proceed to compare the total molecular masses with the stellar masses for these galaxies. Figure \ref{fig:mmolmstars} shows the molecular to stellar mass ratio, M$_{\rm mol}$/M$_*$, across the EDGE sample including upper limits. The stellar masses in CALIFA are based on the STARLIGHT modeling \citep{Cid-Fernandes:13} and assume a Salpeter IMF with a range of stellar masses M$=0.1-72$\,\msun\ \citep{Vazdekis:96}. Other popular IMF choices are Chabrier \citep{Chabrier:01} or Kroupa \citep{Kroupa:01}, which have lower mass-to-light ratios due to the truncation at masses below a Solar mass: converting to those IMFs requires dividing our stellar masses by approximately 1.6 \citep{Bell:03,Calzetti:07,Conroy:13,Madau:14}. The median molecular-to-stellar ratio we measure is 4.9\% (7.8\% for Chabrier or Kroupa IMFs) and it is fairly constant for spiral galaxies of different subtypes. Not surprisingly, there is a marked difference between spirals and early-type galaxies (which have negative morphological indexes). The median molecular-to-stellar ratio for the (poorly represented in CARMA-EDGE) early-type galaxies is an order of magnitude lower, $\sim0.6\%$. Note that although the median for spirals is approximately constant, the dispersion for individual galaxies is extremely large: any given galaxy may deviate by a factor of 4 (0.6 dex) in either direction. Some of this variation, perhaps as much as $\pm0.3$ dex, may be attributable to CO-to-H$_2$ conversion factor changes, but there are very significant galaxy-to-galaxy real variations in the amount of molecular gas per unit stellar mass.

The right panel in Figure \ref{fig:mmolmstars} shows the variation of M$_{\rm mol}$/M$_*$ versus M$_*$. The points are color-coded by presence of bars, rings, or a companion according to HyperLEDA. The stellar mass bins show a sharp decrease in the median molecular-to-stellar mass ratio for M$_*\gtrsim10^{11}$\,\msun. The color-coding also reveals a weak tendency for barred galaxies to have lower M$_{\rm mol}$/M$_*$ than non-barred galaxies. The median molecular-to-stellar mass ratio is 3.9\% for barred galaxies, and 5.1\% for galaxies that are not classified as barred. Comparison of the respective cumulative distributions is tantalizing but unclear about whether this is a significant difference: the Kolmogorov-Smirnov test assigns a probability $P=0.092$ of both datasets being drawn from the same parent distribution, based on the maximum measured cumulative difference of $D=0.223$. If real, this could signal differences in the conversion of atomic to molecular gas related to the presence of a bar, although it is also possible that our choice of a constant CO-to-\htwo\ is too simplistic and there are systematic differences in barred galaxies, although some of the differences suggested by observations go in the opposite direction \citep{Sandstrom:13}.

The right panel in Figure \ref{fig:mmolmstars} also shows the results for COLD GASS \citep{Saintonge:11a}. The histogram corresponds to the median for CO detections in the COLD GASS sample using identical stellar mass bins. The molecular masses in Table 2 of \citet{Saintonge:11a} have been rescaled to adopt our \xco\ value and include the 1.36 correction by mass corresponding to Helium. Similarly, the stellar masses have been scaled from the Chabrier IMF adopted by COLD GASS to the Salpeter IMF used by STARLIGHT and CALIFA, by multiplying them by 1.6 \citep[e.g.,][]{Madau:14}. EDGE spans a wider range of masses than the original COLD GASS sample and its galaxies tend to be slightly more gas rich, but otherwise the results from either survey are very similar. 

\subsection{The Resolved Star Formation Law}

The relation between gas and star formation has been a matter of intense investigation since the pioneering work of \citet{Schmidt:59} and \citet{Kennicutt:98}. Those studies established a link between the surface densities of gas and star formation rate, which is frequently called the ``Star Formation Law". More recent work has pointed out the key role that the molecular component of the gas plays in this relation \citep[e.g.,][]{Wong:02,Bigiel:08,Leroy:08,Leroy:13b,Rahman:12, Kennicutt:12}. In addition, recent analyses have highlighted the role that local physical conditions, including the old stellar disk, play on the formation of molecular clouds and the regulation of the star formation process \citep{Blitz:06,Leroy:08,Krumholz:09c,Eve:10,Schruba:12,Wong:13}.

EDGE provides a unique sample to extend these studies, with the statistics to probe the influence of galaxy parameters on the spatially-resolved relation between molecular gas and star formation activity. The existence of systematic variations in the relation between star formation and molecular gas was explored by \citet{Leroy:13b} for the HERACLES sample, who identify two systematic effects in the molecular depletion time: 1) a trend with dust-to-gas ratio, likely driven by the dependence of \xco\ on that parameter and thus affecting the tracer of \htwo\ and not necessarily the underlying $\Sigma_{\rm SFR}-\Sigma_{\rm mol}$ relation, and 2) a trend where central regions are distinct from disks, possibly driven by the state of the gas in galaxy centers. Forthcoming EDGE papers  explore the molecular depletion time in galaxy centers in our sample (Utomo et al., submitted), and the trends related to galaxy dynamics (Colombo et al., in prep.). Here we simply present a preliminary exploration of the star formation law for the CARMA EDGE sample, looking for these or other trends that may be present in the data. 

Figure \ref{fig:SFRvs} shows three different spatially resolved relations between the star formation rate (SFR) and a second parameter. The points come from a combination of EDGE and CALIFA data, convolved to a common 7\arcsec\ angular resolution and re-sampled on a 3.5\arcsec\ spaced hexagonal grid that produces approximately 3 points per beam, which (being close to the Nyquist sampling rate) allows to appropriately recover spatial information without grossly oversampling the data. Each point is a line-of-sight through an EDGE galaxy. The $\Sigma_{\rm SFR}$ is estimated from the local measured H$\alpha$ intensity assuming a Salpeter IMF and including the correction by nebular extinction inferred from the Balmer decrement (see \S\ref{sec:basicequations}). All measurements are corrected by the inclination of the galaxy to represent physical ``face-on'' surface densities, and objects with $i\geq75^\circ$ have been removed because their inclination-correction is very uncertain. Note that given our typical $3\sigma$ sensitivity of $\Sigma_{\rm mol}\sim11$\,\msunperpcsq\ (before inclination correction; \S\ref{sec:data}), and the fact that typical atomic disks have face-on surface densities of $\Sigma_{\rm HI}\sim10$\,\msunperpcsq, most of our measurements will be in molecular-gas-dominated regions of galactic disks.   
In order to explore dependencies on a third parameter, the measurements are color-coded by the galaxy stellar mass, with galaxies split into two comparable groups (in terms of the resolved measurements available) above and below a mass of $M_*=5\times10^{10}$\,\msun\ (for the assumed Salpeter IMF, about $M_*=3\times10^{10}$\,\msun\ for a Chabrier or Kroupa IMF). This is approximately the median mass for the EDGE sample (see Fig. \ref{fig:surveyphyspar}). Most galaxy properties correlate with galaxy mass, so in this exploration M$_*$ is simply a proxy for what may be a more direct physical driver for changes in the resolved relation. For the panels that include CO measurements, we show in gray the $3\sigma$ upper limits for non-detections. In all panels the black line represents a linear relation through the approximate center of the distribution.

The left side panel of Fig. \ref{fig:SFRvs} is the molecular version of the star formation law. The overlaid contours correspond to the distribution of the lines of sight in HERACLES \citep{Leroy:13b}. These have been corrected from CO $J=2-1$ to $J=1-0$ following Eq. 1 in \citet{Leroy:12}, and adjusted for the difference between the assumed IMFs (Eq. 3 in \citeauthor{Leroy:12} corresponds to a truncated Kroupa IMF, while for these resolved calculations we assume a Salpeter IMF for consistency with {\sc Pipe3D}). The ratio of coefficients applied to the H$\alpha$ luminosity used to correct their $\Sigma_{\rm SFR}$ data to ours is 1.5. The measurements display extremely good agreement despite the difference in the CO transition observed, conclusively showing that on average no significant bias is introduced by using CO $J=2-1$. Indeed, the excitation requirements for CO $J=2-1$ and $J=1-0$ are similar enough that we do not expect a dramatic excitation difference in the typical conditions present in molecular disks, and what difference there is can be taken into account by assuming a typical line ratio $r_{21}=0.7$. Galaxy centers will have a larger dispersion in their line-ratios, as they can frequently be more excited than the disks \citep[e.g.,][]{Israel:01}, but they constitute only a small fraction of the lines-of-sight. Similarly, possible excitation differences between arm and interam regions may be present and add in some small measure to the scatter, but they do not introduce a systematic bias.

The relation is approximately linear, but there is a segregation between the data corresponding to our two stellar mass groups. The rendition in Fig. \ref{fig:SFRvs} somewhat exaggerates the effect, but it is clear that galaxies in the lower mass group are preferentially above the black line, while galaxies in the higher mass group follow the line more closely. This segregation strongly suggests that the relation could be tightened by including a third parameter. The center panel shows the same spatially resolved $\Sigma_{\rm SFR}$ against the stellar mass surface density, $\Sigma_*$, that we used to estimate stellar disk sizes in \S\ref{sec:stargasscalelength}. The relation is, again, approximately linear, showing that there is a well-defined mean specific SFR in the EDGE disks such that $\log[{\rm sSFR/yr^{-1}}]=-10.25\pm0.32$. The approximate linearity of the relation in this and the previous panel is a re-statement of the conclusion in \S\ref{sec:stargasscalelength}: the stellar and molecular disks approximately track each other, which is why their sizes are similar. Once again, however, it is apparent that there is a segregation by galaxy mass. Note that this panel does not include molecular gas measurements, so the segregation cannot be caused by systematic differences in our conversion from CO luminosities to molecular masses. Instead, the separation is attributable to differences in the sSFR associated with the integrated stellar mass. This is the local manifestation of the fact that star-forming galaxies along the ``blue sequence'' have slightly different sSFR depending on their masses. 

The right side panel shows the relation between SFR and a proxy for dynamical equilibrium pressure (${\rm P_{DE}}$). \citet{Wong:02} and \citet{Blitz:06} show that there is a good correlation between the molecular-to-atomic ratio and the mid-plane hydrostatic pressure in galaxy disks, as inferred from the stellar and atomic gas distributions. \citet{Leroy:08} show that the star formation
efficiency for total gas correlates with $\Sigma_*$ and with the mid-plane hydrostatic pressure, finding that pressure is a better predictor of star formation efficiency in atomic-dominated regimes than gas surface density alone. If star formation is self-regulated by feedback, it is expected that $\Sigma_{\rm SFR}$ will be proportional to the dynamical-equilibrium pressure, ${\rm P_{DE}}$ \citep{Eve:10,Eve:11}. Indeed measurements of gas thermal pressure in galaxy disks are in agreement with this general picture \citep{Herrera-Camus:17}. The dynamical equilibrium pressure in a gaseous disk immersed in the stellar potential is \citep{Kim:13}

\begin{equation}
{\rm P_{DE}}=\frac{1}{2}\pi G \Sigma_{\rm gas}^2+\sigma_{\rm gas}\Sigma_{\rm gas}\sqrt{2 G \rho_*}
\end{equation}

\noindent where $\sigma_{\rm gas}$ and $\Sigma_{\rm gas}$ are the gas velocity dispersion and surface density respectively, and $\rho_*$ is the stellar volume density. The second term is usually dominant, and for a constant thickness stellar disk it is proportional to $\Sigma_{\rm gas}\sqrt{\Sigma_*}$. This interpretation relies on assuming that the molecular gas behaves as ``diffuse gas'' over large time- and length-scales, so that its pressure participates in supporting the gaseous disk of the galaxy. The observed relation is also approximately linear. Note, however, that mathematically the three relations cannot be simultaneously exactly linear: their departures from linearity are small compared to the scatter, which is the reason why they are masked by the dynamic range of the measurements. Nonetheless, the point is that the segregation among the two mass ranges persists also in this plot.

\subsection{The Resolved Molecular Depletion Time}

The best way to look for possible third parameters is to remove the mean relation and investigate trends in the residuals. Ideally, a clear dependence on a parameter would result in the points for one mass group and another clustering in areas of the \tdepmol\ vs. parameter diagram with different abscissa and ordinate, with a smooth trend present in the mean relation between both clusters of points. Here we do a first exploration of the trends in the resolved molecular depletion time, \tdepmol, across the CARMA EDGE survey.

The panels in Figure~\ref{fig:tauvs} show the dependence of \tdepmol\ on several parameters of interest. The galaxies are broken in the same two mass groups as in the previous figure. The dotted line indicates the median value of the logarithm of \tdepmol\ across all lines-of-sight detected in CO in the survey in galaxies with $i<75^\circ$, $\log[\tdepmol/{\rm yr}]=9.21\pm0.33$ (where the uncertainty indicates the $1\sigma$ scatter in the measurement), or $\tdepmol=1.64$~Gyr with a factor of $\sim2.1$ scatter.
The thick horizontal bars indicate the medians of the logarithm of \tdepmol\ in bins: the black bars (which are sometimes invisible behind a color bar) show the result for the entire sample, while the color bars show the median for each mass group. Vertical lines crossing the bars show the standard deviation of the average to gauge the significance of the differences. At the bottom of each panel there is a histogram indicating the number of points in each bin, color coded to indicate the two mass groups and, in gray, how many CO upper limits exist in each bin. 

The top left panel of Figure~\ref{fig:tauvs} shows the variation of \tdepmol\ with stellar surface density $\Sigma_*$. The increase at low $\Sigma_*$ is explained by a detection bias, since the lower $\Sigma_*$ bins include the largest fraction of non-detections. The gray symbols illustrate the results of a survival analysis, creating $\Sigma_{\rm SFR}$ points using the $\Sigma_*-\Sigma_{\rm SFR}$ relation with the measured dispersion, generating the corresponding $\Sigma_{\rm mol}$ using the \tdepmol\ measured at high $\Sigma_*$, and imposing sensitivity cuts in $\sigma_{\rm mol}$ similar to those in the data. If anything, this analysis suggests that the real \tdepmol\ likely drops by $\sim0.3$\,dex for our lowest $\Sigma_*$ bin, otherwise we would expect something analogous to the steeper rise seen in the gray points. 

The difference between our two mass groups is apparent in the upper $\Sigma_*$ bins, where CO non-detections play no significant role. The median resolved depletion time for these bins ($\Sigma_*>125$\,\msunperpcsq) is $\tdepmol=1.50$\,Gyr for both mass groups, 1.76~Gyr for the upper mass group, and 1.10~Gyr for the lower mass group. Therefore the trend for a decreasing global depletion time as a function of stellar mass seen in the bottom panel of Figure~\ref{fig:tausSFR} \citep{Saintonge:11b} is also reflected in local measurements, although the change in \tdepmol\ does not appear to be a function of the local $\Sigma_*$. Red points and blue points do not separate clearly in the abscissa, and in the regions where the statistics are good ($\log(\Sigma_*)>2.1$) the binned medians for the two mass groups are different. This suggests that \tdepmol\ is either a function of a local parameter that does not track closely with $\Sigma_*$, or perhaps it is a function of a global galaxy parameter.

The top right panel of Figure~\ref{fig:tauvs} shows a similar plot, but this time plotted against the nebular extinction inferred from the Balmer decrement (which is calculated as the extinction at the H$\alpha$ line, Eq. \ref{eq:Aha}, multiplied by 1.22 to account for the different wavelengths for a Cardelli extinction law). A potential concern is that using the Balmer decrement could under-correct the SFR in regions of high extinction, which would then have longer depletion times than they should have. This occurs because the correction essentially saturates at $\av\sim2$. It is clear that this is not a problem for the data: the trend for the objects with high extinctions and good statistics (the red bars) looks flat for $\av>0.75$. The trends present in this plot for $\av<0.75$ are attributable to a detection bias: the depletion time increases where the fraction of non-detections is very high, that is, for both mass groups at the low \av\ end. The difference in \tdepmol\ for the galaxies in both mass groups, however, is clearly present in the range $0.75\leq\av\leq1.25$ where the statistics are good, and interestingly there is a strong separation between red and blue points in the abscissa. There is also a hint in the binned medians that highly extincted regions (which are very rare for our lower mass group) could have similar depletion times in both mass groups. 

The two bottom panels show \tdepmol\ as a function of metallicity, for two nebular metallicity indicators (see \S\ref{sec:basicequations}). The N2 indicator shows a known saturation behavior at high metallicities \citep{Marino:13}, which results in a lower dynamic range than for the O3N2 indicator. There is a hint of a trend for a decreasing depletion time with decreasing metallicity that appears significant, since the detection bias would tend to push the \tdepmol\ up rather than down. This is in the direction that is expected \citep[as the CO-to-\htwo\ conversion factor increases at lower metallicities, e.g.,][]{Leroy:11,Bolatto:13a}. The separation in the abscissa between both mass groups is present in the N2 indicator (shown by the histogram at the bottom), but it is not clear in the O3N2 indicator. The N2 indicator also shows that the medians for both mass groups agree at the high metallicity end, which is another signature of a physical trend.

In summary, in this preliminary exploration we find evidence for trends of \tdepmol\ with local extinction (parametrized by \av) and possibly metallicity, but no evidence for a strong trend with local stellar surface density. This is qualitatively similar to the effects found by \citet{Leroy:13b}, who found trends for dust-to-gas ratio (a parameter that tracks both with metallicity and extinction). The explanation for the physical cause of such a trend is, in principle, the dependence of the CO-to-\htwo\ conversion factor with metallicity and dust-to-gas ratio. Note, however, that changes in \xco\ translate into changes in the amount of molecular gas present, but would not affect the SFR. Therefore, they cannot be the cause of systematic changes in the local sSFR seen in the central panel of Figure~\ref{fig:SFRvs}.
 
\subsection{Summary and Conclusions}

We present a large interferometric CO and $^{13}$CO \jone\ survey of galaxies in the nearby universe, EDGE. The CARMA EDGE sample comprises 126 galaxies, selected from the IFU CALIFA survey \citep{Sanchez:12} and observed in a combination of the CARMA E and D configurations for typically 4.3 hours each. Although most of our galaxies were selected for their 22\,\micron\ brightness and convenience of scheduling, we show that they constitute a representative sample of the star-forming galaxies in CALIFA. 

Our typical angular resolution is 4.5\arcsec, well-matched to the 2.7\arcsec\ fiber size of CALIFA. Our typical $3\sigma$ surface density sensitivity is approximately $\Sigma_{\rm mol}=11$\,\msunperpcsq\ assuming a Galactic CO-to-\htwo\ conversion factor and a typical linewidth of 30\,\kmpers. The range of spatial structures recovered by the CARMA configurations used is similar to or larger than the optical sizes of the galaxies. Moreover, our integrated CO fluxes follow the relations observed in unresolved single-dish large galaxy samples \citep{Saintonge:11b}.   

In this, the first of many studies that will use the combined EDGE-CALIFA dataset, we explore four topics:
\begin{enumerate}

\item {\em The sizes of the molecular, stellar, and star-forming disks of galaxies}. Using galaxy radial profiles for molecular surface density (from EDGE observations), stellar surface density (from CALIFA optical SED fitting), and SFR surface density (from H$\alpha$ locally extinction-corrected by the Balmer decrement), we find that these three sizes are very comparable. Fitting exponential disk scale-lengths we find that the best fit relations are $l_{\rm mol}=(0.89\pm0.05)\,l_*$, and $l_{\rm mol}=(0.92\pm0.06)\,l_{\rm SFR}$. We also investigate the relation between half-mass radii (a simpler measurement that can be performed even when the distributions are not exponential, or when sources are poorly resolved), and find that $R_{\rm 1/2}^{\rm mol}=(1.00\pm0.03)\,R_{\rm 1/2}^{*}$. Thus these three components track each other very well, a result that has been found before although with smaller samples \citep[e.g.,][]{Regan:01,Leroy:08,Schruba:12}. The significant deviations in the direction of compact molecular gas distributions with $l_{\rm mol}<l_*$ appear to be associated with galaxies undergoing some level of interaction, although bars tend to also add scatter to the relation.

\item {\em The molecular to stellar mass ratio across galaxy types}. We find a median total molecular-to-stellar mass ratio of 4.9\% for a Salpeter IMF (7.8\% for a Chabrier or Kroupa IMF), with a significant dispersion of $\sim0.6$ dex. This ratio is fairly constant for spirals of all types in the sample, but it decreases dramatically for early type galaxies (0.6\% for a Salpeter IMF), and it falls noticeably for galaxies that have $M_*>10^{11}$\,\msun. Our results are very consistent with those found by COLD GASS \citep{Saintonge:11a} over the range of masses where the samples overlap. There is a hint that barred galaxies (as defined by HyperLEDA) may have a slightly lower mass ratio than galaxies that are not barred, but a Kolmogorov-Smirnov test finds the significance of the difference to be low.

\item {\em The resolved star formation law}. We show that there are approximately linear relations between $\Sigma_{\rm SFR}$ and $\Sigma_{\rm mol}$, $\Sigma_*$, and a proxy for dynamical equilibrium pressure for a constant thickness gaseous disk $\Sigma_{\rm mol}\sqrt{\Sigma_*}$. We convincingly reproduce the relation observed by HERACLES \citep{Leroy:13b}, but also find that there is a segregation in all three relations between lines-of-sight corresponding to galaxies above or below $M_*=5\times10^{10}$\,\msun\ for a Salpeter IMF ($M_*=3\times10^{10}$\,\msun\ for a Chabrier or Kroupa IMF,
a mass simply chosen to split the points in two groups of about equal size).  

\item {\em The resolved molecular depletion time}. Following the dependence of the global molecular depletion time on the stellar mass of a galaxy \citep{Saintonge:11b}, we explore the dependence of \tdepmol\ on local physical parameters. We find tentative trends with local nebular extinction and metallicity, which are qualitatively similar to the trends found in HERACLES by \citet{Leroy:13b}. There is no clear trend with $\Sigma_*$ once the censoring of the data at low $\Sigma_*$ is taken into account. The segregation by galaxy mass is apparent in these data. This is a very preliminary exploration of trends, which will be pursued further in follow up papers.  
\end{enumerate}

Although they represent only a fraction of the studies that will be undertaken with EDGE, these results \citep[and others in the literature, for example][]{Galbany:17} already illustrate the excellent promise that the combination of optical IFU and mm-wave interferometric surveys hold to disentangle the physical processes driving galaxy evolution. CALIFA constitutes an ideal sample to follow with interferometric observations because of the combination of field of view, spatial resolution, and surface brightness sensitivity attainable by interferometers. With a few hours per object, CARMA was able to attain interesting sensitivities while tripling the number of galaxies in the largest pre-existing CO interferometric survey (BIMA SONG). Large numbers of objects allow us to explore trends with galaxy type, mass, and other parameters that are otherwise impossible to investigate, adding a statistical dimension to resolved molecular gas surveys that has been previously lacking. In a number of follow-up papers, we study kinematics, dynamical models, and molecular gas depletion time dependencies. Note that many of the CALIFA galaxies are south of $\delta=+35^\circ$, so in a reasonable amount of time it is possible to add to this sample of 126 galaxies another $\sim180$ galaxies using the Atacama Large Millimeter-wave Array (ALMA). A 300 galaxies sample with interferometric molecular gas and IFU observations would allow us to extend our exploration of the physical trends to include early-type and low-mass galaxies, now poorly represented in EDGE. 

We will release the combined EDGE-CALIFA dataset to the public once the quality-assessment is complete. We have constructed a relational database implemented in SQL-lite that we are in the process of validating and testing. This database greatly simplifies the task of generating data such as those employed in Figures \ref{fig:SFRvs} and \ref{fig:tauvs}, and in general allows for a simpler exploitation of the complementarity of the spatially-resolved CALIFA and EDGE observations. We will also release simultaneously the more traditional cubes and maps corresponding to the CO observations.

\acknowledgements
We thank Andrew Harris for useful insight and discussions, and the referee for constructive comments. ADB and RCL acknowledge support from NSF through grants AST-1412419 and AST-1615960. ADB also acknowledges visiting support by the Alexander von Humboldt Foundation. SNV and PT acknowledge support from NSF AST-1615960. TW and YC acknowledge support from NSF through grants AST-1139950 and AST-1616199. DU and LB are supported by the NSF under grants AST-1140063 and AST-1616924. ECO is is supported by NSF grant AST-1312006. AKL is partially supported by the NSF through grants AST-1615105, AST-1615109, and AST-1653300. SFS acknowledges the PAPIIT-DGAPA-IA101217 project and CONACYT-IA-180125. ER is supported by a Discovery Grant from NSERC of Canada. DC acknowledges support by the {\em Deutsche Forschungsgemeinschaft}, DFG through project number SFB956C. RGB acknowledges support from grant AYA2016-77846-P. HD acknowledges financial support from the Spanish Ministry of Economy and Competitiveness (MINECO) under the 2014 Ram\'on y Cajal program MINECO RYC-2014-15686.
Support for CARMA construction was derived from the Gordon and Betty Moore Foundation, the Eileen and Kenneth Norris Foundation, the Caltech Associates, the states of California, Illinois, and Maryland, and the NSF. Funding for CARMA development and operations were supported by NSF and the CARMA partner universities. 

\newpage

\begin{deluxetable*}{lcccccrrrrr}
\tablewidth{0pt}
\tabletypesize{\footnotesize}
\tablecaption{EDGE Survey Galaxy Properties\label{tab:table1}}
\tablehead{
\colhead{Name} & \colhead{R.A.} & \colhead{Dec.} & \colhead{Type} & \colhead{Notes} & \colhead{V$_{\rm LSR}$} & \colhead{D$_{25}$} & \colhead{Inc} & \colhead{PA} & \colhead{Dist} \\ 
\colhead{} & \colhead{(J2000)} & \colhead{(J2000)} & \colhead{} & \colhead{} & \colhead{(\kmpers)} & \colhead{($\arcsec$)} & \colhead{($^\circ$)} & \colhead{($^\circ$)} & \colhead{(Mpc)} 
}
\startdata
ARP220 & $15^{\rm h}34^{\rm m}57\fs3$ & $+23^\circ30\arcmin09\farcs7$ & Sm &  & 5247 & 66 & 30 & 338 & 78.0 \\ 
IC0480 & $07^{\rm h}55^{\rm m}23\fs1$ & $+26^\circ44\arcmin34\farcs0$ & Sbc &  & 4595 & 105 & 77 & 168 & 66.3 \\ 
IC0540 & $09^{\rm h}30^{\rm m}10\fs2$ & $+07^\circ54\arcmin09\farcs3$ & Sb &  & 2022 & 64 & 68 & 350 & 29.9 \\ 
IC0944 & $13^{\rm h}51^{\rm m}30\fs9$ & $+14^\circ05\arcmin31\farcs2$ & Sa &  & 6907 & 92 & 75 & 106 & 100.8 \\ 
IC1151 & $15^{\rm h}58^{\rm m}32\fs5$ & $+17^\circ26\arcmin29\farcs4$ & SBc & B & 2192 & 134 & 68 & 204 & 30.8 \\ 
IC1199 & $16^{\rm h}10^{\rm m}34\fs3$ & $+10^\circ02\arcmin24\farcs3$ & Sbc &  & 4686 & 71 & 64 & 339 & 68.3 \\ 
IC1683 & $01^{\rm h}22^{\rm m}38\fs8$ & $+34^\circ26\arcmin13\farcs1$ & Sb &  & 4820 & 79 & 55 & 16 & 69.7 \\ 
IC2247 & $08^{\rm h}15^{\rm m}59\fs0$ & $+23^\circ11\arcmin58\farcs5$ & Sbc &  & 4254 & 109 & 78 & 328 & 62.0 \\ 
IC2487 & $09^{\rm h}30^{\rm m}09\fs2$ & $+20^\circ05\arcmin27\farcs2$ & Sb &  & 4310 & 110 & 78 & 163 & 62.3 \\ 
IC4566 & $15^{\rm h}36^{\rm m}42\fs1$ & $+43^\circ32\arcmin21\farcs8$ & SABb & B & 5537 & 81 & 54 & 145 & 80.7 \\ 
IC5376 & $00^{\rm h}01^{\rm m}19\fs7$ & $+34^\circ31\arcmin32\farcs5$ & Sab &  & 4979 & 98 & 72 & 3 & 72.9 \\ 
NGC0444 & $01^{\rm h}15^{\rm m}49\fs6$ & $+31^\circ04\arcmin48\farcs7$ & Sc &  & 4776 & 94 & 75 & 159 & 70.1 \\ 
NGC0447 & $01^{\rm h}15^{\rm m}37\fs5$ & $+33^\circ04\arcmin03\farcs3$ & S0-a & B & 5552 & 133 & 29 & 227 & 79.7 \\ 
NGC0477 & $01^{\rm h}21^{\rm m}20\fs3$ & $+40^\circ29\arcmin17\farcs5$ & Sc & B & 5796 & 93 & 60 & 140 & 85.4 \\ 
NGC0496 & $01^{\rm h}23^{\rm m}11\fs5$ & $+33^\circ31\arcmin44\farcs0$ & Sbc &  & 5958 & 53 & 57 & 36 & 87.5 \\ 
NGC0523 & $01^{\rm h}25^{\rm m}20\fs7$ & $+34^\circ01\arcmin29\farcs0$ & Sbc & M & 4760 & 151 & 72 & 277 & 67.9 \\ 
NGC0528 & $01^{\rm h}25^{\rm m}33\fs5$ & $+33^\circ40\arcmin17\farcs4$ & S0 &  & 4638 & 106 & 61 & 58 & 68.8 \\ 
NGC0551 & $01^{\rm h}27^{\rm m}40\fs6$ & $+37^\circ10\arcmin58\farcs4$ & SBbc & B & 5141 & 89 & 64 & 315 & 74.5 \\ 
NGC1167 & $03^{\rm h}01^{\rm m}42\fs3$ & $+35^\circ12\arcmin20\farcs5$ & S0 &  & 4797 & 109 & 40 & 88 & 70.9 \\ 
NGC2253 & $06^{\rm h}43^{\rm m}41\fs8$ & $+65^\circ12\arcmin22\farcs3$ & Sc &  & 3545 & 86 & 47 & 300 & 51.2 \\ 
NGC2347 & $07^{\rm h}16^{\rm m}03\fs9$ & $+64^\circ42\arcmin38\farcs8$ & Sb & R & 4387 & 99 & 50 & 189 & 63.7 \\ 
NGC2410 & $07^{\rm h}35^{\rm m}02\fs2$ & $+32^\circ49\arcmin19\farcs5$ & Sb & B & 4642 & 130 & 72 & 217 & 67.5 \\ 
NGC2480 & $07^{\rm h}57^{\rm m}10\fs4$ & $+23^\circ46\arcmin47\farcs2$ & SBd & BM & 2287 & 79 & 55 & 343 & 33.1 \\ 
NGC2486 & $07^{\rm h}57^{\rm m}56\fs4$ & $+25^\circ09\arcmin38\farcs8$ & Sa &  & 4569 & 89 & 56 & 93 & 67.5 \\ 
NGC2487 & $07^{\rm h}58^{\rm m}20\fs4$ & $+25^\circ08\arcmin57\farcs1$ & Sb & BR & 4795 & 104 & 31 & 118 & 70.5 \\ 
NGC2623 & $08^{\rm h}38^{\rm m}24\fs0$ & $+25^\circ45\arcmin14\farcs7$ & Sb & M & 5454 & 45 & 46 & 255 & 80.2 \\ 
NGC2639 & $08^{\rm h}43^{\rm m}38\fs0$ & $+50^\circ12\arcmin19\farcs4$ & Sa & R & 3168 & 97 & 50 & 314 & 45.7 \\ 
NGC2730 & $09^{\rm h}02^{\rm m}15\fs7$ & $+16^\circ50\arcmin17\farcs8$ & Sd & B & 3727 & 87 & 28 & 261 & 54.8 \\ 
NGC2880 & $09^{\rm h}29^{\rm m}34\fs5$ & $+62^\circ29\arcmin26\farcs1$ & E-S0 & B & 1530 & 140 & 50 & 323 & 22.7 \\ 
NGC2906 & $09^{\rm h}32^{\rm m}06\fs2$ & $+08^\circ26\arcmin29\farcs7$ & Sc &  & 2133 & 81 & 56 & 262 & 37.7 \\ 
NGC2916 & $09^{\rm h}34^{\rm m}57\fs5$ & $+21^\circ42\arcmin18\farcs7$ & Sb &  & 3620 & 144 & 50 & 200 & 53.2 \\ 
NGC2918 & $09^{\rm h}35^{\rm m}44\fs0$ & $+31^\circ42\arcmin19\farcs8$ & E &  & 6569 & 93 & 46 & 75 & 96.6 \\ 
NGC3303 & $10^{\rm h}37^{\rm m}00\fs1$ & $+18^\circ08\arcmin08\farcs5$ & Sa & M & 6040 & 66 & 60 & 160 & 89.8 \\ 
NGC3381 & $10^{\rm h}48^{\rm m}24\fs8$ & $+34^\circ42\arcmin41\farcs0$ & SBb & B & 1625 & 121 & 31 & 333 & 23.4 \\ 
NGC3687 & $11^{\rm h}28^{\rm m}00\fs4$ & $+29^\circ30\arcmin39\farcs5$ & Sbc & BR & 2497 & 84 & 20 & 326 & 36.0 \\ 
NGC3811 & $11^{\rm h}41^{\rm m}16\fs8$ & $+47^\circ41\arcmin26\farcs8$ & SBc & BR & 3073 & 122 & 40 & 352 & 44.3 \\ 
NGC3815 & $11^{\rm h}41^{\rm m}39\fs1$ & $+24^\circ48\arcmin01\farcs4$ & Sab &  & 3686 & 86 & 60 & 68 & 53.6 \\ 
NGC3994 & $11^{\rm h}57^{\rm m}36\fs7$ & $+32^\circ16\arcmin38\farcs2$ & Sc & RM & 3097 & 51 & 60 & 188 & 44.7 \\ 
NGC4047 & $12^{\rm h}02^{\rm m}50\fs6$ & $+48^\circ38\arcmin10\farcs3$ & Sb &  & 3419 & 92 & 42 & 104 & 49.1 \\ 
NGC4149 & $12^{\rm h}10^{\rm m}32\fs8$ & $+58^\circ18\arcmin14\farcs7$ & SABc & B & 3050 & 82 & 66 & 85 & 44.1 \\ 
NGC4185 & $12^{\rm h}13^{\rm m}22\fs0$ & $+28^\circ30\arcmin39\farcs5$ & SBbc & BR & 3874 & 111 & 48 & 344 & 55.9 \\ 
NGC4210 & $12^{\rm h}15^{\rm m}15\fs8$ & $+65^\circ59\arcmin07\farcs4$ & Sb & BR & 2714 & 116 & 41 & 278 & 38.8 \\ 
NGC4211NED02 & $12^{\rm h}15^{\rm m}37\fs4$ & $+28^\circ10\arcmin10\farcs5$ & S0-a & M & 6605 & 62 & 30 & 25 & 96.9 \\ 
NGC4470 & $12^{\rm h}29^{\rm m}37\fs6$ & $+07^\circ49\arcmin26\farcs0$ & Sa &  & 2338 & 77 & 48 & 350 & 33.4 \\ 
NGC4644 & $12^{\rm h}42^{\rm m}42\fs8$ & $+55^\circ08\arcmin43\farcs7$ & Sb & BM & 4915 & 91 & 73 & 57 & 71.6 \\ 
NGC4676A & $12^{\rm h}46^{\rm m}10\fs2$ & $+30^\circ43\arcmin55\farcs5$ & S0-a & BM & 6541 & 130 & 50 & 185 & 96.6 \\ 
NGC4711 & $12^{\rm h}48^{\rm m}45\fs7$ & $+35^\circ19\arcmin57\farcs7$ & SBb & B & 4044 & 72 & 58 & 215 & 58.8 \\ 
NGC4961 & $13^{\rm h}05^{\rm m}47\fs7$ & $+27^\circ44\arcmin02\farcs0$ & SBc & B & 2521 & 67 & 47 & 100 & 36.6 \\ 
NGC5000 & $13^{\rm h}09^{\rm m}47\fs5$ & $+28^\circ54\arcmin24\farcs4$ & Sbc & BRM & 5557 & 77 & 20 & 31 & 80.8 \\ 
NGC5016 & $13^{\rm h}12^{\rm m}06\fs8$ & $+24^\circ05\arcmin41\farcs0$ & SABb & B & 2581 & 94 & 40 & 57 & 36.9 \\ 
NGC5056 & $13^{\rm h}16^{\rm m}12\fs3$ & $+30^\circ57\arcmin00\farcs7$ & Sc &  & 5550 & 97 & 61 & 178 & 81.1 \\ 
NGC5205 & $13^{\rm h}30^{\rm m}03\fs5$ & $+62^\circ30\arcmin41\farcs3$ & Sbc &  & 1762 & 95 & 50 & 169 & 25.1 \\ 
NGC5218 & $13^{\rm h}32^{\rm m}10\fs3$ & $+62^\circ46\arcmin04\farcs0$ & SBb & B & 2888 & 110 & 30 & 236 & 41.7 \\ 
NGC5394 & $13^{\rm h}58^{\rm m}33\fs6$ & $+37^\circ27\arcmin12\farcs5$ & SBb & BM & 3431 & 158 & 70 & 189 & 49.5 \\ 
NGC5406 & $14^{\rm h}00^{\rm m}20\fs1$ & $+38^\circ54\arcmin55\farcs4$ & Sbc & B & 5350 & 102 & 45 & 111 & 77.8 \\ 
NGC5480 & $14^{\rm h}06^{\rm m}21\fs5$ & $+50^\circ43\arcmin30\farcs3$ & Sc &  & 1882 & 100 & 42 & 178 & 27.0 \\ 
NGC5485 & $14^{\rm h}07^{\rm m}11\fs2$ & $+55^\circ00\arcmin05\farcs7$ & S0 &  & 1893 & 151 & 47 & 74 & 26.9 \\ 
NGC5520 & $14^{\rm h}12^{\rm m}22\fs6$ & $+50^\circ20\arcmin54\farcs6$ & Sb &  & 1870 & 96 & 59 & 245 & 26.7 \\ 
NGC5614 & $14^{\rm h}24^{\rm m}07\fs5$ & $+34^\circ51\arcmin32\farcs0$ & Sab & RM & 3859 & 146 & 36 & 270 & 55.7 \\ 
NGC5633 & $14^{\rm h}27^{\rm m}28\fs4$ & $+46^\circ08\arcmin47\farcs0$ & Sb & R & 2319 & 65 & 42 & 17 & 33.4 \\ 
NGC5657 & $14^{\rm h}30^{\rm m}43\fs5$ & $+29^\circ10\arcmin50\farcs5$ & Sb & BR & 3860 & 105 & 68 & 349 & 56.3 \\ 
NGC5682 & $14^{\rm h}34^{\rm m}45\fs1$ & $+48^\circ40\arcmin10\farcs1$ & Sb & B & 2242 & 39 & 76 & 311 & 32.6 \\ 
NGC5732 & $14^{\rm h}40^{\rm m}39\fs0$ & $+38^\circ38\arcmin15\farcs7$ & Sbc &  & 3723 & 74 & 58 & 43 & 54.0 \\ 
NGC5784 & $14^{\rm h}54^{\rm m}16\fs5$ & $+42^\circ33\arcmin28\farcs0$ & S0 &  & 5427 & 89 & 45 & 255 & 79.4 \\ 
NGC5876 & $15^{\rm h}09^{\rm m}31\fs6$ & $+54^\circ30\arcmin23\farcs4$ & SBab & BR & 3240 & 148 & 66 & 51 & 46.9 \\ 
NGC5908 & $15^{\rm h}16^{\rm m}43\fs3$ & $+55^\circ24\arcmin33\farcs8$ & Sb &  & 3294 & 202 & 77 & 153 & 47.1 \\ 
NGC5930 & $15^{\rm h}26^{\rm m}07\fs7$ & $+41^\circ40\arcmin33\farcs9$ & SABa & BM & 2637 & 111 & 45 & 155 & 37.2 \\ 
NGC5934 & $15^{\rm h}28^{\rm m}12\fs7$ & $+42^\circ55\arcmin47\farcs6$ & Sa & M & 5566 & 37 & 55 & 5 & 82.7 \\ 
NGC5947 & $15^{\rm h}30^{\rm m}36\fs7$ & $+42^\circ43\arcmin01\farcs9$ & SBbc & B & 5898 & 70 & 32 & 249 & 86.1 \\ 
NGC5953 & $15^{\rm h}34^{\rm m}32\fs5$ & $+15^\circ11\arcmin37\farcs6$ & S0-a & M & 1988 & 88 & 26 & 48 & 28.4 \\ 
NGC5980 & $15^{\rm h}41^{\rm m}30\fs4$ & $+15^\circ47\arcmin15\farcs3$ & Sbc &  & 4060 & 98 & 66 & 15 & 59.4 \\ 
NGC6004 & $15^{\rm h}50^{\rm m}22\fs5$ & $+18^\circ56\arcmin21\farcs1$ & Sc & B & 3818 & 114 & 37 & 272 & 55.2 \\ 
NGC6021 & $15^{\rm h}57^{\rm m}30\fs6$ & $+15^\circ57\arcmin21\farcs5$ & E &  & 4673 & 93 & 43 & 157 & 69.1 \\ 
NGC6027 & $15^{\rm h}59^{\rm m}12\fs4$ & $+20^\circ45\arcmin47\farcs8$ & S0-a & BM & 4338 & 137 & 31 & 231 & 62.9 \\ 
NGC6060 & $16^{\rm h}05^{\rm m}52\fs0$ & $+21^\circ29\arcmin05\farcs6$ & SABc & BR & 4398 & 114 & 64 & 102 & 63.2 \\ 
NGC6063 & $16^{\rm h}07^{\rm m}13\fs0$ & $+07^\circ58\arcmin43\farcs9$ & Sc &  & 2807 & 98 & 56 & 332 & 40.7 \\ 
NGC6081 & $16^{\rm h}12^{\rm m}56\fs8$ & $+09^\circ52\arcmin01\farcs3$ & S0 &  & 4978 & 93 & 66 & 308 & 73.5 \\ 
NGC6125 & $16^{\rm h}19^{\rm m}11\fs6$ & $+57^\circ59\arcmin02\farcs7$ & E &  & 4522 & 83 & 17 & 5 & 68.0 \\ 
NGC6146 & $16^{\rm h}25^{\rm m}10\fs1$ & $+40^\circ53\arcmin33\farcs3$ & E &  & 8693 & 100 & 41 & 78 & 128.7 \\ 
NGC6155 & $16^{\rm h}26^{\rm m}08\fs1$ & $+48^\circ22\arcmin00\farcs4$ & Sc &  & 2418 & 80 & 45 & 130 & 34.6 \\ 
NGC6168 & $16^{\rm h}31^{\rm m}21\fs0$ & $+20^\circ11\arcmin07\farcs8$ & Sd &  & 2540 & 99 & 77 & 111 & 36.1 \\ 
NGC6186 & $16^{\rm h}34^{\rm m}25\fs6$ & $+21^\circ32\arcmin27\farcs2$ & Sa & B & 2940 & 94 & 71 & 70 & 42.4 \\ 
NGC6301 & $17^{\rm h}08^{\rm m}32\fs6$ & $+42^\circ20\arcmin20\farcs3$ & Sc & R & 8222 & 106 & 53 & 288 & 121.4 \\ 
NGC6310 & $17^{\rm h}07^{\rm m}57\fs3$ & $+60^\circ59\arcmin24\farcs3$ & Sb &  & 3459 & 101 & 74 & 70 & 48.7 \\ 
NGC6314 & $17^{\rm h}12^{\rm m}38\fs8$ & $+23^\circ16\arcmin12\farcs7$ & Sa &  & 6551 & 82 & 58 & 356 & 95.9 \\ 
NGC6361 & $17^{\rm h}18^{\rm m}41\fs0$ & $+60^\circ36\arcmin29\farcs1$ & Sb &  & 3789 & 123 & 75 & 47 & 54.9 \\ 
NGC6394 & $17^{\rm h}30^{\rm m}21\fs6$ & $+59^\circ38\arcmin23\farcs6$ & SBb & B & 8444 & 69 & 60 & 232 & 124.3 \\ 
NGC6478 & $17^{\rm h}48^{\rm m}38\fs1$ & $+51^\circ09\arcmin25\farcs9$ & Sc &  & 6797 & 87 & 73 & 29 & 97.4 \\ 
NGC7738 & $23^{\rm h}44^{\rm m}02\fs0$ & $+00^\circ31\arcmin00\farcs1$ & Sb & B & 6682 & 71 & 66 & 235 & 97.8 \\ 
NGC7819 & $00^{\rm h}04^{\rm m}24\fs4$ & $+31^\circ28\arcmin19\farcs2$ & Sb & B & 4918 & 86 & 54 & 270 & 71.6 \\ 
UGC00809 & $01^{\rm h}15^{\rm m}51\fs8$ & $+33^\circ48\arcmin38\farcs5$ & Sc &  & 4171 & 71 & 79 & 19 & 60.4 \\ 
UGC03253 & $05^{\rm h}19^{\rm m}41\fs6$ & $+84^\circ03\arcmin08\farcs0$ & Sb & BR & 4040 & 82 & 58 & 268 & 59.5 \\ 
UGC03539 & $06^{\rm h}48^{\rm m}53\fs9$ & $+66^\circ15\arcmin40\farcs6$ & Sbc & B & 3278 & 110 & 72 & 303 & 47.1 \\ 
UGC03969 & $07^{\rm h}41^{\rm m}14\fs3$ & $+27^\circ36\arcmin50\farcs7$ & Sc &  & 8037 & 76 & 70 & 134 & 118.4 \\ 
UGC03973 & $07^{\rm h}42^{\rm m}32\fs5$ & $+49^\circ48\arcmin35\farcs2$ & Sb & B & 6594 & 80 & 39 & 144 & 95.9 \\ 
UGC04029 & $07^{\rm h}48^{\rm m}19\fs0$ & $+34^\circ19\arcmin55\farcs9$ & Sbc & B & 4389 & 108 & 78 & 58 & 63.5 \\ 
UGC04132 & $07^{\rm h}59^{\rm m}13\fs0$ & $+32^\circ54\arcmin53\farcs6$ & Sbc &  & 5151 & 74 & 72 & 213 & 75.4 \\ 
UGC04280 & $08^{\rm h}14^{\rm m}33\fs3$ & $+54^\circ47\arcmin58\farcs2$ & Sa &  & 3500 & 79 & 72 & 184 & 50.9 \\ 
UGC04461 & $08^{\rm h}33^{\rm m}22\fs6$ & $+52^\circ31\arcmin56\farcs2$ & Sbc &  & 4941 & 83 & 70 & 223 & 72.3 \\ 
UGC05108 & $09^{\rm h}35^{\rm m}26\fs3$ & $+29^\circ48\arcmin45\farcs3$ & SBab & B & 8015 & 66 & 66 & 136 & 118.4 \\ 
UGC05111 & $09^{\rm h}36^{\rm m}52\fs4$ & $+66^\circ47\arcmin18\farcs2$ & Sbc &  & 6660 & 90 & 73 & 118 & 98.2 \\ 
UGC05244 & $09^{\rm h}48^{\rm m}48\fs1$ & $+64^\circ10\arcmin04\farcs8$ & Sc &  & 2974 & 90 & 78 & 33 & 43.7 \\ 
UGC05359 & $09^{\rm h}58^{\rm m}51\fs6$ & $+19^\circ12\arcmin53\farcs0$ & SABb & B & 8344 & 93 & 72 & 94 & 123.2 \\ 
UGC05498NED01 & $10^{\rm h}12^{\rm m}03\fs6$ & $+23^\circ05\arcmin07\farcs4$ & Sa & M & 6250 & 94 & 81 & 62 & 91.8 \\ 
UGC05598 & $10^{\rm h}22^{\rm m}14\fs1$ & $+20^\circ35\arcmin21\farcs8$ & Sbc &  & 5591 & 76 & 75 & 216 & 81.1 \\ 
UGC06312 & $11^{\rm h}18^{\rm m}00\fs0$ & $+07^\circ50\arcmin40\farcs7$ & Sa & B & 6266 & 71 & 69 & 225 & 90.0 \\ 
UGC07012 & $12^{\rm h}02^{\rm m}03\fs1$ & $+29^\circ50\arcmin53\farcs1$ & SBc & B & 3052 & 65 & 60 & 184 & 44.3 \\ 
UGC08107 & $12^{\rm h}59^{\rm m}39\fs8$ & $+53^\circ20\arcmin28\farcs6$ & IB & B & 8201 & 137 & 71 & 228 & 121.6 \\ 
UGC08250 & $13^{\rm h}10^{\rm m}20\fs2$ & $+32^\circ28\arcmin57\farcs3$ & Sc &  & 5169 & 79 & 76 & 12 & 76.0 \\ 
UGC08267 & $13^{\rm h}11^{\rm m}11\fs4$ & $+43^\circ43\arcmin35\farcs4$ & Sb &  & 7159 & 67 & 75 & 223 & 103.7 \\ 
UGC09067 & $14^{\rm h}10^{\rm m}45\fs4$ & $+15^\circ12\arcmin33\farcs1$ & Sab &  & 7740 & 49 & 62 & 15 & 114.5 \\ 
UGC09476 & $14^{\rm h}41^{\rm m}31\fs9$ & $+44^\circ30\arcmin45\farcs7$ & SABc & B & 3243 & 90 & 48 & 307 & 46.6 \\ 
UGC09537 & $14^{\rm h}48^{\rm m}26\fs6$ & $+34^\circ59\arcmin52\farcs7$ & Sb &  & 8662 & 134 & 72 & 136 & 130.1 \\ 
UGC09542 & $14^{\rm h}49^{\rm m}01\fs2$ & $+42^\circ27\arcmin50\farcs3$ & Sc &  & 5413 & 86 & 73 & 214 & 79.7 \\ 
UGC09665 & $15^{\rm h}01^{\rm m}32\fs5$ & $+48^\circ19\arcmin11\farcs2$ & Sbc &  & 2561 & 102 & 74 & 138 & 36.5 \\ 
UGC09759 & $15^{\rm h}10^{\rm m}40\fs8$ & $+55^\circ21\arcmin01\farcs4$ & Sb &  & 3394 & 80 & 67 & 50 & 49.2 \\ 
UGC09873 & $15^{\rm h}29^{\rm m}50\fs6$ & $+42^\circ37\arcmin44\farcs3$ & Sc &  & 5575 & 80 & 75 & 129 & 91.5 \\ 
UGC09892 & $15^{\rm h}32^{\rm m}52\fs0$ & $+41^\circ11\arcmin29\farcs0$ & Sb &  & 5591 & 78 & 72 & 101 & 82.2 \\ 
UGC09919 & $15^{\rm h}35^{\rm m}39\fs4$ & $+12^\circ36\arcmin22\farcs6$ & Sc &  & 3160 & 89 & 78 & 349 & 47.1 \\ 
UGC10043 & $15^{\rm h}48^{\rm m}41\fs4$ & $+21^\circ52\arcmin10\farcs1$ & Sbc &  & 2154 & 132 & 90 & 328 & 31.0 \\ 
UGC10123 & $15^{\rm h}59^{\rm m}02\fs7$ & $+51^\circ18\arcmin16\farcs5$ & Sab &  & 3738 & 69 & 70 & 235 & 53.8 \\ 
UGC10205 & $16^{\rm h}06^{\rm m}40\fs3$ & $+30^\circ05\arcmin56\farcs4$ & Sa &  & 6491 & 87 & 52 & 129 & 94.9 \\ 
UGC10331 & $16^{\rm h}17^{\rm m}21\fs1$ & $+59^\circ19\arcmin12\farcs3$ & Sb &  & 4415 & 87 & 76 & 141 & 64.8 \\ 
UGC10380 & $16^{\rm h}25^{\rm m}49\fs7$ & $+16^\circ34\arcmin33\farcs9$ & Sb &  & 8624 & 93 & 78 & 288 & 110.8 \\ 
UGC10384 & $16^{\rm h}26^{\rm m}46\fs6$ & $+11^\circ34\arcmin48\farcs7$ & Sab &  & 4929 & 70 & 70 & 278 & 71.8 \\ 
UGC10710 & $17^{\rm h}06^{\rm m}52\fs5$ & $+43^\circ07\arcmin19\farcs5$ & Sb &  & 8228 & 100 & 70 & 330 & 121.7 \\ 
\enddata
\end{deluxetable*}

\begin{deluxetable*}{lrrcrrrrr}
\tablewidth{0pt}
\tabletypesize{\footnotesize}
\tablecaption{EDGE Survey Observations and Photometry\label{tab:table2}}
\tablehead{
\colhead{Name} & \colhead{$S_{\rm CO}\Delta v$} & \colhead{$\theta$\tablenotemark{a}} & \colhead{RMS\tablenotemark{b}} & \colhead{Log[M$_{\rm mol}$]\tablenotemark{c}} & \colhead{$g$\tablenotemark{d}} & \colhead{$r$\tablenotemark{d}} & \colhead{$W1$} & \colhead{$W4$} \\ 
\colhead{} & \colhead{(Jy\,km\,s$^{-1}$)} & \colhead{($^{\prime\prime}$)} & \colhead{(mJy\,beam$^{-1}$)} & \colhead{(\msun)} & \colhead{(mag)} & \colhead{(mag)} & \colhead{(mag)} & \colhead{(mag)} 
}
\startdata
ARP220 & $456.0\pm5.2$ & 4.2 & 7.9 & $9.72\pm0.00$\tablenotemark{e} & 14.11 & 13.43 & 9.50 & 0.58 \\ 
IC0480 & $78.1\pm4.4$ & 4.4 & 11.2 & $9.55\pm0.02$ & 15.05 & 14.48 & 10.32 & 4.46 \\ 
IC0540 & $26.2\pm4.0$ & 4.8 & 16.2 & $8.39\pm0.06$ & 14.48 & 13.80 & 10.34 & 6.50 \\ 
IC0944 & $96.5\pm5.6$ & 4.7 & 10.1 & $10.00\pm0.02$ & 14.22 & 13.48 & 9.55 & 5.24 \\ 
IC1151 & $8.6\pm3.2$ & 4.5 & 16.9 & $7.93\pm0.14$ & 14.02 & 13.62 & 10.19 & 5.00 \\ 
IC1199 & $47.0\pm4.4$ & 4.6 & 10.9 & $9.35\pm0.04$ & 14.14 & 13.54 & 10.18 & 5.22 \\ 
IC1683 & $96.1\pm4.1$ & 4.3 & 9.8 & $9.68\pm0.02$ & 14.25 & 13.57 & 10.18 & 4.03 \\ 
IC2247 & $74.2\pm4.3$ & 4.3 & 8.7 & $9.47\pm0.02$ & 14.59 & 13.87 & 9.72 & 4.65 \\ 
IC2487 & $54.4\pm4.7$ & 4.9 & 11.9 & $9.34\pm0.04$ & 14.50 & 13.89 & 10.12 & 5.23 \\ 
IC4566 & $55.9\pm4.6$ & 4.8 & 10.5 & $9.57\pm0.03$ & 14.24 & 13.54 & 10.02 & 5.70 \\ 
IC5376 & $30.9\pm3.7$ & 4.3 & 10.4 & $9.23\pm0.05$ & 14.51 & 13.77 & 10.16 & 6.53 \\ 
NGC0444 & $<16.6$ & 4.5 & 12.9 & $<8.93$ & 15.15 & 14.71 & 11.44 & 6.67 \\ 
NGC0447 & $32.5\pm3.6$ & 4.1 & 8.2 & $9.33\pm0.05$ & 13.88 & 13.21 & 9.58 & 5.76 \\ 
NGC0477 & $46.3\pm5.2$ & 4.5 & 11.0 & $9.54\pm0.05$ & 14.26 & 13.76 & 10.26 & 5.23 \\ 
NGC0496 & $38.7\pm3.6$ & 4.3 & 9.3 & $9.48\pm0.04$ & 14.49 & 14.03 & 10.53 & 5.13 \\ 
NGC0523 & $94.3\pm5.5$ & 4.2 & 11.1 & $9.65\pm0.02$ & 14.10 & 13.52 & 9.44 & 3.98 \\ 
NGC0528 & $4.7\pm1.6$ & 4.4 & 10.4 & $8.36\pm0.13$ & 13.51 & 12.82 & 10.13 & 6.74 \\ 
NGC0551 & $42.6\pm4.4$ & 4.3 & 9.1 & $9.39\pm0.04$ & 14.18 & 13.61 & 10.02 & 5.25 \\ 
NGC1167 & $36.3\pm5.5$ & 4.2 & 10.0 & $9.28\pm0.06$ & 13.04 & 12.38 & 8.54 & 5.63 \\ 
NGC2253 & $153.9\pm6.6$ & 4.8 & 11.7 & $9.62\pm0.02$ & 13.46 & 12.88 & 9.27 & 3.48 \\ 
NGC2347 & $86.3\pm4.7$ & 4.8 & 10.2 & $9.56\pm0.02$ & 13.42 & 12.83 & 9.23 & 4.26 \\ 
NGC2410 & $91.8\pm6.1$ & 4.5 & 11.8 & $9.64\pm0.03$ & 13.83 & 13.15 & 9.23 & 3.51 \\ 
NGC2480 & $16.0\pm2.9$ & 4.6 & 12.8 & $8.26\pm0.07$ & 14.65 & 14.30 & 11.25 & 5.30 \\ 
NGC2486 & $<23.9$ & 4.2 & 17.2 & $<9.05$ & 11.91 & 11.14 & 10.11 & 6.72 \\ 
NGC2487 & $57.6\pm6.3$ & 4.4 & 12.7 & $9.47\pm0.05$ & 14.17 & 13.49 & 9.62 & 5.39 \\ 
NGC2623 & $125.8\pm3.1$ & 4.5 & 6.6 & $9.92\pm0.01$ & 14.03 & 13.58 & 10.15 & 2.18 \\ 
NGC2639 & $104.2\pm6.1$ & 4.7 & 12.3 & $9.36\pm0.02$ & 12.82 & 12.15 & 8.34 & 4.34 \\ 
NGC2730 & $32.2\pm5.0$ & 4.7 & 11.8 & $9.00\pm0.06$ & 14.64 & 14.20 & 10.56 & 5.05 \\ 
NGC2880 & $<15.8$ & 4.5 & 14.1 & $<7.93$ & 12.75 & 12.07 & 8.65 & 8.43 \\ 
NGC2906 & $87.7\pm6.3$ & 5.1 & 14.3 & $9.11\pm0.03$ & 13.38 & 12.73 & 8.95 & 4.43 \\ 
NGC2916 & $38.4\pm5.3$ & 4.2 & 11.2 & $9.05\pm0.06$ & 13.37 & 12.85 & 9.05 & 4.46 \\ 
NGC2918 & $<14.9$ & 4.7 & 12.4 & $<9.15$ & 13.50 & 12.90 & 9.44 & 7.66 \\ 
NGC3303 & $44.8\pm4.2$ & 4.6 & 11.0 & $9.57\pm0.04$ & 14.08 & 13.35 & 9.84 & 7.28 \\ 
NGC3381 & $22.5\pm4.5$ & 4.4 & 14.6 & $8.11\pm0.08$ & 13.93 & 13.56 & 9.92 & 4.45 \\ 
NGC3687 & $<19.3$ & 4.6 & 15.5 & $<8.42$ & 13.70 & 13.12 & 9.55 & 5.38 \\ 
NGC3811 & $93.7\pm6.5$ & 4.5 & 11.6 & $9.28\pm0.03$ & 13.74 & 13.12 & 9.46 & 3.98 \\ 
NGC3815 & $48.6\pm4.6$ & 4.4 & 11.4 & $9.16\pm0.04$ & 13.94 & 13.37 & 10.04 & 5.09 \\ 
NGC3994 & $86.9\pm5.2$ & 4.7 & 12.5 & $9.26\pm0.03$ & 13.25 & 12.69 & 9.38 & 3.46 \\ 
NGC4047 & $184.6\pm6.6$ & 4.5 & 12.7 & $9.66\pm0.02$ & 13.18 & 12.65 & 9.01 & 3.55 \\ 
NGC4149 & $78.7\pm4.4$ & 4.4 & 11.3 & $9.20\pm0.02$ & 14.03 & 13.37 & 10.59 & 5.77 \\ 
NGC4185 & $37.4\pm6.4$ & 4.2 & 12.8 & $9.08\pm0.07$ & 14.16 & 13.56 & 9.42 & 5.03 \\ 
NGC4210 & $46.5\pm5.6$ & 3.1 & 7.3 & $8.86\pm0.05$ & 13.98 & 13.42 & 9.60 & 5.05 \\ 
NGC4211NED02 & $20.3\pm2.9$ & 4.6 & 12.4 & $9.29\pm0.06$ & 15.34 & 14.72 & 10.03 & 4.73 \\ 
NGC4470 & $33.5\pm5.0$ & 4.8 & 14.5 & $8.59\pm0.06$ & 13.31 & 12.95 & 9.79 & 4.42 \\ 
NGC4644 & $29.7\pm3.3$ & 4.6 & 9.1 & $9.20\pm0.05$ & 14.44 & 13.82 & 10.50 & 5.92 \\ 
NGC4676A & $78.9\pm4.5$ & 4.8 & 12.2 & $9.88\pm0.02$ & 14.79 & 14.13 & 10.59 & 4.08 \\ 
NGC4711 & $41.8\pm4.6$ & 4.6 & 12.5 & $9.18\pm0.05$ & 14.21 & 13.67 & 10.25 & 5.13 \\ 
NGC4961 & $18.5\pm3.5$ & 4.4 & 12.8 & $8.41\pm0.08$ & 13.98 & 13.63 & 10.49 & 4.97 \\ 
NGC5000 & $41.8\pm3.9$ & 4.1 & 9.1 & $9.45\pm0.04$ & 14.46 & 13.83 & 10.17 & 4.84 \\ 
NGC5016 & $56.1\pm5.5$ & 4.7 & 12.8 & $8.90\pm0.04$ & 13.56 & 13.06 & 9.57 & 4.46 \\ 
NGC5056 & $41.4\pm3.8$ & 5.0 & 10.3 & $9.45\pm0.04$ & 14.00 & 12.94 & 10.25 & 4.70 \\ 
NGC5205 & $35.5\pm6.2$ & 4.3 & 14.5 & $8.37\pm0.07$ & 13.87 & 13.31 & 9.84 & 5.49 \\ 
NGC5218 & $399.5\pm7.0$ & 4.5 & 12.2 & $9.86\pm0.01$ & 13.45 & 12.77 & 9.00 & 2.64 \\ 
NGC5394 & $164.9\pm4.0$ & 4.6 & 13.3 & $9.62\pm0.01$ & 13.84 & 13.28 & 9.59 & 2.49 \\ 
NGC5406 & $77.6\pm6.8$ & 4.5 & 11.5 & $9.69\pm0.04$ & 13.62 & 13.03 & 9.21 & 4.93 \\ 
NGC5480 & $109.9\pm7.2$ & 4.0 & 11.6 & $8.92\pm0.03$ & 13.53 & 13.09 & 9.21 & 3.34 \\ 
NGC5485 & $<16.2$ & 4.4 & 13.4 & $<8.09$ & 12.84 & 12.14 & 8.32 & 6.23 \\ 
NGC5520 & $62.8\pm5.0$ & 4.3 & 12.5 & $8.67\pm0.03$ & 13.50 & 13.02 & 9.65 & 4.03 \\ 
NGC5614 & $217.3\pm6.1$ & 5.1 & 13.1 & $9.84\pm0.01$ & 12.96 & 12.24 & 8.41 & 4.36 \\ 
NGC5633 & $120.1\pm6.0$ & 4.4 & 13.0 & $9.14\pm0.02$ & 13.17 & 12.67 & 9.28 & 3.63 \\ 
NGC5657 & $38.9\pm4.2$ & 4.4 & 12.8 & $9.11\pm0.04$ & 14.20 & 13.57 & 10.07 & 4.31 \\ 
NGC5682 & $<17.6$ & 4.2 & 12.7 & $<8.29$ & 14.90 & 14.52 & 11.39 & 6.00 \\ 
NGC5732 & $22.1\pm3.7$ & 4.8 & 13.4 & $8.82\pm0.07$ & 14.46 & 14.02 & 10.87 & 5.67 \\ 
NGC5784 & $38.8\pm3.5$ & 4.3 & 8.6 & $9.40\pm0.04$ & 13.40 & 12.75 & 9.13 & 5.39 \\ 
NGC5876 & $<16.0$ & 4.7 & 13.0 & $<8.56$ & 13.59 & 12.88 & 9.39 & 6.37 \\ 
NGC5908 & $376.3\pm7.1$ & 4.1 & 8.5 & $9.94\pm0.01$ & 13.43 & 12.61 & 8.00 & 3.16 \\ 
NGC5930 & $147.3\pm5.5$ & 4.6 & 12.3 & $9.33\pm0.02$ & 13.44 & 12.77 & 8.64 & 1.97 \\ 
NGC5934 & $92.0\pm4.1$ & 4.4 & 9.1 & $9.81\pm0.02$ & 13.99 & 13.27 & 9.69 & 4.89 \\ 
NGC5947 & $23.8\pm3.5$ & 4.6 & 9.2 & $9.26\pm0.06$ & 14.39 & 13.86 & 10.40 & 5.34 \\ 
NGC5953 & $363.7\pm8.1$ & 4.5 & 11.8 & $9.49\pm0.01$ & 12.95 & 12.31 & 8.95 & 2.67 \\ 
NGC5980 & $136.6\pm5.2$ & 4.4 & 9.3 & $9.70\pm0.02$ & 13.51 & 12.88 & 9.16 & 3.46 \\ 
NGC6004 & $68.0\pm7.2$ & 4.6 & 12.4 & $9.33\pm0.04$ & 14.01 & 13.44 & 10.07 & 5.30 \\ 
NGC6021 & $<19.4$ & 4.3 & 14.2 & $<8.98$ & 13.78 & 13.06 & 9.71 & 7.50 \\ 
NGC6027 & $2.5\pm1.7$ & 4.4 & 12.4 & $8.01\pm0.22$ & 13.95 & 13.33 & 10.66 & 6.59 \\ 
NGC6060 & $115.2\pm6.8$ & 4.2 & 10.0 & $9.68\pm0.03$ & 13.83 & 13.15 & 9.07 & 4.00 \\ 
NGC6063 & $<19.7$ & 4.8 & 16.4 & $<8.53$ & 14.40 & 13.84 & 10.26 & 5.75 \\ 
NGC6081 & $<17.5$ & 4.5 & 13.5 & $<8.99$ & 13.93 & 13.16 & 9.59 & 6.26 \\ 
NGC6125 & $<14.2$ & 4.6 & 11.1 & $<8.83$ & 13.15 & 12.44 & 8.83 & 7.23 \\ 
NGC6146 & $<13.5$ & 4.5 & 10.4 & $<9.36$ & 13.50 & 12.83 & 9.30 & 7.14 \\ 
NGC6155 & $70.6\pm5.8$ & 4.6 & 12.4 & $8.94\pm0.03$ & 13.62 & 13.11 & 9.66 & 4.11 \\ 
NGC6168 & $33.3\pm4.9$ & 4.5 & 13.7 & $8.65\pm0.06$ & 14.58 & 14.07 & 10.28 & 4.50 \\ 
NGC6186 & $154.8\pm6.8$ & 4.5 & 13.7 & $9.46\pm0.02$ & 13.70 & 13.01 & 9.36 & 3.79 \\ 
NGC6301 & $60.2\pm4.8$ & 4.5 & 9.5 & $9.96\pm0.03$ & 14.31 & 13.82 & 9.79 & 5.24 \\ 
NGC6310 & $10.6\pm2.6$ & 4.4 & 12.4 & $8.42\pm0.09$ & 14.14 & 13.43 & 9.72 & 6.10 \\ 
NGC6314 & $39.1\pm3.1$ & 4.1 & 7.3 & $9.57\pm0.03$ & 13.82 & 13.16 & 9.58 & 5.60 \\ 
NGC6361 & $367.2\pm6.2$ & 4.7 & 9.9 & $10.06\pm0.01$ & 14.10 & 13.28 & 8.81 & 3.27 \\ 
NGC6394 & $45.9\pm4.2$ & 4.7 & 10.7 & $9.86\pm0.04$ & 14.61 & 13.99 & 10.54 & 4.82 \\ 
NGC6478 & $142.2\pm5.8$ & 4.4 & 10.5 & $10.14\pm0.02$ & 13.76 & 13.18 & 9.45 & 4.28 \\ 
NGC7738 & $98.8\pm2.9$ & 4.0 & 5.9 & $9.99\pm0.01$ & 14.23 & 13.48 & 9.57 & 3.25 \\ 
NGC7819 & $35.0\pm3.4$ & 4.1 & 9.4 & $9.27\pm0.04$ & 14.58 & 14.11 & 10.61 & 4.77 \\ 
UGC00809 & $22.0\pm3.7$ & 3.9 & 10.5 & $8.92\pm0.07$ & 15.12 & 14.68 & 11.26 & 5.95 \\ 
UGC03253 & $20.7\pm3.1$ & 5.5 & 13.1 & $8.88\pm0.06$ & 14.04 & 13.46 & 9.90 & 5.25 \\ 
UGC03539 & $55.6\pm4.2$ & 5.0 & 12.9 & $9.11\pm0.03$ & 15.14 & 14.37 & 10.29 & 4.97 \\ 
UGC03969 & $52.2\pm4.2$ & 4.7 & 10.4 & $9.87\pm0.03$ & 15.26 & 14.54 & 10.70 & 5.51 \\ 
UGC03973 & $34.3\pm4.2$ & 4.6 & 10.1 & $9.51\pm0.05$ & 14.10 & 13.50 & 8.81 & 2.59 \\ 
UGC04029 & $56.7\pm4.2$ & 4.4 & 10.1 & $9.37\pm0.03$ & 14.78 & 14.12 & 10.09 & 4.50 \\ 
UGC04132 & $179.2\pm6.0$ & 4.7 & 11.5 & $10.02\pm0.01$ & 13.95 & 13.31 & 9.30 & 3.55 \\ 
UGC04280 & $16.9\pm3.0$ & 4.6 & 13.3 & $8.66\pm0.07$ & 14.26 & 13.63 & 10.37 & 5.36 \\ 
UGC04461 & $32.4\pm3.6$ & 4.5 & 11.9 & $9.24\pm0.05$ & 14.60 & 14.13 & 10.66 & 4.93 \\ 
UGC05108 & $39.3\pm3.5$ & 4.9 & 9.4 & $9.75\pm0.04$ & 14.41 & 13.78 & 10.36 & 4.77 \\ 
UGC05111 & $91.7\pm4.5$ & 5.1 & 10.5 & $9.96\pm0.02$ & 14.73 & 13.93 & 10.05 & 5.08 \\ 
UGC05244 & $4.6\pm1.9$ & 4.9 & 15.2 & $7.96\pm0.15$ & 15.19 & 14.75 & 11.67 & 6.96 \\ 
UGC05359 & $28.9\pm3.4$ & 5.7 & 9.9 & $9.65\pm0.05$ & 14.98 & 14.39 & 11.03 & 6.18 \\ 
UGC05498NED01 & $<17.4$ & 4.5 & 13.8 & $<9.18$ & 14.71 & 13.98 & 10.23 & 6.05 \\ 
UGC05598 & $21.6\pm3.3$ & 4.9 & 14.1 & $9.17\pm0.06$ & 14.99 & 14.44 & 10.96 & 5.73 \\ 
UGC06312 & $<14.5$ & 4.6 & 11.6 & $<9.08$ & 14.40 & 13.69 & 10.19 & 6.57 \\ 
UGC07012 & $11.0\pm2.9$ & 4.3 & 11.1 & $8.35\pm0.10$ & 14.52 & 14.18 & 11.36 & 5.96 \\ 
UGC08107 & $85.8\pm4.0$ & 4.7 & 8.6 & $10.11\pm0.02$ & 14.44 & 13.78 & 9.87 & 5.04 \\ 
UGC08250 & $<13.7$ & 4.4 & 10.2 & $<8.91$ & 15.49 & 14.97 & 11.52 & 6.12 \\ 
UGC08267 & $51.7\pm3.9$ & 4.3 & 10.2 & $9.76\pm0.03$ & 15.07 & 14.35 & 10.51 & 5.30 \\ 
UGC09067 & $51.0\pm4.4$ & 5.0 & 12.7 & $9.83\pm0.04$ & 14.49 & 13.95 & 10.69 & 5.51 \\ 
UGC09476 & $62.4\pm6.3$ & 4.5 & 12.7 & $9.15\pm0.04$ & 14.31 & 13.83 & 10.01 & 4.88 \\ 
UGC09537 & $52.7\pm4.4$ & 4.3 & 8.7 & $9.96\pm0.03$ & 14.45 & 13.79 & 9.71 & 5.31 \\ 
UGC09542 & $31.5\pm3.8$ & 4.3 & 10.5 & $9.31\pm0.05$ & 14.99 & 14.40 & 10.79 & 5.72 \\ 
UGC09665 & $63.2\pm5.1$ & 4.1 & 11.8 & $8.94\pm0.03$ & 14.44 & 13.80 & 9.90 & 4.34 \\ 
UGC09759 & $46.8\pm4.0$ & 4.3 & 10.8 & $9.07\pm0.04$ & 14.86 & 14.15 & 10.37 & 5.31 \\ 
UGC09873 & $14.0\pm2.3$ & 4.2 & 9.8 & $9.08\pm0.07$ & 15.50 & 14.94 & 11.41 & 6.04 \\ 
UGC09892 & $21.4\pm2.8$ & 4.5 & 9.8 & $9.17\pm0.05$ & 15.13 & 14.56 & 11.07 & 6.38 \\ 
UGC09919 & $12.6\pm2.8$ & 4.7 & 12.5 & $8.46\pm0.09$ & 15.27 & 14.74 & 11.19 & 5.86 \\ 
UGC10043 & $72.9\pm5.6$ & 4.3 & 13.7 & $8.86\pm0.03$ & 15.08 & 14.36 & 10.04 & 4.91 \\ 
UGC10123 & $101.4\pm5.3$ & 4.1 & 10.6 & $9.48\pm0.02$ & 14.69 & 13.93 & 9.69 & 4.35 \\ 
UGC10205 & $43.5\pm4.2$ & 4.8 & 11.1 & $9.60\pm0.04$ & 14.14 & 13.43 & 9.61 & 5.79 \\ 
UGC10331 & $<20.5$ & 5.1 & 18.8 & $<8.95$ & 15.13 & 14.61 & 11.14 & 4.53 \\ 
UGC10380 & $20.9\pm3.1$ & 4.6 & 10.7 & $9.42\pm0.06$ & 14.70 & 13.98 & 10.57 & 6.43 \\ 
UGC10384 & $77.4\pm4.1$ & 4.6 & 9.9 & $9.61\pm0.02$ & 14.76 & 14.10 & 10.26 & 4.17 \\ 
UGC10710 & $50.2\pm4.7$ & 4.5 & 10.9 & $9.88\pm0.04$ & 14.42 & 13.83 & 10.27 & 5.92 \\ 
\enddata
\tablenotetext{a}{Equivalent round synthesized beam}
\tablenotetext{b}{RMS in 10\,\kmpers\ channels}
\tablenotetext{c}{Molecular mass in \msun\ using $\alpha_{\rm CO}=4.36$\,\acounits}
\tablenotetext{d}{Integrated magnitudes corrected by Galactic extinction}
\tablenotetext{e}{Using $\alpha_{\rm CO}=0.8$\,\acounits, appropriate for ULIRGs}
\end{deluxetable*}

\begin{deluxetable*}{lrrrrrrr}
\tablewidth{0pt}
\tabletypesize{\footnotesize}
\tablecaption{EDGE Survey Galaxy Parameters and Scale Lengths\label{tab:table3}}
\tablehead{
\colhead{Name} & \colhead{Log[M$_*$]\tablenotemark{a}} & \colhead{Log[SFR]\tablenotemark{b}} & \colhead{Log[O/H]\tablenotemark{c}} & \colhead{$\Rhmol$} & \colhead{$l_{\rm mol}$} & \colhead{$l_*$} & \colhead{$l_{\rm SFR}$} \\ 
\colhead{} & \colhead{(\msun)} & \colhead{(M$_\odot$\,yr$^{-1}$)} & \colhead{} & \colhead{(kpc)} & \colhead{(kpc)} & \colhead{(kpc)} & \colhead{(kpc)} 
}
\startdata
ARP220 & $10.91\pm0.09$ & $1.57\pm0.18$ & \nodata & $1.34\pm0.23$ & $1.59\pm0.69$ & $2.55\pm0.20$ & $1.24\pm0.79$ \\ 
IC0480 & $10.27\pm0.13$ & $0.11\pm0.10$ & $8.49\pm0.05$ & $3.30\pm0.09$ & $2.23\pm0.43$ & $3.08\pm0.32$ & $2.58\pm0.41$ \\ 
IC0540 & $9.84\pm0.12$ & $-1.09\pm0.17$ & \nodata & \nodata & \nodata & \nodata & \nodata \\ 
IC0944 & $11.26\pm0.10$ & $0.41\pm0.15$ & $8.52\pm0.06$ & $6.92\pm0.04$ & $5.16\pm0.90$ & $5.06\pm0.15$ & $8.70\pm0.79$ \\ 
IC1151 & $10.02\pm0.10$ & $-0.20\pm0.06$ & $8.41\pm0.04$ & \nodata & \nodata & \nodata & \nodata \\ 
IC1199 & $10.78\pm0.10$ & $0.16\pm0.07$ & $8.53\pm0.06$ & $4.23\pm0.09$ & $4.21\pm1.75$ & $2.58\pm0.23$ & $2.55\pm0.62$ \\ 
IC1683 & $10.76\pm0.11$ & $0.54\pm0.07$ & $8.57\pm0.05$ & $1.99\pm0.12$ & $1.65\pm0.42$ & $2.38\pm0.09$ & $1.33\pm0.21$ \\ 
IC2247 & $10.44\pm0.11$ & $0.23\pm0.15$ & $8.51\pm0.04$ & $4.42\pm0.07$ & $2.91\pm0.79$ & $2.62\pm0.13$ & $2.79\pm0.46$ \\ 
IC2487 & $10.59\pm0.12$ & $0.17\pm0.08$ & $8.52\pm0.05$ & $4.92\pm0.07$ & $3.82\pm1.03$ & $3.83\pm0.09$ & $5.36\pm0.54$ \\ 
IC4566 & $11.02\pm0.10$ & $0.15\pm0.14$ & $8.55\pm0.07$ & $6.27\pm0.08$ & \nodata & $3.22\pm0.07$ & \nodata \\ 
IC5376 & $10.66\pm0.11$ & $0.01\pm0.11$ & $8.53\pm0.05$ & \nodata & \nodata & \nodata & \nodata \\ 
NGC0444 & $10.25\pm0.12$ & $-0.18\pm0.09$ & $8.41\pm0.05$ & \nodata & \nodata & \nodata & \nodata \\ 
NGC0447 & $11.43\pm0.10$ & $-0.03\pm0.28$ & $8.54\pm0.07$ & \nodata & \nodata & \nodata & \nodata \\ 
NGC0477 & $10.90\pm0.12$ & $0.49\pm0.08$ & $8.49\pm0.06$ & $9.07\pm0.08$ & $13.12\pm8.77$ & $5.28\pm0.40$ & \nodata \\ 
NGC0496 & $10.85\pm0.13$ & $0.66\pm0.07$ & $8.46\pm0.06$ & $3.74\pm0.07$ & $3.31\pm1.04$ & $4.30\pm0.11$ & $2.32\pm0.23$ \\ 
NGC0523 & $10.89\pm0.09$ & $0.58\pm0.06$ & $8.50\pm0.05$ & $6.99\pm0.06$ & $13.46\pm12.49$ & $3.90\pm0.13$ & \nodata \\ 
NGC0528 & $11.06\pm0.10$ & $-0.16\pm0.33$ & \nodata & \nodata & \nodata & \nodata & \nodata \\ 
NGC0551 & $10.95\pm0.11$ & $0.31\pm0.07$ & $8.59\pm0.04$ & $6.74\pm0.07$ & \nodata & $3.75\pm0.07$ & \nodata \\ 
NGC1167 & $11.48\pm0.09$ & $0.46\pm0.21$ & $8.55\pm0.06$ & \nodata & \nodata & \nodata & \nodata \\ 
NGC2253 & $10.81\pm0.11$ & $0.50\pm0.06$ & $8.59\pm0.04$ & $3.64\pm0.09$ & $2.83\pm0.85$ & $2.48\pm0.18$ & $1.82\pm0.52$ \\ 
NGC2347 & $11.04\pm0.10$ & $0.54\pm0.07$ & $8.57\pm0.04$ & $3.23\pm0.09$ & $2.45\pm0.68$ & $2.16\pm0.06$ & $1.37\pm0.35$ \\ 
NGC2410 & $11.03\pm0.10$ & $0.55\pm0.11$ & $8.52\pm0.05$ & $5.27\pm0.06$ & $4.09\pm1.29$ & $3.22\pm0.13$ & $3.43\pm0.19$ \\ 
NGC2480 & $9.62\pm0.13$ & $-0.35\pm0.08$ & $8.42\pm0.04$ & \nodata & \nodata & \nodata & \nodata \\ 
NGC2486 & $10.79\pm0.09$ & $-0.05\pm0.26$ & $8.47\pm0.07$ & \nodata & \nodata & \nodata & \nodata \\ 
NGC2487 & $11.06\pm0.10$ & $0.25\pm0.13$ & $8.54\pm0.07$ & \nodata & \nodata & $3.91\pm0.42$ & \nodata \\ 
NGC2623 & $10.66\pm0.11$ & $0.74\pm0.13$ & \nodata & \nodata & \nodata & $1.75\pm0.22$ & $1.41\pm0.11$ \\ 
NGC2639 & $11.17\pm0.09$ & $0.42\pm0.10$ & $8.58\pm0.04$ & $3.05\pm0.16$ & $1.96\pm1.22$ & $1.71\pm0.07$ & \nodata \\ 
NGC2730 & $10.13\pm0.09$ & $0.23\pm0.06$ & $8.45\pm0.04$ & \nodata & \nodata & \nodata & $11.61\pm4.11$ \\ 
NGC2880 & $10.56\pm0.08$ & $-2.28\pm0.66$ & \nodata & \nodata & \nodata & \nodata & \nodata \\ 
NGC2906 & $10.59\pm0.09$ & $-0.10\pm0.06$ & $8.60\pm0.04$ & \nodata & \nodata & $1.75\pm1.48$ & $0.87\pm0.34$ \\ 
NGC2916 & $10.96\pm0.08$ & $0.35\pm0.07$ & $8.55\pm0.06$ & \nodata & \nodata & \nodata & \nodata \\ 
NGC2918 & $11.32\pm0.10$ & $-0.79\pm0.59$ & \nodata & \nodata & \nodata & \nodata & \nodata \\ 
NGC3303 & $11.17\pm0.10$ & $0.41\pm0.14$ & $8.53\pm0.05$ & $2.89\pm0.09$ & $2.46\pm0.77$ & $3.13\pm0.14$ & \nodata \\ 
NGC3381 & $9.88\pm0.09$ & $-0.41\pm0.06$ & $8.52\pm0.04$ & \nodata & \nodata & \nodata & \nodata \\ 
NGC3687 & $10.51\pm0.11$ & $-0.33\pm0.07$ & $8.54\pm0.05$ & \nodata & \nodata & \nodata & \nodata \\ 
NGC3811 & $10.64\pm0.11$ & $0.35\pm0.07$ & $8.56\pm0.04$ & $3.10\pm0.04$ & \nodata & $2.10\pm0.13$ & $2.96\pm1.80$ \\ 
NGC3815 & $10.53\pm0.09$ & $0.05\pm0.07$ & $8.55\pm0.03$ & $2.61\pm0.17$ & \nodata & $1.36\pm0.05$ & $1.39\pm0.14$ \\ 
NGC3994 & $10.59\pm0.11$ & $0.57\pm0.05$ & $8.50\pm0.03$ & $1.80\pm0.16$ & $1.31\pm1.03$ & $1.31\pm0.08$ & \nodata \\ 
NGC4047 & $10.87\pm0.10$ & $0.56\pm0.06$ & $8.57\pm0.04$ & $2.27\pm0.16$ & \nodata & $2.11\pm0.04$ & $1.05\pm0.13$ \\ 
NGC4149 & $10.45\pm0.11$ & $-0.36\pm0.12$ & $8.53\pm0.04$ & \nodata & \nodata & $1.55\pm1.01$ & $1.05\pm0.27$ \\ 
NGC4185 & $10.86\pm0.11$ & $0.05\pm0.09$ & $8.56\pm0.06$ & \nodata & \nodata & \nodata & \nodata \\ 
NGC4210 & $10.51\pm0.10$ & $-0.21\pm0.07$ & $8.57\pm0.04$ & \nodata & \nodata & $2.50\pm0.11$ & \nodata \\ 
NGC4211NED02 & $10.53\pm0.13$ & $0.22\pm0.19$ & $8.46\pm0.09$ & \nodata & \nodata & $1.10\pm1.07$ & \nodata \\ 
NGC4470 & $10.23\pm0.09$ & $-0.01\pm0.05$ & $8.46\pm0.03$ & \nodata & \nodata & $1.62\pm0.40$ & $0.58\pm0.54$ \\ 
NGC4644 & $10.68\pm0.11$ & $0.09\pm0.09$ & $8.59\pm0.04$ & $5.69\pm0.07$ & $7.18\pm3.37$ & $2.64\pm0.18$ & $5.26\pm0.80$ \\ 
NGC4676A & $10.86\pm0.10$ & $0.52\pm0.09$ & \nodata & $1.76\pm0.18$ & $0.85\pm0.65$ & $2.04\pm0.09$ & \nodata \\ 
NGC4711 & $10.58\pm0.09$ & $0.08\pm0.07$ & $8.60\pm0.04$ & $4.11\pm0.12$ & $5.59\pm5.41$ & $3.01\pm0.11$ & $3.13\pm0.68$ \\ 
NGC4961 & $9.98\pm0.10$ & $-0.16\pm0.06$ & $8.42\pm0.05$ & \nodata & \nodata & \nodata & $1.83\pm0.98$ \\ 
NGC5000 & $10.94\pm0.10$ & $0.37\pm0.14$ & $8.55\pm0.06$ & \nodata & \nodata & $2.95\pm0.10$ & \nodata \\ 
NGC5016 & $10.47\pm0.09$ & $-0.00\pm0.10$ & $8.56\pm0.06$ & \nodata & \nodata & \nodata & $1.76\pm0.82$ \\ 
NGC5056 & $10.85\pm0.09$ & $0.57\pm0.06$ & $8.49\pm0.03$ & $4.78\pm0.05$ & $4.37\pm1.60$ & $2.96\pm0.08$ & $4.68\pm0.59$ \\ 
NGC5205 & $9.98\pm0.09$ & $-0.81\pm0.12$ & $8.55\pm0.07$ & \nodata & \nodata & \nodata & \nodata \\ 
NGC5218 & $10.64\pm0.09$ & $0.28\pm0.09$ & $8.55\pm0.05$ & $1.20\pm0.16$ & \nodata & \nodata & $2.77\pm1.04$ \\ 
NGC5394 & $10.38\pm0.11$ & $0.53\pm0.06$ & $8.57\pm0.06$ & $2.08\pm0.11$ & $1.60\pm0.26$ & $1.95\pm0.17$ & $1.63\pm0.13$ \\ 
NGC5406 & $11.27\pm0.09$ & $0.44\pm0.08$ & $8.60\pm0.06$ & $9.11\pm0.07$ & \nodata & $4.31\pm0.33$ & \nodata \\ 
NGC5480 & $10.18\pm0.08$ & $0.15\pm0.05$ & $8.58\pm0.04$ & \nodata & \nodata & \nodata & $1.58\pm1.03$ \\ 
NGC5485 & $10.75\pm0.08$ & $-1.60\pm0.34$ & \nodata & \nodata & \nodata & \nodata & \nodata \\ 
NGC5520 & $10.07\pm0.11$ & $-0.07\pm0.05$ & $8.51\pm0.04$ & \nodata & \nodata & \nodata & \nodata \\ 
NGC5614 & $11.22\pm0.09$ & $0.20\pm0.11$ & $8.55\pm0.06$ & $2.26\pm0.16$ & $1.04\pm0.50$ & $2.31\pm0.21$ & $3.04\pm1.04$ \\ 
NGC5633 & $10.40\pm0.11$ & $0.18\pm0.05$ & $8.61\pm0.02$ & \nodata & \nodata & \nodata & \nodata \\ 
NGC5657 & $10.50\pm0.10$ & $0.21\pm0.07$ & $8.48\pm0.05$ & \nodata & \nodata & $2.17\pm0.14$ & $2.31\pm0.19$ \\ 
NGC5682 & $9.59\pm0.11$ & $-0.50\pm0.06$ & $8.35\pm0.04$ & \nodata & \nodata & \nodata & \nodata \\ 
NGC5732 & $10.23\pm0.11$ & $-0.03\pm0.07$ & $8.46\pm0.05$ & \nodata & \nodata & $2.09\pm0.17$ & \nodata \\ 
NGC5784 & \nodata & \nodata & \nodata & $2.54\pm0.15$ & \nodata & $2.96\pm0.21$ & $1.11\pm0.28$ \\ 
NGC5876 & $10.78\pm0.10$ & $-0.72\pm0.24$ & $8.52\pm0.06$ & \nodata & \nodata & \nodata & \nodata \\ 
NGC5908 & $10.95\pm0.10$ & $0.36\pm0.08$ & $8.54\pm0.05$ & $4.96\pm0.07$ & $3.25\pm0.48$ & $3.21\pm0.07$ & $2.32\pm0.24$ \\ 
NGC5930 & $10.61\pm0.11$ & $0.41\pm0.06$ & $8.46\pm0.05$ & \nodata & \nodata & \nodata & \nodata \\ 
NGC5934 & $10.87\pm0.09$ & $0.46\pm0.15$ & \nodata & $2.36\pm0.14$ & $1.50\pm0.25$ & $1.78\pm0.35$ & $1.63\pm0.26$ \\ 
NGC5947 & $10.87\pm0.10$ & $0.32\pm0.07$ & $8.55\pm0.04$ & $3.48\pm0.12$ & $6.64\pm4.10$ & $2.79\pm0.16$ & $7.50\pm0.80$ \\ 
NGC5953 & $10.38\pm0.11$ & $0.45\pm0.06$ & $8.50\pm0.05$ & \nodata & \nodata & \nodata & \nodata \\ 
NGC5980 & $10.81\pm0.10$ & $0.71\pm0.06$ & $8.58\pm0.03$ & $3.87\pm0.08$ & $2.60\pm0.60$ & $2.37\pm0.05$ & $1.87\pm0.30$ \\ 
NGC6004 & $10.87\pm0.08$ & $0.21\pm0.07$ & $8.58\pm0.06$ & $3.01\pm0.14$ & \nodata & $4.22\pm0.28$ & \nodata \\ 
NGC6021 & $10.98\pm0.10$ & $-0.54\pm0.35$ & \nodata & \nodata & \nodata & \nodata & \nodata \\ 
NGC6027 & $11.02\pm0.10$ & $0.05\pm0.17$ & \nodata & \nodata & \nodata & \nodata & \nodata \\ 
NGC6060 & $10.99\pm0.09$ & $0.62\pm0.14$ & $8.50\pm0.08$ & $6.75\pm0.05$ & $6.09\pm1.77$ & $3.90\pm0.21$ & $5.31\pm1.07$ \\ 
NGC6063 & $10.36\pm0.12$ & $-0.27\pm0.08$ & $8.49\pm0.06$ & \nodata & \nodata & \nodata & \nodata \\ 
NGC6081 & $11.04\pm0.09$ & $-0.33\pm0.28$ & \nodata & \nodata & \nodata & \nodata & \nodata \\ 
NGC6125 & $11.36\pm0.09$ & $-1.27\pm0.58$ & \nodata & \nodata & \nodata & \nodata & \nodata \\ 
NGC6146 & $11.72\pm0.09$ & $-0.10\pm0.34$ & \nodata & \nodata & \nodata & \nodata & \nodata \\ 
NGC6155 & $10.38\pm0.10$ & $0.18\pm0.05$ & $8.57\pm0.03$ & \nodata & \nodata & $1.90\pm0.14$ & $0.53\pm0.21$ \\ 
NGC6168 & $9.94\pm0.11$ & $-0.07\pm0.06$ & $8.40\pm0.03$ & $2.94\pm0.16$ & \nodata & $2.42\pm0.40$ & $1.68\pm0.53$ \\ 
NGC6186 & $10.62\pm0.09$ & $0.30\pm0.06$ & $8.59\pm0.04$ & $2.43\pm0.09$ & $2.25\pm0.45$ & $2.43\pm0.11$ & $1.66\pm0.40$ \\ 
NGC6301 & $11.18\pm0.12$ & $0.93\pm0.19$ & $8.53\pm0.07$ & $9.77\pm0.05$ & $14.95\pm7.98$ & $7.17\pm0.37$ & \nodata \\ 
NGC6310 & $10.69\pm0.11$ & $-0.36\pm0.11$ & $8.56\pm0.05$ & \nodata & \nodata & $3.60\pm0.37$ & \nodata \\ 
NGC6314 & $11.21\pm0.09$ & $0.00\pm0.28$ & $8.49\pm0.06$ & $2.47\pm0.11$ & $2.25\pm0.80$ & $3.77\pm0.21$ & $0.97\pm0.18$ \\ 
NGC6361 & $10.73\pm0.11$ & $0.72\pm0.06$ & $8.57\pm0.04$ & $4.79\pm0.06$ & $2.95\pm0.37$ & $3.28\pm0.29$ & $3.39\pm0.69$ \\ 
NGC6394 & $11.11\pm0.10$ & $0.61\pm0.12$ & $8.54\pm0.05$ & $6.73\pm0.06$ & $6.37\pm1.83$ & $3.51\pm0.15$ & \nodata \\ 
NGC6478 & $11.27\pm0.10$ & $1.00\pm0.07$ & $8.56\pm0.04$ & $8.76\pm0.05$ & $6.60\pm1.13$ & $6.23\pm0.27$ & $15.99\pm4.00$ \\ 
NGC7738 & $11.21\pm0.11$ & $1.18\pm0.09$ & $8.56\pm0.06$ & $2.18\pm0.15$ & $1.68\pm0.54$ & $2.30\pm0.24$ & $1.14\pm0.20$ \\ 
NGC7819 & $10.61\pm0.09$ & $0.41\pm0.07$ & $8.47\pm0.07$ & \nodata & \nodata & $3.75\pm0.32$ & $4.21\pm1.07$ \\ 
UGC00809 & $10.00\pm0.13$ & $-0.14\pm0.08$ & $8.41\pm0.03$ & $3.56\pm0.13$ & $6.14\pm3.15$ & $3.84\pm0.16$ & $2.99\pm0.36$ \\ 
UGC03253 & $10.63\pm0.11$ & $0.23\pm0.11$ & $8.51\pm0.07$ & $3.84\pm0.09$ & $5.14\pm1.58$ & $2.42\pm0.09$ & $3.16\pm1.03$ \\ 
UGC03539 & $9.84\pm0.13$ & $-0.17\pm0.09$ & $8.39\pm0.07$ & $2.41\pm0.14$ & $1.58\pm1.03$ & $1.46\pm0.02$ & $1.62\pm0.15$ \\ 
UGC03969 & $10.74\pm0.10$ & $0.55\pm0.08$ & $8.50\pm0.05$ & $6.05\pm0.05$ & $4.86\pm0.95$ & $2.78\pm0.24$ & $4.68\pm0.26$ \\ 
UGC03973 & $10.94\pm0.08$ & $0.92\pm0.11$ & $8.49\pm0.05$ & $4.47\pm0.06$ & $4.50\pm1.35$ & $3.86\pm0.34$ & \nodata \\ 
UGC04029 & $10.38\pm0.10$ & $0.18\pm0.09$ & $8.48\pm0.08$ & $4.65\pm0.06$ & $4.03\pm0.97$ & $3.38\pm0.16$ & $4.33\pm0.34$ \\ 
UGC04132 & $10.94\pm0.12$ & $0.96\pm0.07$ & $8.54\pm0.04$ & $4.86\pm0.06$ & $3.13\pm0.62$ & $3.63\pm0.16$ & $4.42\pm0.49$ \\ 
UGC04280 & $10.30\pm0.09$ & $-0.20\pm0.10$ & $8.50\pm0.03$ & \nodata & \nodata & $1.63\pm0.04$ & $1.95\pm0.19$ \\ 
UGC04461 & $10.37\pm0.12$ & $0.34\pm0.07$ & $8.41\pm0.08$ & $3.31\pm0.08$ & $3.28\pm1.06$ & $3.14\pm0.07$ & $3.65\pm0.38$ \\ 
UGC05108 & $11.11\pm0.11$ & $0.66\pm0.12$ & $8.50\pm0.06$ & $3.41\pm0.09$ & $2.75\pm0.80$ & $3.79\pm0.10$ & $2.72\pm0.28$ \\ 
UGC05111 & $10.82\pm0.12$ & $0.59\pm0.11$ & $8.53\pm0.05$ & $6.43\pm0.06$ & $4.19\pm0.78$ & $3.41\pm0.17$ & $5.36\pm0.39$ \\ 
UGC05244 & $9.67\pm0.15$ & $-0.41\pm0.09$ & $8.35\pm0.05$ & \nodata & \nodata & \nodata & \nodata \\ 
UGC05359 & $10.86\pm0.13$ & $0.30\pm0.21$ & $8.47\pm0.11$ & $8.47\pm0.05$ & $9.21\pm2.78$ & $4.24\pm0.06$ & $8.08\pm3.49$ \\ 
UGC05498NED01 & $10.76\pm0.11$ & $-0.04\pm0.13$ & $8.44\pm0.08$ & \nodata & \nodata & \nodata & \nodata \\ 
UGC05598 & $10.40\pm0.12$ & $0.15\pm0.09$ & $8.45\pm0.05$ & $2.99\pm0.11$ & $2.68\pm0.72$ & $3.09\pm0.21$ & $4.59\pm0.51$ \\ 
UGC06312 & $10.93\pm0.12$ & $0.02\pm0.23$ & \nodata & \nodata & \nodata & $3.31\pm0.04$ & $5.16\pm0.50$ \\ 
UGC07012 & $9.90\pm0.11$ & $-0.10\pm0.06$ & $8.40\pm0.05$ & \nodata & \nodata & \nodata & $2.41\pm0.81$ \\ 
UGC08107 & $11.20\pm0.10$ & $0.89\pm0.08$ & $8.45\pm0.06$ & $5.22\pm0.05$ & $3.61\pm0.51$ & $6.23\pm0.23$ & $7.77\pm0.46$ \\ 
UGC08250 & $10.06\pm0.15$ & $-0.17\pm0.11$ & $8.39\pm0.05$ & \nodata & \nodata & \nodata & \nodata \\ 
UGC08267 & $10.78\pm0.13$ & $0.48\pm0.15$ & $8.55\pm0.04$ & $4.27\pm0.08$ & $3.02\pm0.29$ & $3.09\pm0.21$ & $3.29\pm0.41$ \\ 
UGC09067 & $10.96\pm0.12$ & $0.70\pm0.07$ & $8.54\pm0.04$ & $4.66\pm0.08$ & $3.15\pm0.90$ & $2.97\pm0.05$ & $3.99\pm0.74$ \\ 
UGC09476 & $10.43\pm0.11$ & $0.05\pm0.06$ & $8.53\pm0.04$ & $3.60\pm0.10$ & \nodata & $2.93\pm0.20$ & $6.63\pm1.90$ \\ 
UGC09537 & $11.23\pm0.08$ & $0.55\pm0.26$ & $8.47\pm0.07$ & $7.60\pm0.05$ & $8.28\pm3.08$ & $4.66\pm0.18$ & \nodata \\ 
UGC09542 & $10.53\pm0.13$ & $0.27\pm0.09$ & $8.49\pm0.05$ & $4.91\pm0.07$ & $5.44\pm2.24$ & $3.45\pm0.10$ & $5.96\pm1.05$ \\ 
UGC09665 & $9.99\pm0.10$ & $-0.11\pm0.07$ & $8.46\pm0.03$ & $3.12\pm0.11$ & $2.38\pm0.82$ & \nodata & $2.50\pm0.36$ \\ 
UGC09759 & $10.02\pm0.10$ & $-0.34\pm0.18$ & $8.39\pm0.08$ & \nodata & \nodata & $1.66\pm0.21$ & $2.68\pm1.46$ \\ 
UGC09873 & $10.21\pm0.10$ & $0.10\pm0.09$ & $8.46\pm0.05$ & $2.88\pm0.10$ & $2.86\pm0.94$ & $3.69\pm0.14$ & $2.97\pm0.27$ \\ 
UGC09892 & $10.48\pm0.10$ & $-0.03\pm0.08$ & $8.48\pm0.05$ & $5.06\pm0.07$ & $5.72\pm2.05$ & $2.90\pm0.12$ & $4.78\pm0.61$ \\ 
UGC09919 & $9.74\pm0.08$ & $-0.33\pm0.09$ & $8.40\pm0.06$ & \nodata & \nodata & $2.41\pm0.24$ & $2.04\pm0.32$ \\ 
UGC10043 & $9.68\pm0.09$ & $-0.62\pm0.10$ & $8.40\pm0.04$ & \nodata & \nodata & \nodata & \nodata \\ 
UGC10123 & $10.30\pm0.10$ & $0.21\pm0.07$ & $8.54\pm0.03$ & $3.24\pm0.10$ & $2.23\pm0.59$ & $1.62\pm0.11$ & $2.19\pm0.20$ \\ 
UGC10205 & $11.08\pm0.10$ & $0.38\pm0.20$ & $8.49\pm0.04$ & $3.57\pm0.08$ & $2.94\pm0.84$ & $3.12\pm0.09$ & $2.01\pm0.06$ \\ 
UGC10331 & $10.27\pm0.10$ & $0.54\pm0.05$ & $8.38\pm0.03$ & \nodata & \nodata & $3.53\pm0.21$ & $3.76\pm0.18$ \\ 
UGC10380 & $10.92\pm0.11$ & $0.68\pm0.22$ & $8.52\pm0.07$ & $4.32\pm0.07$ & $3.33\pm0.57$ & $4.33\pm0.11$ & $6.63\pm1.39$ \\ 
UGC10384 & $10.33\pm0.14$ & $0.65\pm0.06$ & $8.50\pm0.05$ & $2.93\pm0.12$ & $1.77\pm0.29$ & $1.53\pm0.10$ & $1.84\pm0.16$ \\ 
UGC10710 & $10.92\pm0.09$ & $0.50\pm0.10$ & $8.52\pm0.05$ & $5.38\pm0.06$ & $4.39\pm0.96$ & $5.15\pm0.42$ & $4.62\pm0.55$ \\ 
\enddata
\tablenotetext{a}{Stellar mass in \msun, Salpeter IMF}
\tablenotetext{b}{Total SFR from H$\alpha$ corrected by Balmer-decrement inferred extinction}
\tablenotetext{c}{Metallicity as 12+Log(O/H) at the effective radius}
\end{deluxetable*}

\newpage

\appendix

\section{Additional Multi-panel Images}

\begin{figure*}[t] 
\begin{center}
\includegraphics[width=\textwidth]{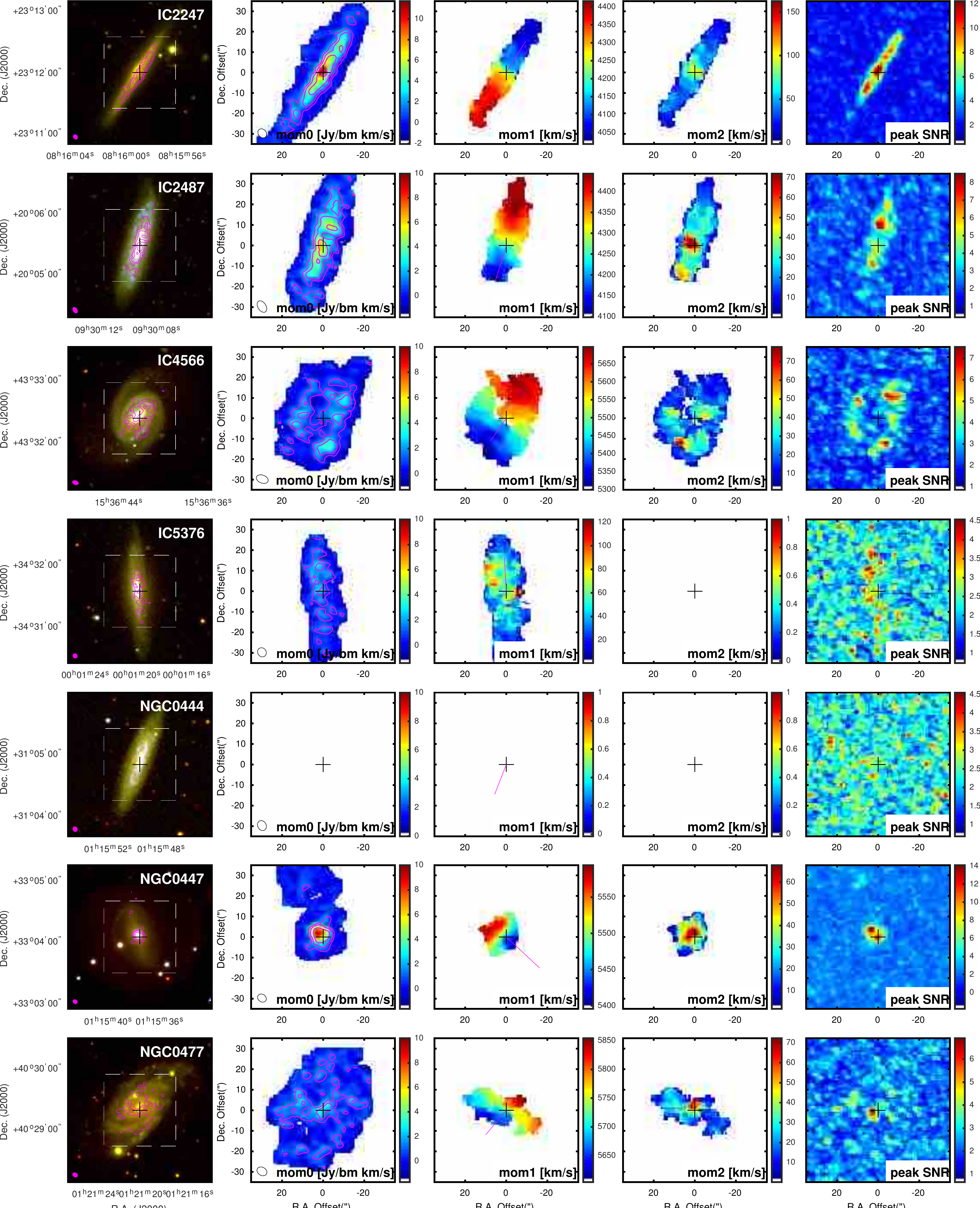}
\end{center}
\caption{Images for CARMA EDGE galaxies. See caption in Figure \ref{fig:multipanel}} 
\end{figure*}

\begin{figure*}[t] 
\begin{center}
\includegraphics[width=\textwidth]{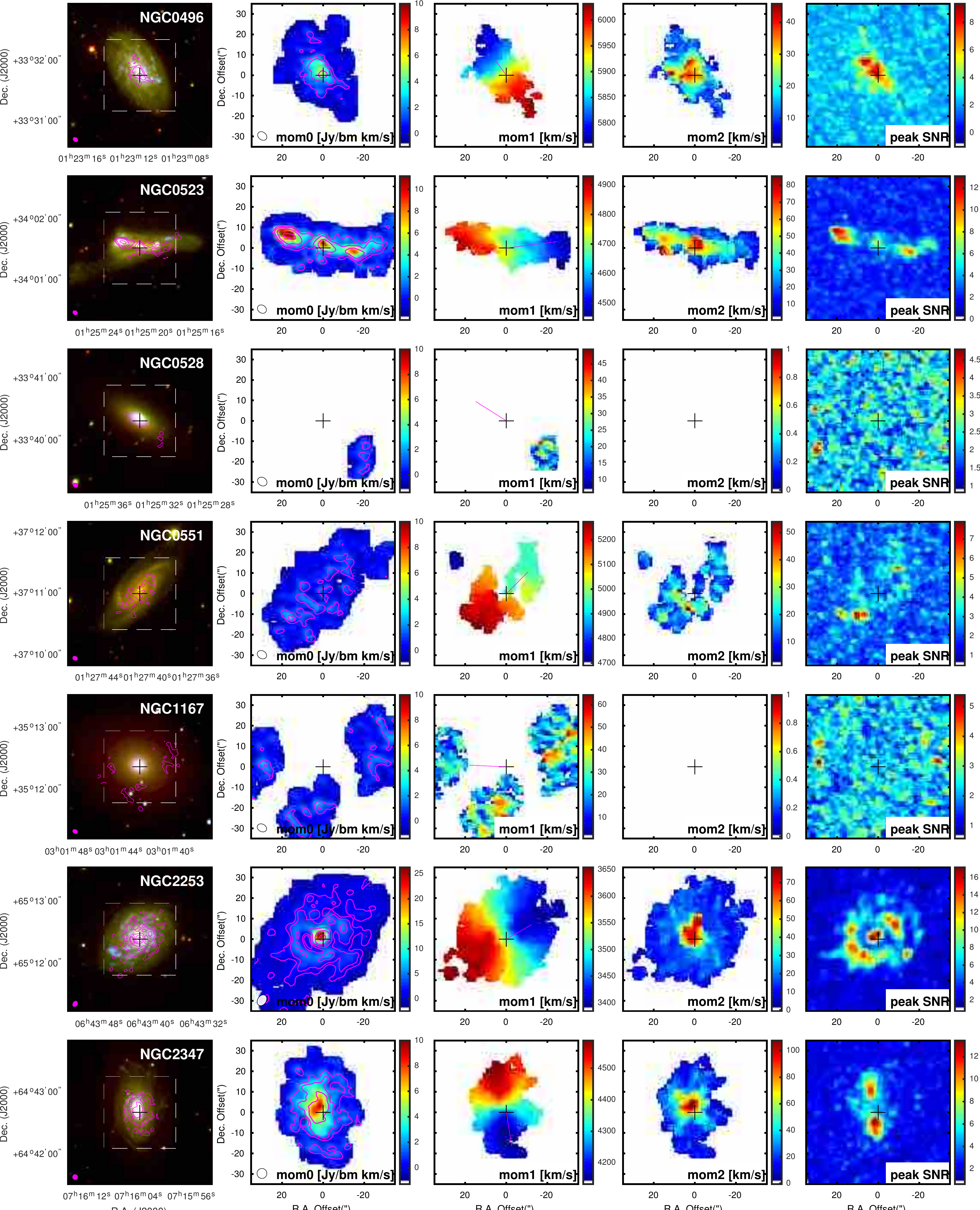}
\end{center}
\caption{Images for CARMA EDGE galaxies. See caption in Figure \ref{fig:multipanel}} 
\end{figure*}

\begin{figure*}[t] 
\begin{center}
\includegraphics[width=\textwidth]{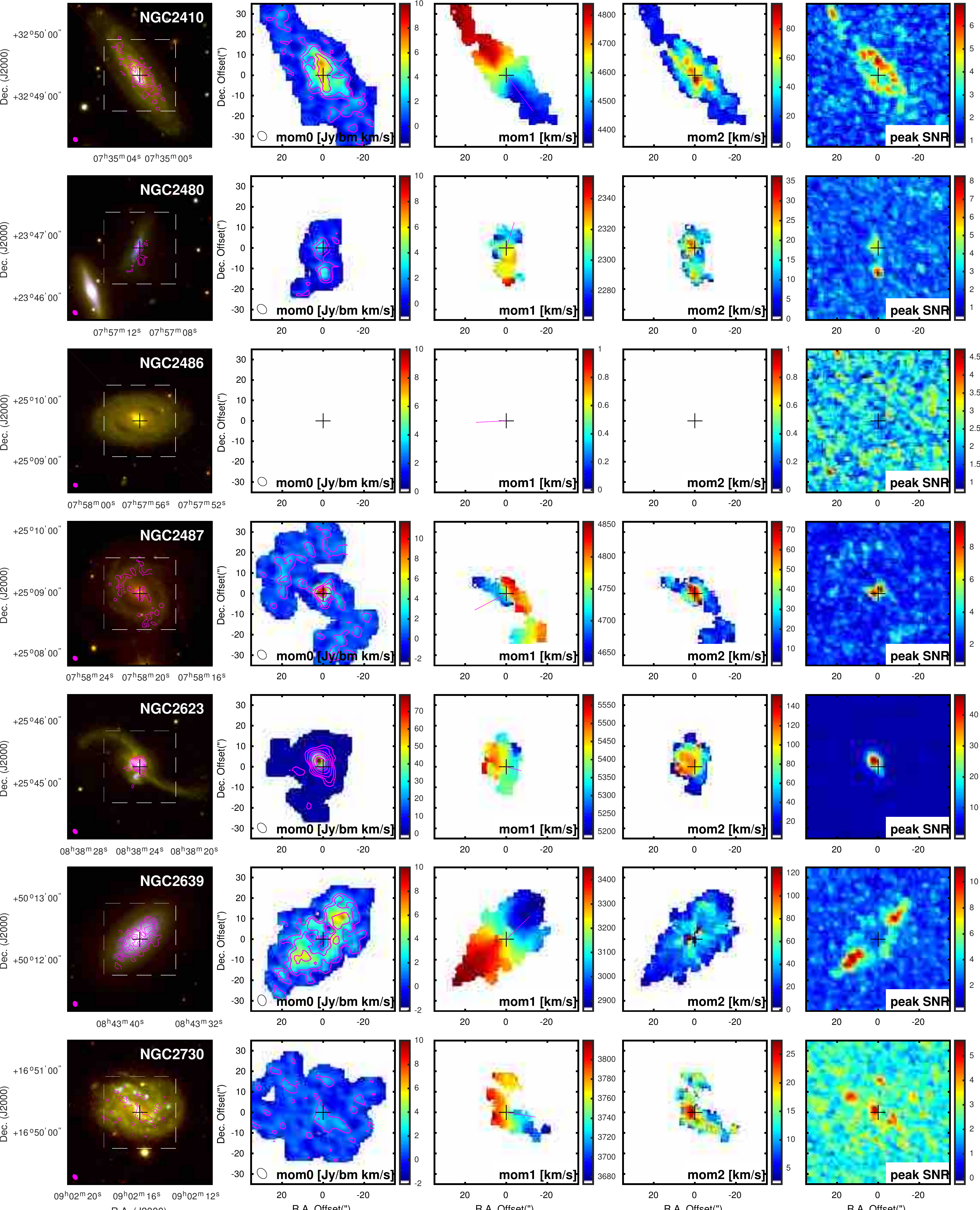}
\end{center}
\caption{Images for CARMA EDGE galaxies. See caption in Figure \ref{fig:multipanel}} 
\end{figure*}

\begin{figure*}[t] 
\begin{center}
\includegraphics[width=\textwidth]{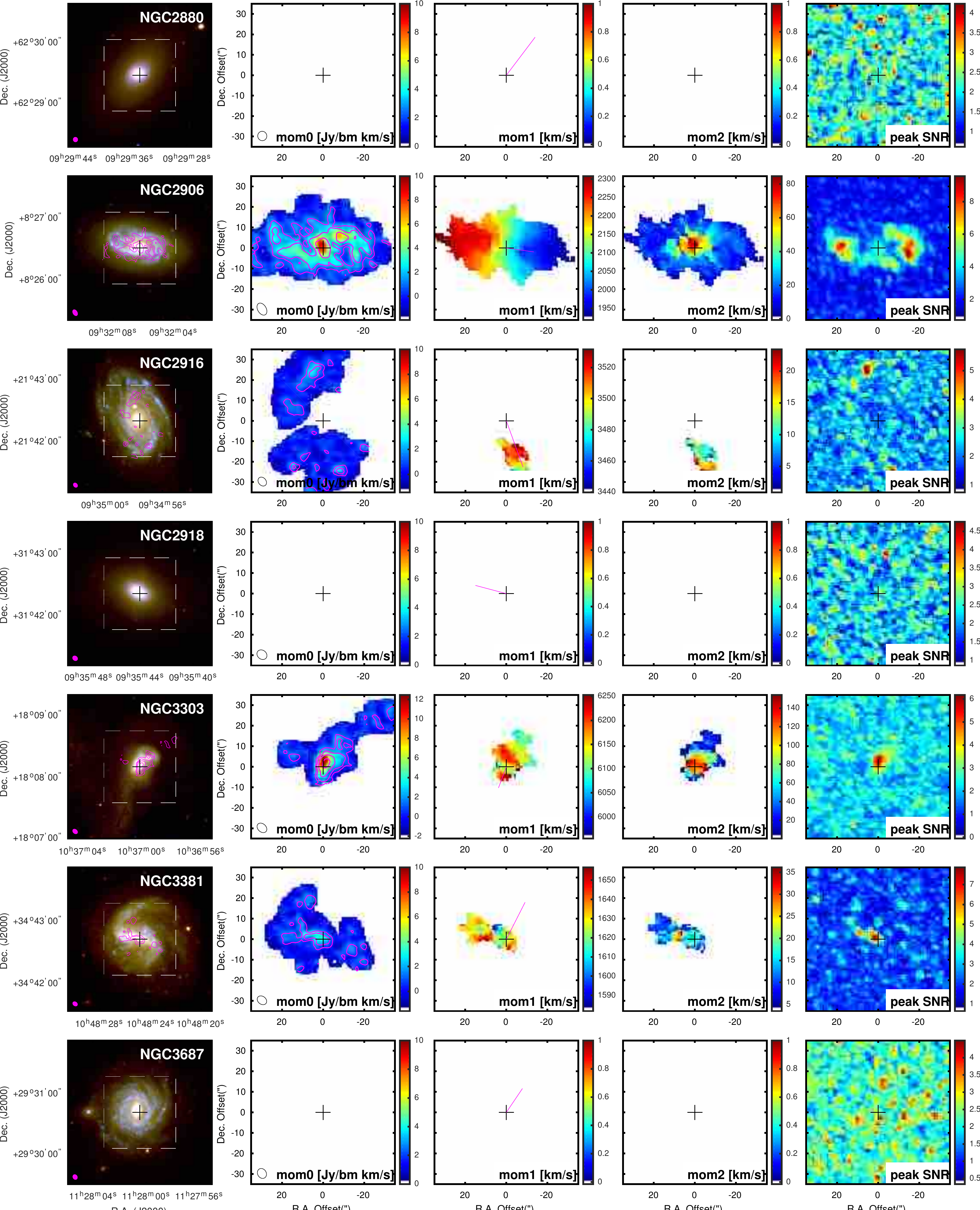}
\end{center}
\caption{Images for CARMA EDGE galaxies. See caption in Figure \ref{fig:multipanel}} 
\end{figure*}

\begin{figure*}[t] 
\begin{center}
\includegraphics[width=\textwidth]{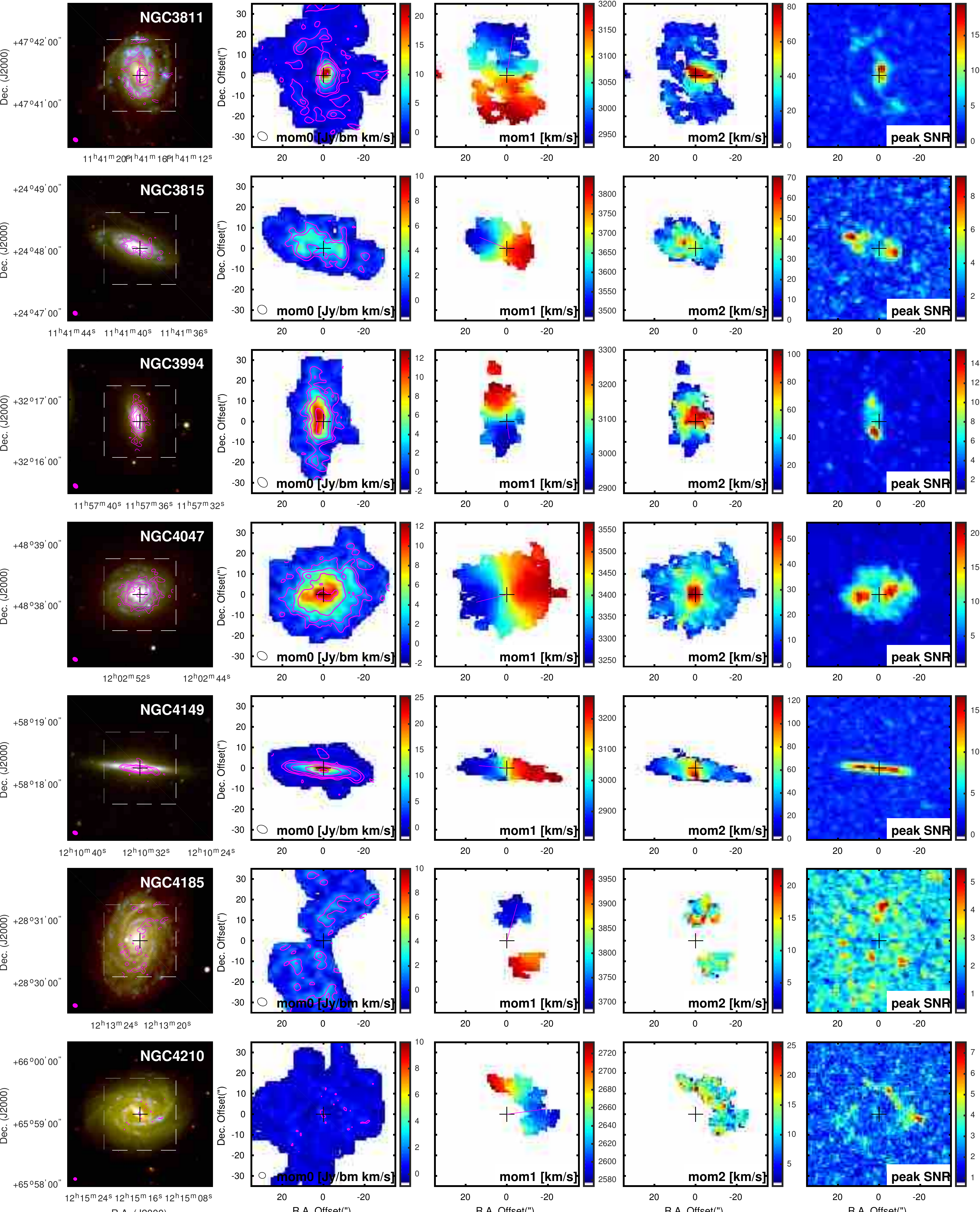}
\end{center}
\caption{Images for CARMA EDGE galaxies. See caption in Figure \ref{fig:multipanel}} 
\end{figure*}

\begin{figure*}[t] 
\begin{center}
\includegraphics[width=\textwidth]{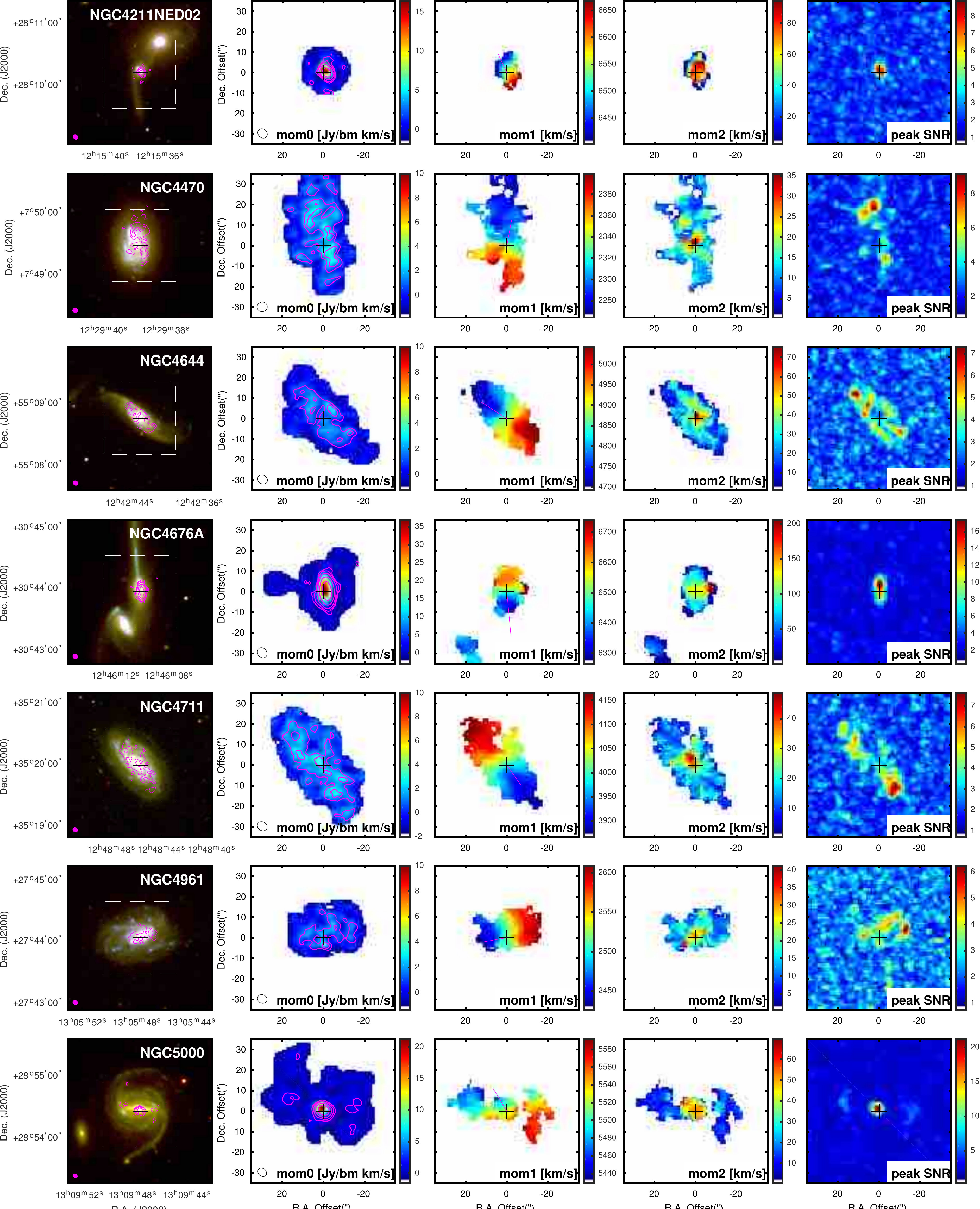}
\end{center}
\caption{Images for CARMA EDGE galaxies. See caption in Figure \ref{fig:multipanel}} 
\end{figure*}

\begin{figure*}[t] 
\begin{center}
\includegraphics[width=\textwidth]{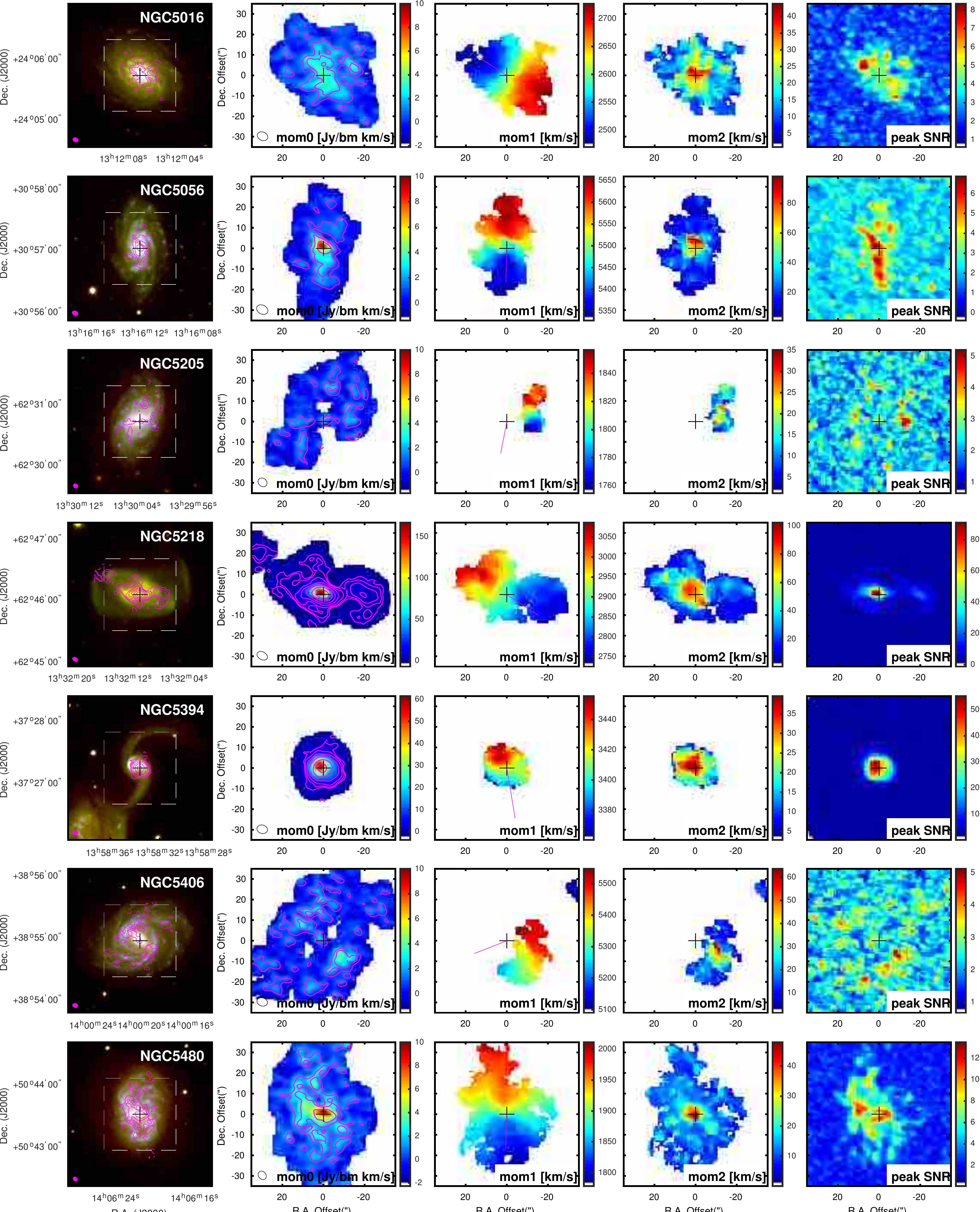}
\end{center}
\caption{Images for CARMA EDGE galaxies. See caption in Figure \ref{fig:multipanel}} 
\end{figure*}

\begin{figure*}[t] 
\begin{center}
\includegraphics[width=\textwidth]{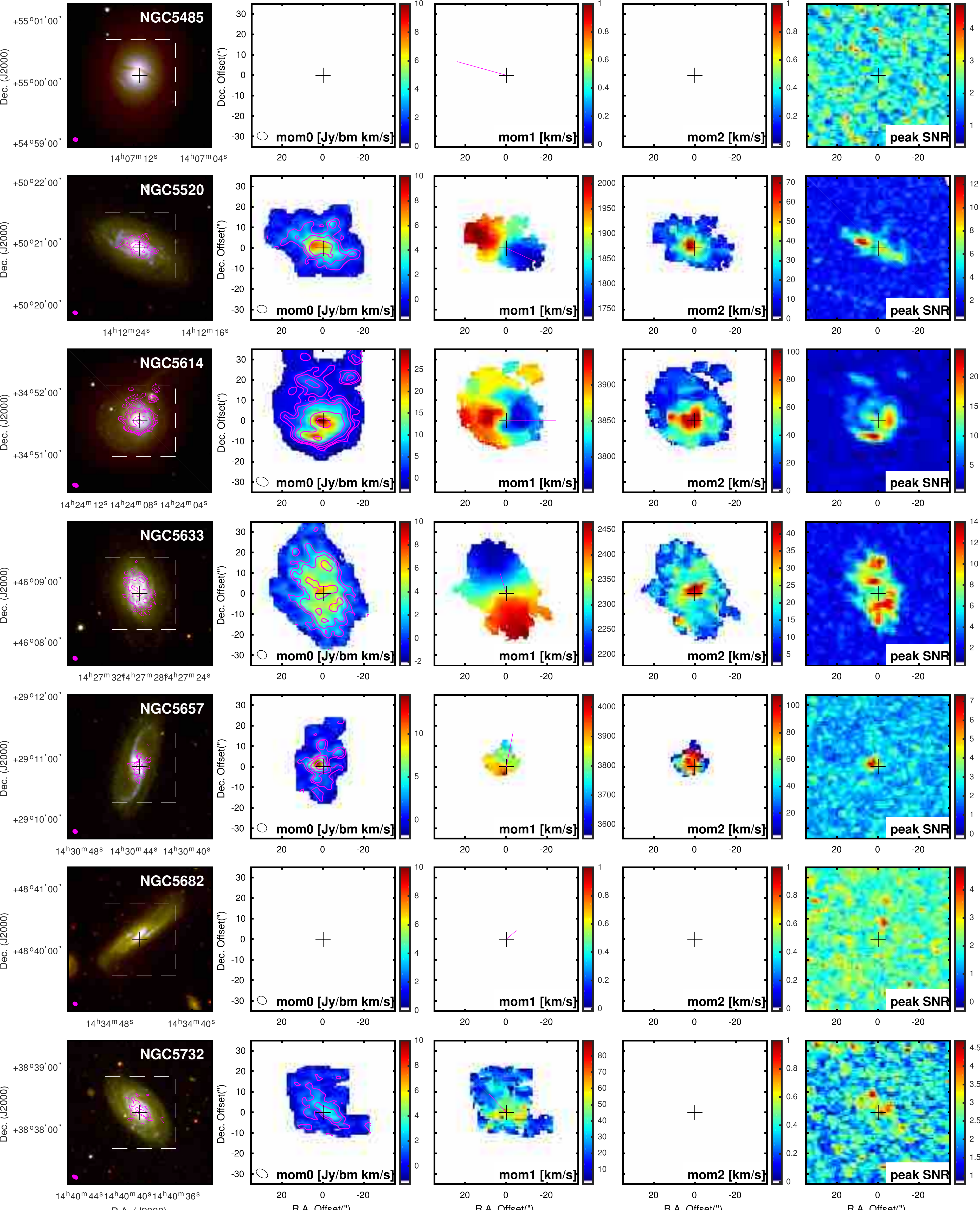}
\end{center}
\caption{Images for CARMA EDGE galaxies. See caption in Figure \ref{fig:multipanel}} 
\end{figure*}

\begin{figure*}[t] 
\begin{center}
\includegraphics[width=\textwidth]{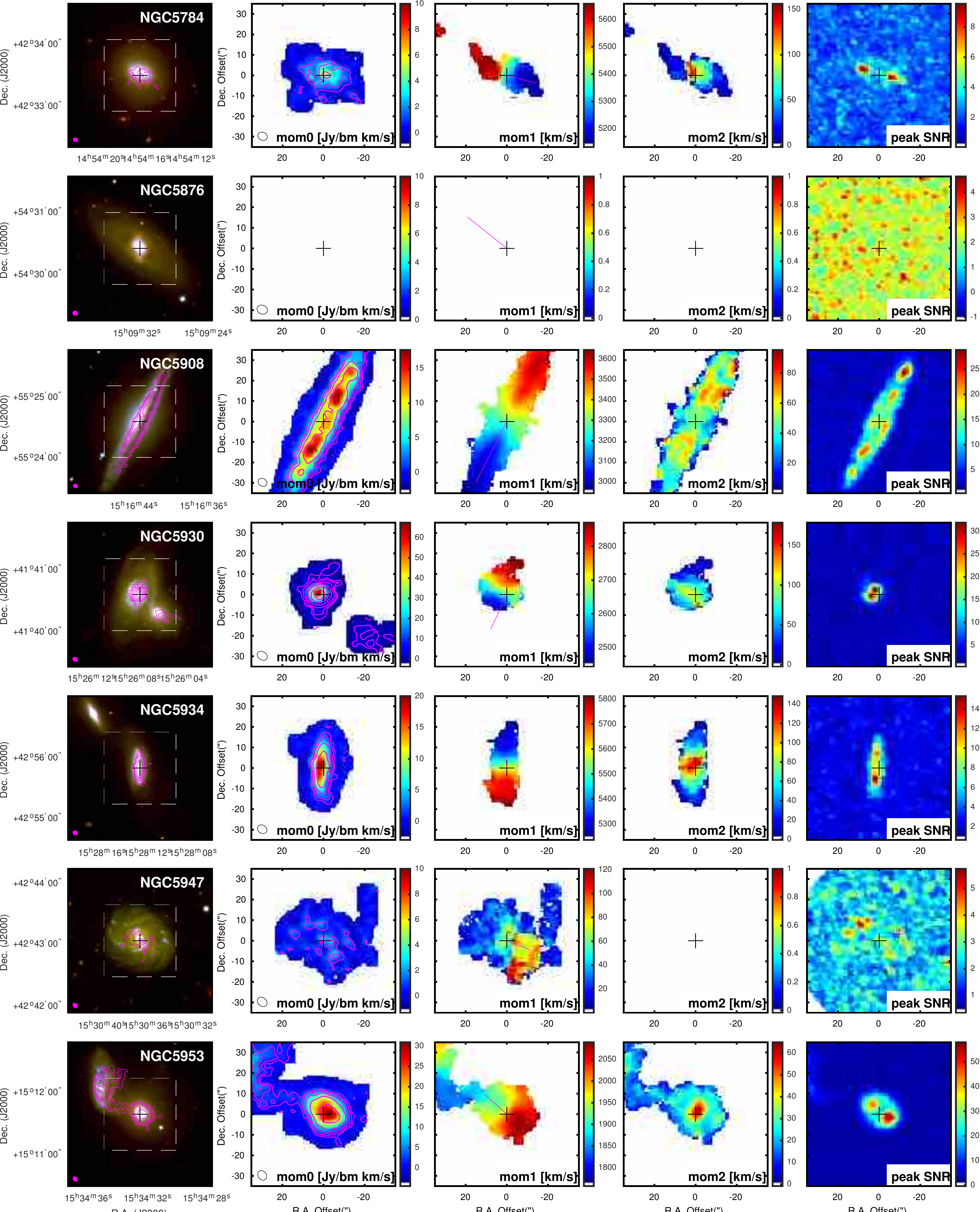}
\end{center}
\caption{Images for CARMA EDGE galaxies. See caption in Figure \ref{fig:multipanel}} 
\end{figure*}

\begin{figure*}[t] 
\begin{center}
\includegraphics[width=\textwidth]{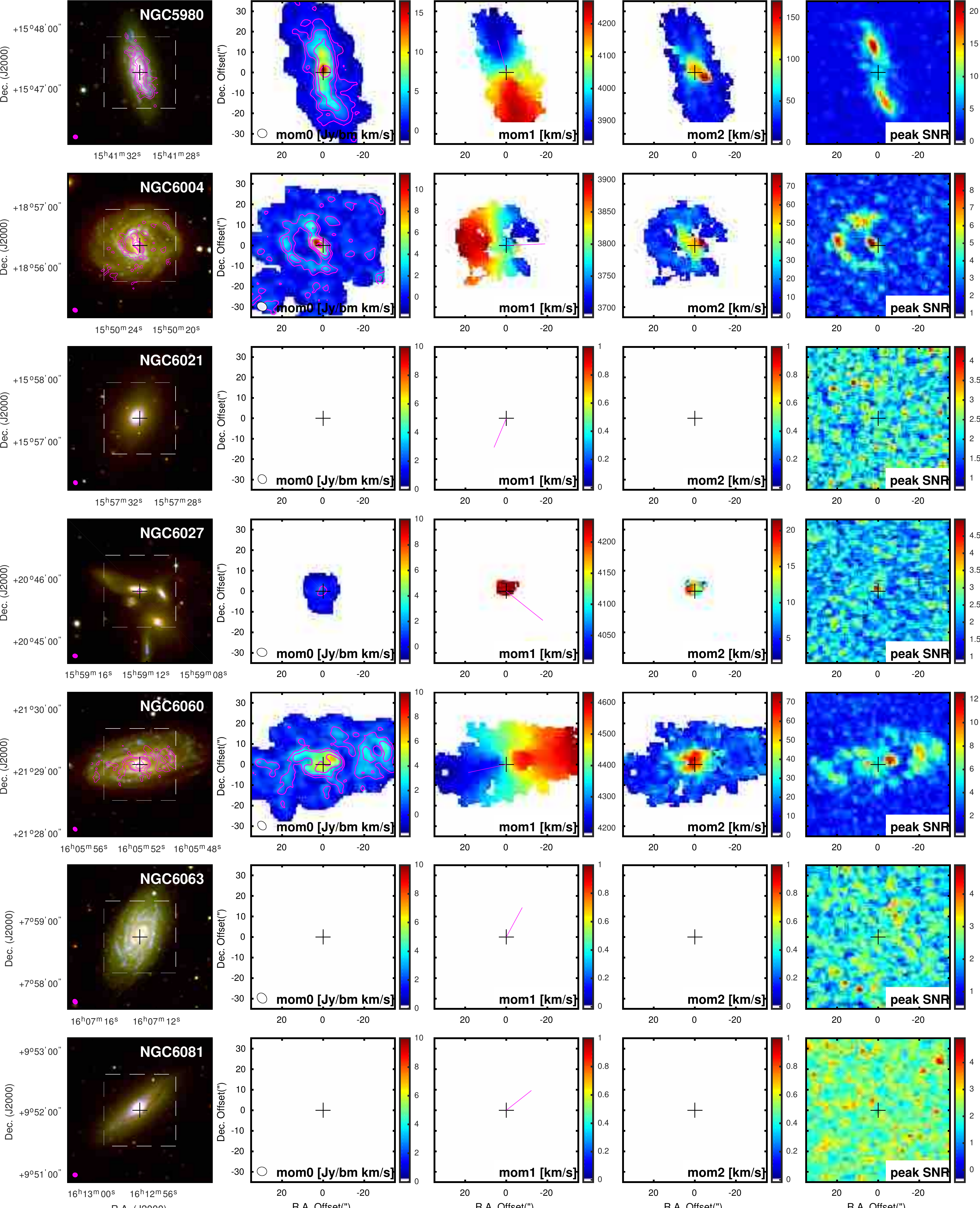}
\end{center}
\caption{Images for CARMA EDGE galaxies. See caption in Figure \ref{fig:multipanel}} 
\end{figure*}

\begin{figure*}[t] 
\begin{center}
\includegraphics[width=\textwidth]{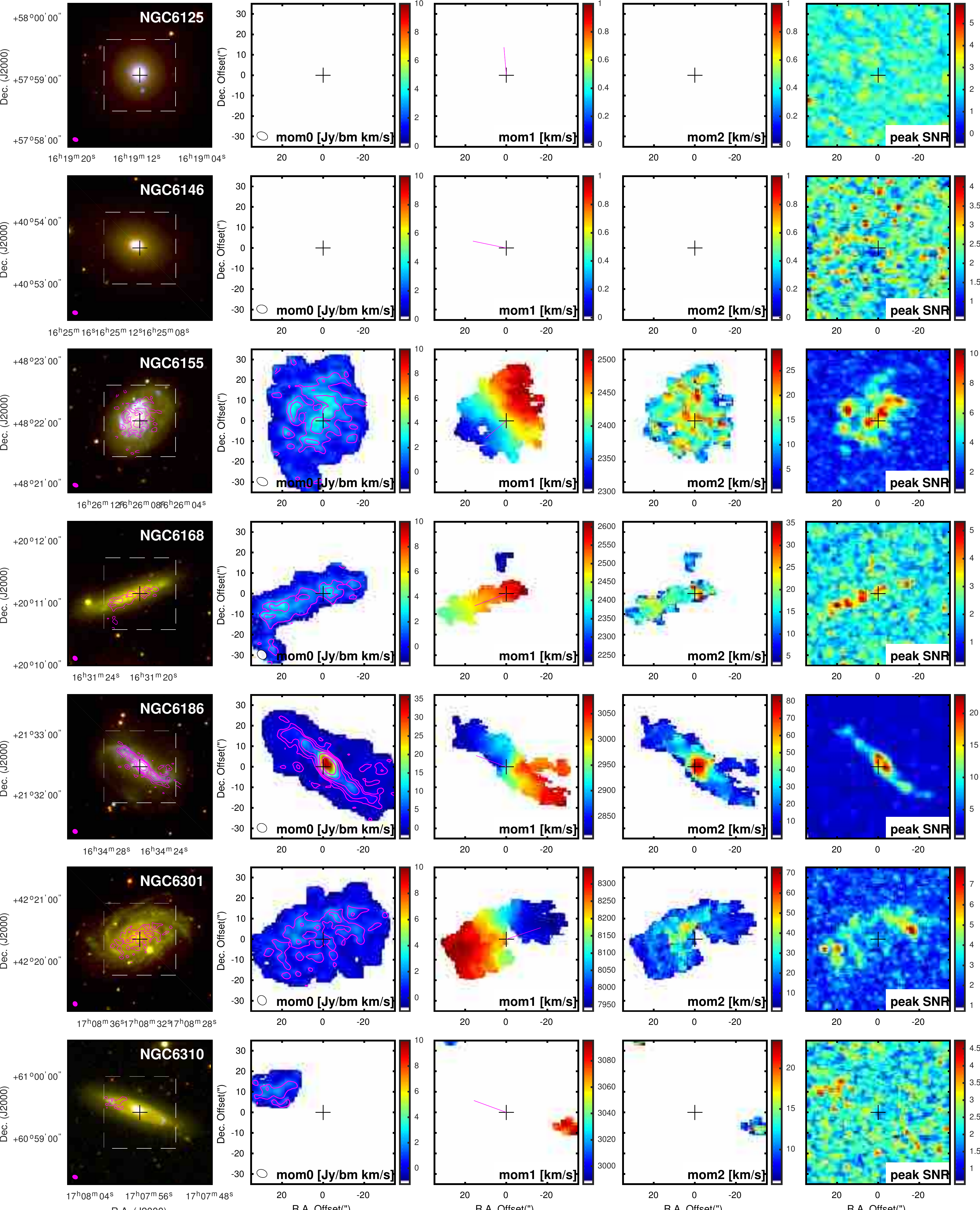}
\end{center}
\caption{Images for CARMA EDGE galaxies. See caption in Figure \ref{fig:multipanel}} 
\end{figure*}

\begin{figure*}[t] 
\begin{center}
\includegraphics[width=\textwidth]{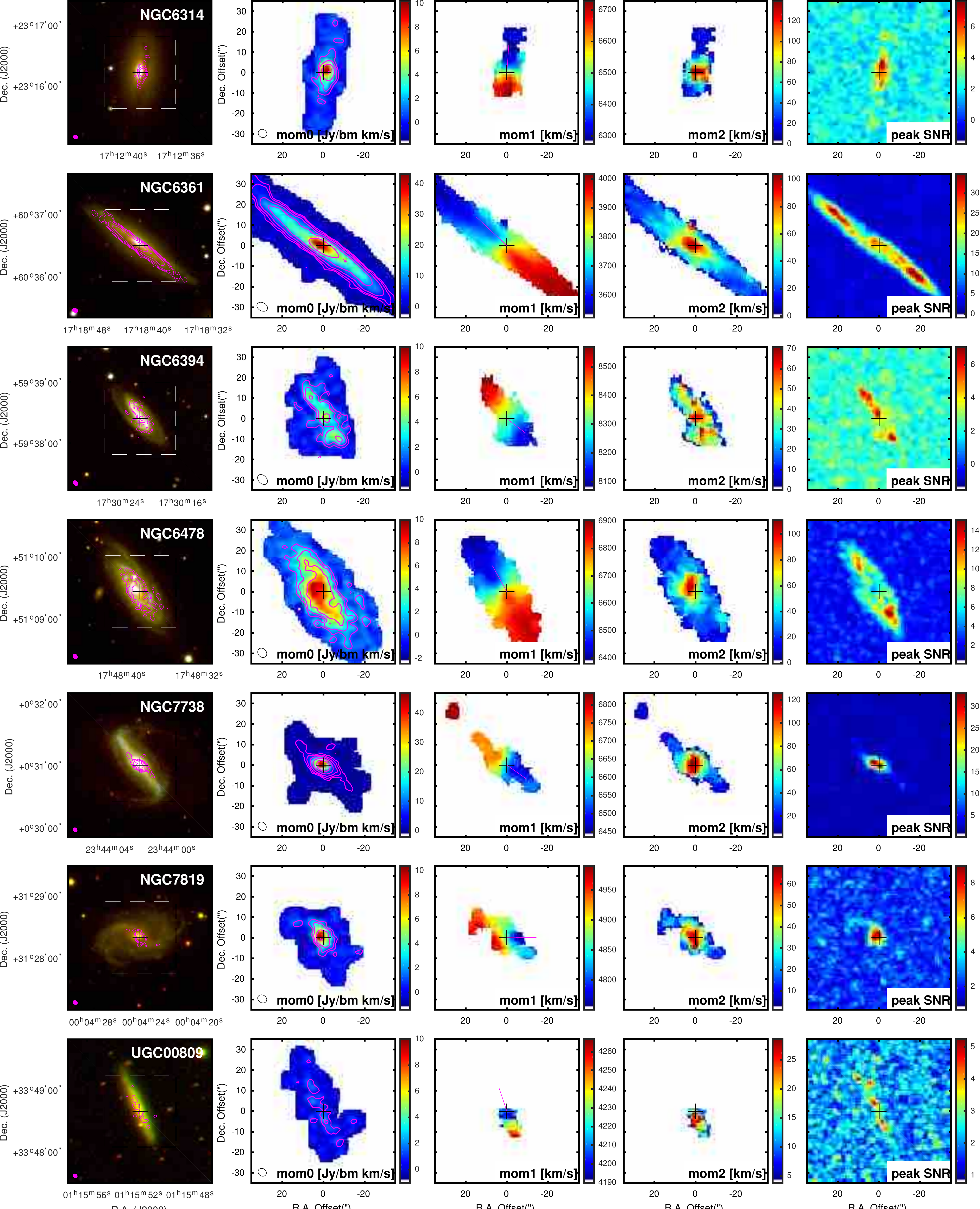}
\end{center}
\caption{Images for CARMA EDGE galaxies. See caption in Figure \ref{fig:multipanel}} 
\end{figure*}

\begin{figure*}[t] 
\begin{center}
\includegraphics[width=\textwidth]{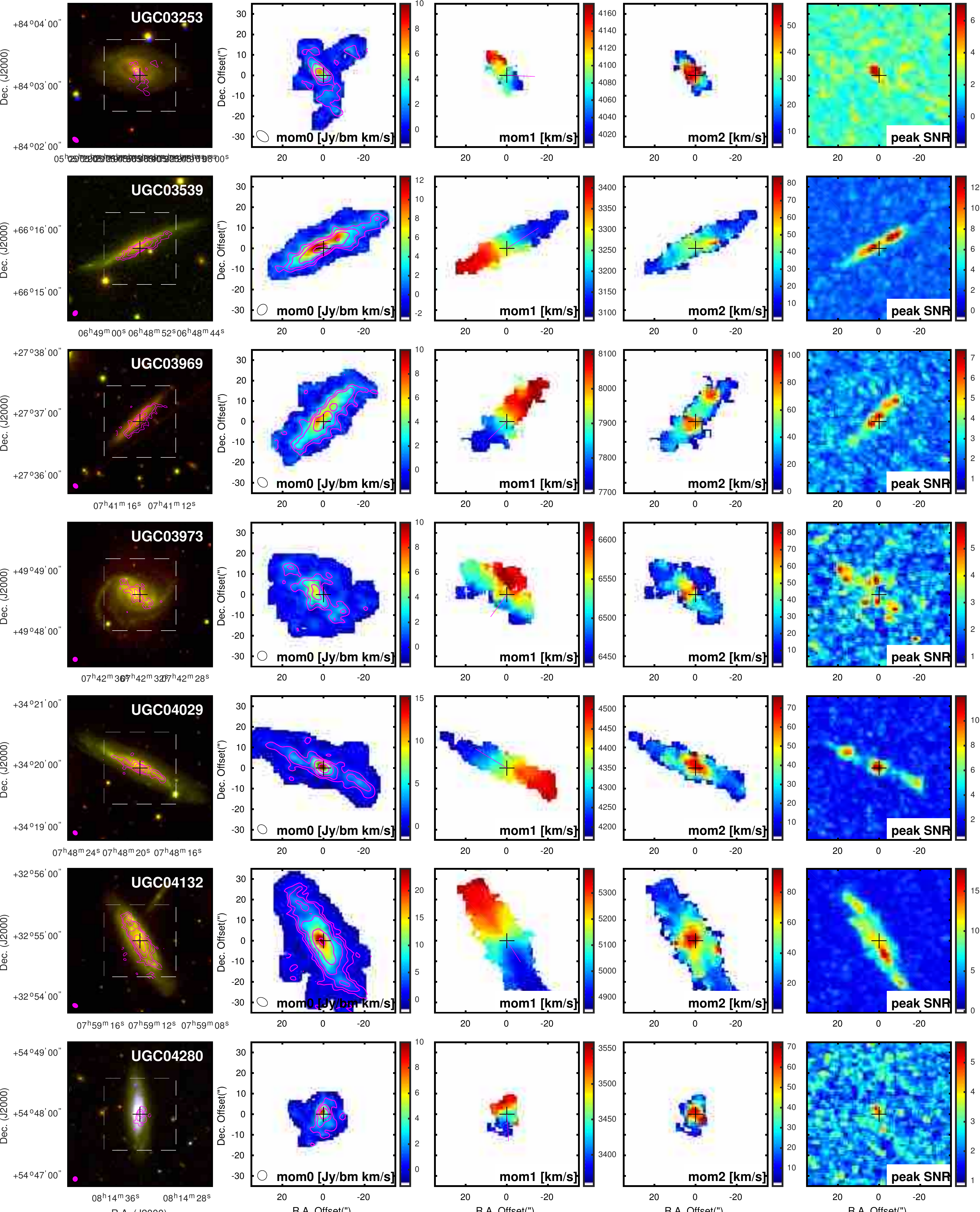}
\end{center}
\caption{Images for CARMA EDGE galaxies. See caption in Figure \ref{fig:multipanel}} 
\end{figure*}

\begin{figure*}[t] 
\begin{center}
\includegraphics[width=\textwidth]{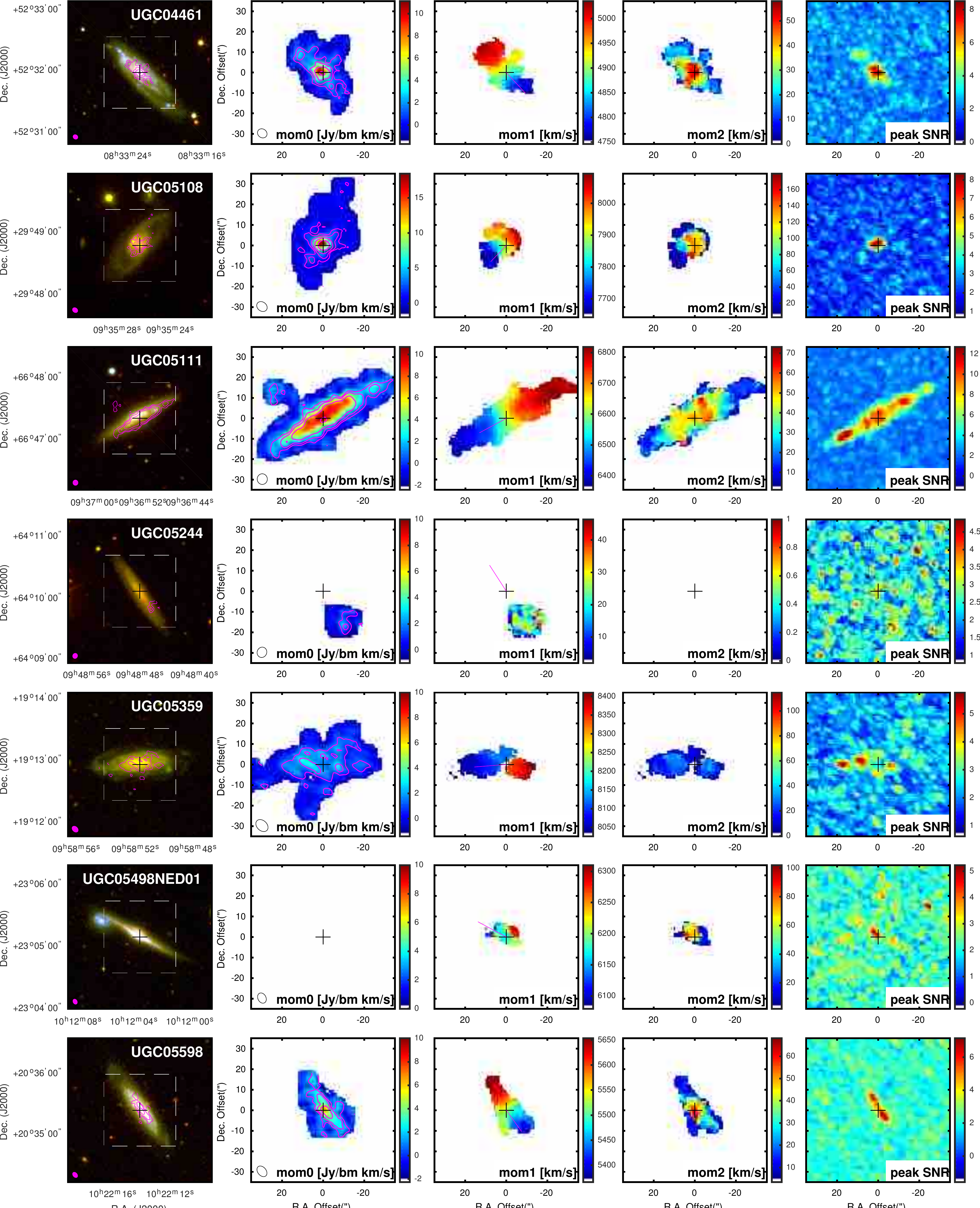}
\end{center}
\caption{Images for CARMA EDGE galaxies. See caption in Figure \ref{fig:multipanel}} 
\end{figure*}

\begin{figure*}[t] 
\begin{center}
\includegraphics[width=\textwidth]{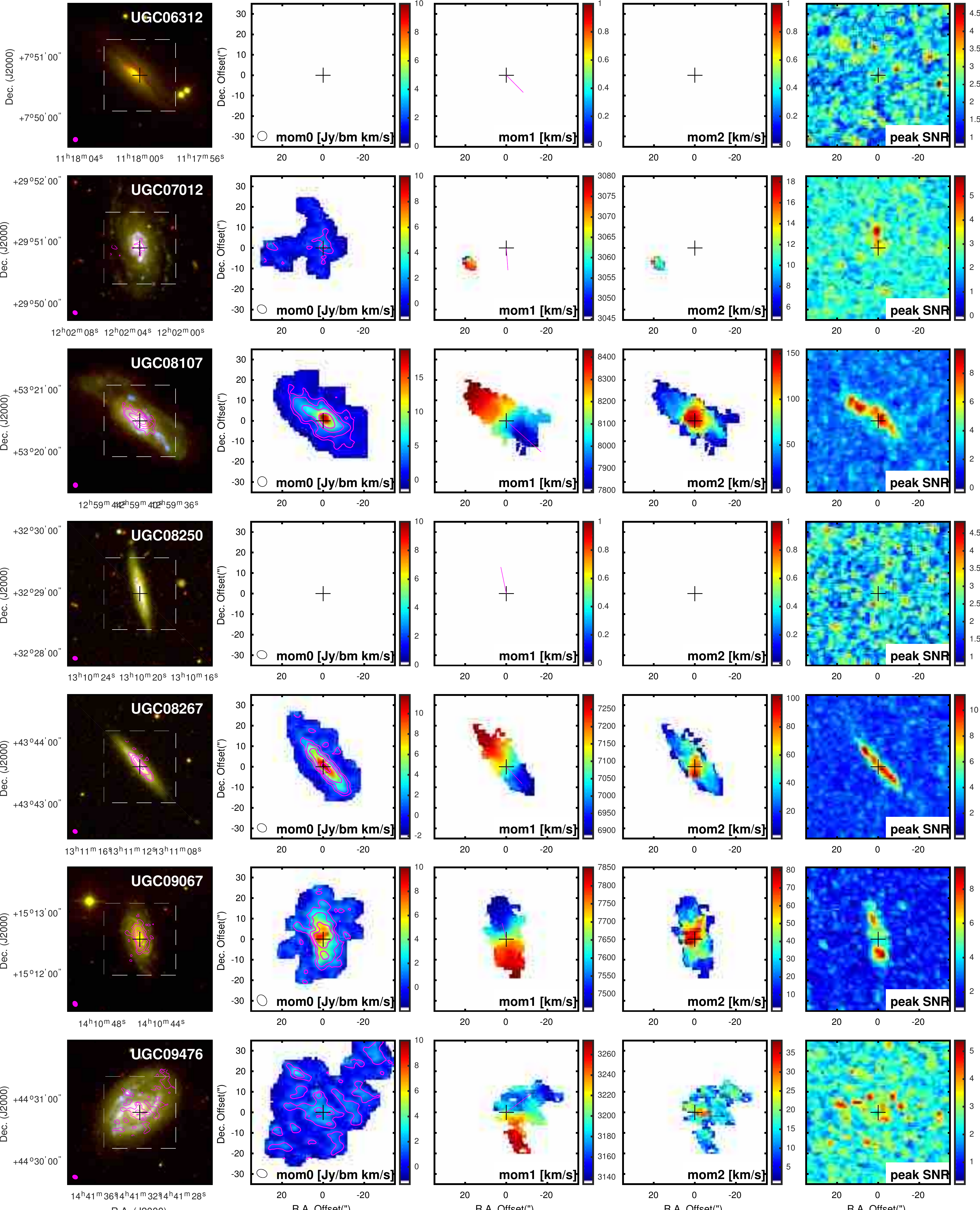}
\end{center}
\caption{Images for CARMA EDGE galaxies. See caption in Figure \ref{fig:multipanel}} 
\end{figure*}

\begin{figure*}[t] 
\begin{center}
\includegraphics[width=\textwidth]{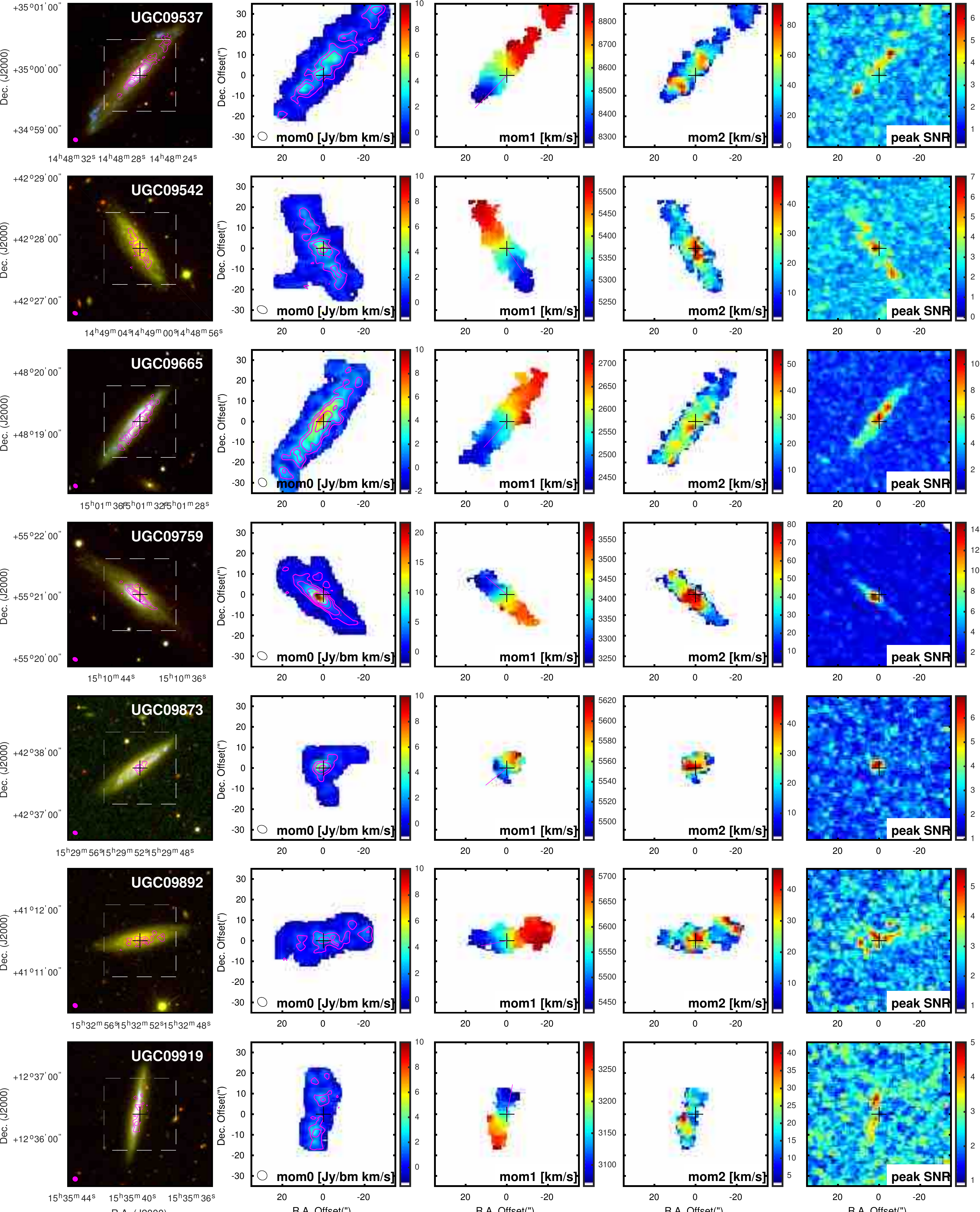}
\end{center}
\caption{Images for CARMA EDGE galaxies. See caption in Figure \ref{fig:multipanel}} 
\end{figure*}

\begin{figure*}[t] 
\begin{center}
\includegraphics[width=\textwidth]{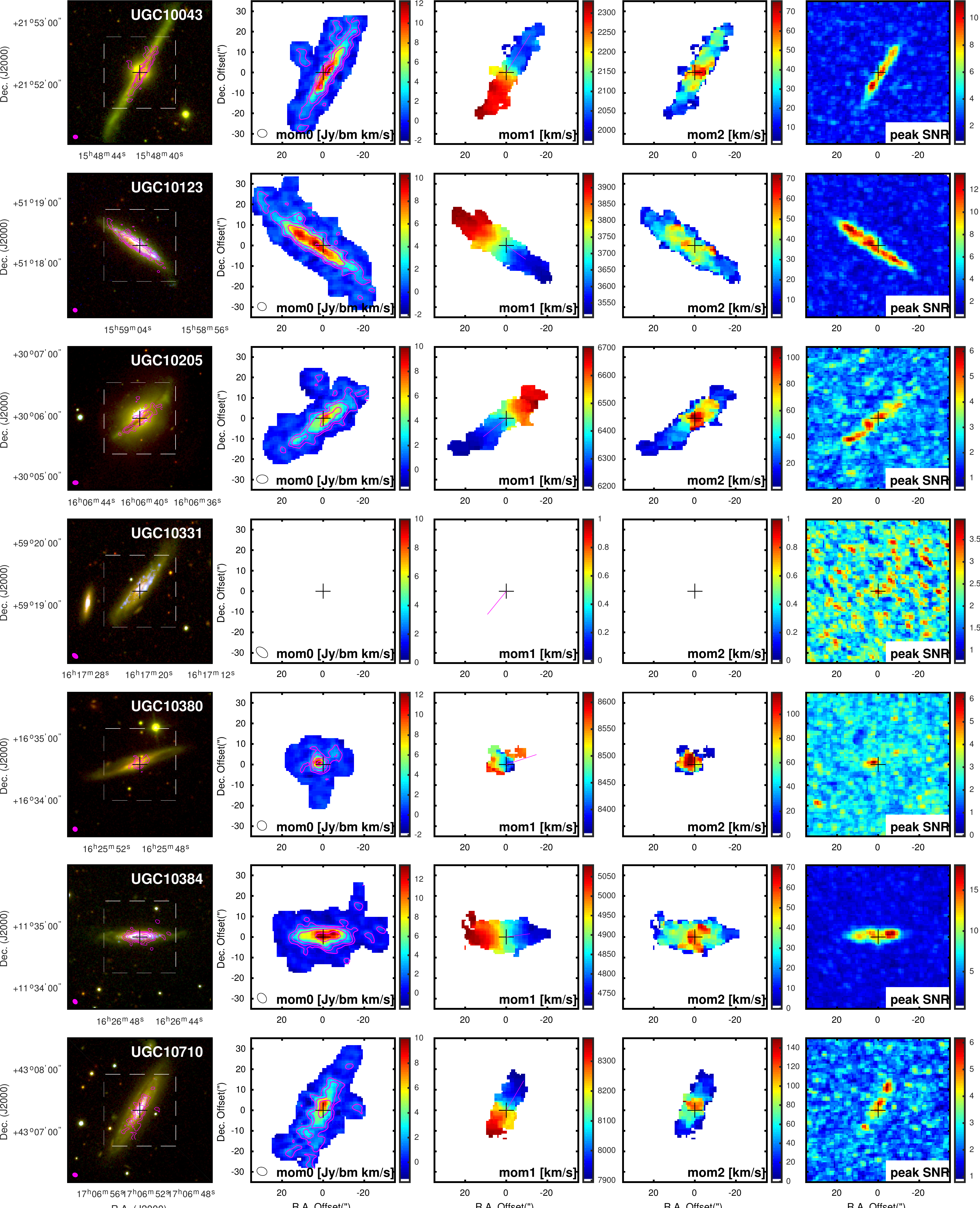}
\end{center}
\caption{Images for CARMA EDGE galaxies. See caption in Figure \ref{fig:multipanel}} 
\end{figure*}

\clearpage

\bibliographystyle{apj}
\bibliography{bibtw,bolatto,sfsanchez}
\end{document}